\documentclass[
10pt,
pra,
twocolumn,
amsmath,
amssymb,
amsfonts,
aps,
superscriptaddress
]{revtex4-2}

\usepackage{graphicx}
\usepackage{dcolumn}
\usepackage{bm}
\usepackage[colorlinks=true,linkcolor=blue,citecolor=blue,urlcolor=blue]{hyperref}
\usepackage{braket}
\usepackage{comment}
\usepackage{amsthm}

\theoremstyle{definition}
\usepackage{enumitem}

\usepackage{tikz}
\usetikzlibrary{quantikz2}
\usetikzlibrary{calc}
\usepackage{booktabs}
\usepackage{multirow}
\usepackage{nicematrix}
\usepackage{orcidlink}
\usepackage{dsfont}
\usepackage{subcaption}
\newtheorem{theorem}{Theorem}
\usepackage{makecell}
\usepackage{ragged2e}                                
\DeclareCaptionJustification{justified}{\justifying}
\ifdefined\pdfobjcompresslevel \pdfobjcompresslevel=0 \fi

\begin{document}

\title{Development and experimental implementation of noise characterisation techniques on a superconducting quantum processor}

\author{Nelson~M.~M. Fernández\,\orcidlink{0009-0006-4001-9429}}
\altaffiliation{These authors contributed equally to this work.}
\affiliation{Algorithmiq Oy, Kasarmikatu 42, FI-00130 Helsinki, Finland}

\author{Francesco Grieco\,\orcidlink{https://orcid.org/0009-0001-3051-6986}}
\altaffiliation{These authors contributed equally to this work.}
\affiliation{Algorithmiq Oy, Kasarmikatu 42, FI-00130 Helsinki, Finland}
\affiliation{Department of Physics, University of Helsinki, P.O. Box 43, FI-00014 Helsinki, Finland}

\author{Hetta Vappula\,\orcidlink{0009-0006-6732-6744}}
\altaffiliation{These authors contributed equally to this work.}
\affiliation{Algorithmiq Oy, Kasarmikatu 42, FI-00130 Helsinki, Finland}
\affiliation{Department of Physics, University of Helsinki, P.O. Box 43, FI-00014 Helsinki, Finland}

\author{Stefano Mangini\,\orcidlink{0000-0002-0056-0660}}
\affiliation{Algorithmiq Oy, Kasarmikatu 42, FI-00130 Helsinki, Finland}
\affiliation{Department of Physics, University of Helsinki, P.O. Box 43, FI-00014 Helsinki, Finland}

\author{Boris Sokolov}
\affiliation{Algorithmiq Oy, Kasarmikatu 42, FI-00130 Helsinki, Finland}

\author{Elsi-Mari Borrelli}
\affiliation{Algorithmiq Oy, Kasarmikatu 42, FI-00130 Helsinki, Finland}

\author{Matteo~A.~C. Rossi\,\orcidlink{0000-0003-4665-9284}}
\affiliation{Algorithmiq Oy, Kasarmikatu 42, FI-00130 Helsinki, Finland}

\author{Sergey~N. Filippov\,\orcidlink{0000-0001-6414-2137}}
\altaffiliation{These authors contributed equally to this work.}
\affiliation{Algorithmiq Oy, Kasarmikatu 42, FI-00130 Helsinki, Finland}

\author{N. Walter Talarico\,\orcidlink{0000-0001-6617-7798}}
\altaffiliation{These authors contributed equally to this work.}
\affiliation{Algorithmiq Oy, Kasarmikatu 42, FI-00130 Helsinki, Finland}

\begin{abstract}
Noise characterisation not only enables error mitigation but also provides valuable information about the intrinsic performance of a quantum device, identifying dominant error contributions and bounding noise stability well beyond the aggregated calibration data. Deploying such methods on a specific hardware topology must combine efficiency and accuracy, assess noise stability, and validate the reconstructed noise model. Here we develop and implement these methods on a superconducting processor with square-lattice topology. First, we revise the cycle-benchmarking protocol and reduce the bias arising from unlearnable degrees of freedom via a regularised cost function. Second, we extend a previously developed tensor-network tomography method to match the device layout and show how its accuracy can be improved by repeated layer execution. Both methods allow approximate sublayer noise models to be extracted from parent ones. Using noise characterisation as a tool, we map the noise structure, track drift, and validate the adequacy of the learnt noise models against experimentally observed noisy estimations in benchmark physical and chemistry circuits. These results inform further improvements in the underlying technology and provide estimates of the size of noise-mitigated circuits that the device can support.
\end{abstract}

\maketitle

\section{\label{section-introduction} Introduction}

Recent years have seen an inflection point in quantum computing: hardware demonstrations in solving scientific problems involving the estimation of target observables have become progressively more challenging for purely classical simulations~\cite{googlequantum2025otoc,alam2025fermionicdynamicstrappedionquantum,king2025beyondclassical,filippov2025ole,hartnett2026fermihubbard,Haghshenas2026,leviatan2026prethermal,barron2026observable} if addressable classically at all within a reasonable runtime~\cite{quantumadvantagetracker2026}. These advances in pre-fault-tolerant quantum computing are enabled by the development of error-suppression and error-mitigation techniques, whose primary goals are to reduce the overall noise during circuit execution, shape the noise structure if necessary, and make allowance for the residual errors~\cite{endo-2018,endo2021hybrid,cai_quantum_2023,filippov2024scalability,zimboras2025myths,li2026practical}. While some error-mitigation techniques are problem-specific and less sensitive to the detailed distribution of errors (e.g., in the echo-type circuits~\cite{googlequantum2025otoc,barron2026observable} and symmetry-preserving circuits~\cite{bonetmonroig2018low,mcclean2020decoding,yoshioka2022generalized,dubovitskii2026discretetime}), many others are universally applicable to all types of quantum circuits (see, e.g., Refs.~\cite{li2017efficient,temme2017error,huggins2021virtual,czarnik2021error,berg_probabilistic_2023,kim2023scalable,kim2023evidence,filippov2023scalable,fischer2026dynamical,bertini2026mitigated}). The latter techniques vary in the level of rigour of their predictions for error-mitigated observable estimates, with the least heuristic ones relying on precise noise characterisation for the main sources of errors in the circuit. 

Noise-aware error-mitigation techniques such as tensor-network error mitigation~\cite{filippov2023scalable,fischer2026dynamical}, zero-noise extrapolation with probabilistic error amplification~\cite{kim2023evidence}, and probabilistic error cancellation~\cite{temme2017error,berg_probabilistic_2023,kim2023scalable} exploit the learnt noise profiles of the noisiest circuit layers to provide error-mitigated values together with estimates of statistical and systematic errors. Although these techniques vary in the complexity of their hardware/post-processing implementation and exhibit noticeably different sampling overheads affecting the statistical errors~\cite{filippov2024scalability}, they are all also subject to systematic errors originating, e.g., from ambiguous noise components~\cite{chen_learnability_2023,chen_disambiguating_2026,cheng2026characterization}, non-Markovian effects~\cite{cheng2026characterization,white2020demonstration,figueroaromero2022towards,white2022characterization,oda2026sparse}, and drift of the noise parameters over time~\cite{klimov2018fluctuations,proctor2020detecting,kim2025error}. Addressing these systematic errors, along with noise characterisation methods, enables one to establish error bounds on error-mitigated observable estimations in general quantum circuits that are as reliable as possible~\cite{filippov2024scalability}. 

The role of noise characterisation in advancing quantum computation is therefore twofold: on one hand, it serves as a prerequisite for efficient error mitigation; on the other hand, it serves as a probing tool for identifying significant error contributions and specific decoherence channels as well as for setting bounds on the noise stability, thus providing important information for further improvements in technology (well beyond periodically updated noise calibration data, which are often presented in aggregated form). In practical applications of noise characterisation for error mitigation, these roles are two sides of the same coin and imply an unavoidable trade-off between the effectiveness of noise-model extraction and model accuracy. For example, noise learning through cycle benchmarking~\cite{berg_probabilistic_2023,kim2023scalable,jaloveckas2023efficient,berg_techniques_2024} is effective and robust to state preparation and measurement errors due to noise amplification, but cannot reliably disambiguate some noise components, whereas noise learning through quantum tomography~\cite{mangini2024tensor,torlai-2023,belkin2026better} is generally more demanding in terms of the measurement-shot budget and additionally requires detector tomography but is able to capture longer-range noise correlations and address noise components unambiguously. For these reasons, the deployment of error-characterisation methods on quantum hardware with a specific topology has to address the problem of combining efficiency and accuracy as well as assessing noise stability and the adequacy of the reconstructed noise model.

In this study, we fill the gap in the experimental analysis of noise in the superconducting quantum processor VTT Q50~\cite{noauthor_vtt_2026} (Espoo, Finland), which has a square-lattice topology. We deploy revised versions of cycle-benchmarking-based noise learning and tensor-network noise characterisation equipped with a number of novel improvements: (i) we reduce the ambiguity in learning the parameters of the sparse Pauli-Lindblad noise model by modifying the optimisation cost function; (ii) we extend tensor-network noise characterisation~\cite{mangini2024tensor} from one- to two-dimensional topology and show how its accuracy can be increased with the help of repeated execution of the noisy layer; (iii) for scenarios prioritising effectiveness over precision, we develop an approximate method for learning the noise of circuit sublayers through a parent layer. We check the adequacy of the learnt noise models in specific applications (many-body dynamics with controllable magic, preparation of quantum-chemical ground states) by comparing the actual noisy signals with those predicted by the noise model. 

We exploit the noise characterisation machinery in the high-accuracy regime to assess noise stability over time, since the drift of noise parameters is responsible for systematic errors in all noise-mitigation protocols. Based on the best performance across the range of qubits and assuming that it is scalable, our findings show that the total estimation error (comprising both statistical and systematic components) in optimally noise-mitigated observables can be below $13\%$ for circuits with $400$ two-qubit entangling gates (depth $L=15$, width $N=50$ for highly nonlocal observables) and $10^8$ measurement shots, provided the optimal error mitigation tool is used~\cite{filippov2024scalability}.

The paper is organised as follows. In Sec.~\ref{section-device}, we review the basic specifications of the superconducting processor used. In Sec.~\ref{section-cb}, we review the theory behind cycle benchmarking as a noise-learning tool, present its modified version with a change in the cost function that results in higher precision of the learnt sparse Pauli–Lindblad noise model, and support the results with numerical simulations. In Sec.~\ref{section-tn}, we review the tensor-network channel tomography as well as extend it to two-dimensional topologies of qubit layouts and repeated applications of the noisy layer. In Sec.~\ref{section-circuits}, we introduce the benchmark circuits and the corresponding layers in one- and two-dimensional topologies that are later used for validating the learnt noise models. In Sec.~\ref{sec:noise analysis}, we report the experimental results on the noise structure and error-rate intensity in typical one- and two-dimensional layers, including the study of sublayer trimming and drift over time. In Sec.~\ref{section-experimental-results}, we validate the experimentally learnt noise models for benchmark layers by comparing the resulting noise-model predictions against measured circuit outcomes. Sec.~\ref{section-conclusions} concludes the paper with a review of the results obtained.

\section{\label{section-device} Basic device specifications}
\begin{figure}
    \centering
    \includegraphics[width=\linewidth]{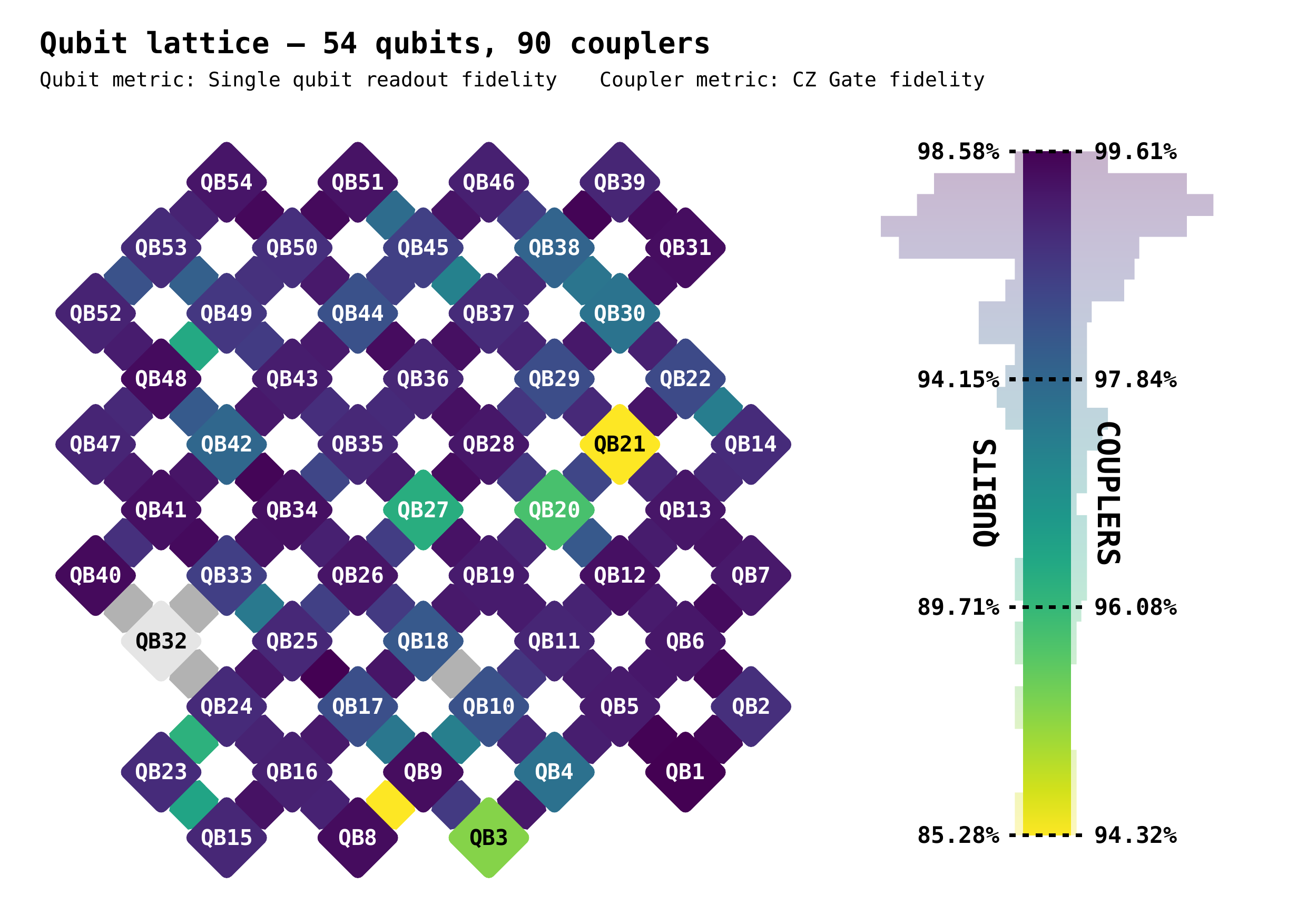}
    \caption{Coupling map and typical readout and CZ-gate fidelities of VTT Q50~\cite{noauthor_vtt_2026}. Although the physical hardware topology labels start at $1$, the software implementation adopts 0-based indexing. The qubit labels are otherwise contiguous, except for the missing QB32. Accordingly, the Qiskit backend reports 53 rather than 54 qubits, accounting for the deactivated qubit.
}
    \label{fig: vtt-coupling-map}
\end{figure}

The noise characterisation experiments presented in this work were performed on VTT Q50 quantum computer, a $53$-qubit superconducting processor co-developed and built by IQM and VTT with a square-lattice topology illustrated in Fig.~\ref{fig: vtt-coupling-map}. VTT Q50 is a transmon-based device featuring flux-tunable qubits and tunable couplers, enabling the implementation of high-fidelity single-qubit and two-qubit gates \cite{noauthor_vtt_2026, marxer_long-distance_2023}.

VTT Q50 device natively supports the following quantum operations: 

(i) Single-qubit phased X-rotation (PRX) gates of the form $R_{\phi}(\theta) = e^{-i(X\cos\phi + Y\sin\phi)\theta/2}$ that implements a rotation of the qubit's Bloch sphere around any axis in the XY-plane, with a flexibility in the choice of microwave control pulses~\cite{noauthor_vtt_2026}.

(ii) Two-qubit controlled-Z (CZ) entangling Clifford gates for every active connection in the coupling map (Fig.~\ref{fig: vtt-coupling-map}). $\mathrm{CZ}$ gate is implemented by applying a flux-pulse to the tunable coupler that mediates the interaction between control and target qubits. Subsequently, a second flux-pulse is applied to one of the qubits, in order to detune its transition frequency to the desired interaction point, thereby enabling the CZ operation \cite{noauthor_vtt_2026, marxer_long-distance_2023}.

(iii) Qubit-wise measurements in the computational Z-basis. 

\par
\begin{figure}
    \centering
    \includegraphics[width=1\linewidth]{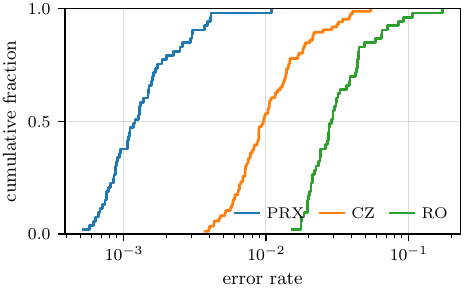}
    \caption{Cumulative distribution of PRX-gate, CZ-gate, and readout (RO) infidelities across VTT Q50~\cite{noauthor_vtt_2026}. Each curve is the empirical distribution over all calibrated qubits, or all calibrated couplers, of the calibration set of 2026-08-21; the medians are $1.2\times10^{-3}$, $9.6\times10^{-3}$ and $2.9\times10^{-2}$ respectively.}
    \label{fig: calibration-data}
\end{figure}

 Fig.~\ref{fig: calibration-data} depicts the typical distribution of gate and readout infidelities across the device at one of the calibrations (exploiting the Clifford randomised benchmarking for PRX gates~\cite{magesan2012characterizing} and the CZ interleaved randomized benchmarking~\cite{magesan2012efficient}). Since single-qubit error rates are about one order of magnitude less than two-qubit error rates, the latter represent the main target for noise mitigation and noise characterisation. Many protocols addressing two-qubit errors treat the single-qubit gates as noiseless (e.g., at the stages of noise shaping and basis rotations for state preparation and measurement), which implies in practice that the learnt noise model for entangling layers composed of two-qubit gates indirectly contains information also about the average single-qubit gate errors as long as the latter are involved in the noise learning protocol. This takes place for the noise-learning methods considered in this study. 

\par Figure~\ref{fig: calibration-data} also indicates that the readout errors are typically the largest, which makes it necessary to resort to detector tomography in the tomographic approaches to noise learning~\cite{mangini2024tensor,torlai-2023}. Although the noise learning through cycle benchmarking is insensitive to the readout errors, the signal degradation due to measurement errors should still be taken into account at the noise-model validation stage, when the actual noisy signal is compared with one predicted on the basis of the learnt noise model. In the noisy-signal prediction, we either use the reconstructed noisy Positive Operator-Valued Measure (POVM) effects or make allowance for the Twirled Readout Error eXtinction (TREX) correction~\cite{vandenberg2022modelfree}. 

\section{\label{section-cb} Noise learning through cycle benchmarking}
\par The first method we deploy on VTT Q50 assumes a sparse
Pauli--Lindblad (SPL) model for the noise of the entangling gates
\cite{berg_probabilistic_2023}. Writing
$\mathsf{P}^n=\{I,X,Y,Z\}^{\otimes n}$ and letting
$\mathcal{K}\subset\mathsf{P}^n$ be the set of weight-one Pauli operators supported on the model qubits, together with weight-two Paulis supported on pairs of physically connected qubits, the layer noise channel is modelled as
\begin{equation}
\Lambda
=
\exp
\left[
\sum_{P_k\in\mathcal{K}}
\lambda_k
\left(
\mathcal{P}_k - \mathcal{I}
\right)
\right],
\qquad
\lambda_k \geq 0,
\label{eq: spl-model}
\end{equation}
where $\mathcal{I}$ is the identity map and $\mathcal{P}_k$ is the
conjugation superoperator $\mathcal{P}_k(\rho) = P_k\rho P_k$, so that
the $4^n-1$ parameters of a general Pauli channel are replaced by the
$O(n)$ generator rates $\{\lambda_k\}$ allowed by the device
connectivity. Since $\mathcal{P}_k(P_a) = (-1)^{\langle a,k\rangle_{\textsf{sp}}}P_a$,
the Pauli operators are eigenvectors of every generator and the
corresponding Pauli fidelities read
\begin{equation}
f_a
=
\exp
\left[
-2
\sum_{P_k\in\mathcal{K}}
\langle a,k\rangle_{\textsf{sp}}\,
\lambda_k
\right],
\label{eq: spl-fidelities}
\end{equation}
where $\langle a,k\rangle_{\textsf{sp}}$ denotes the symplectic inner
product between $P_a$ and $P_k$, equal to zero if $[P_a,P_k]=0$ and to
one otherwise. Equation~\eqref{eq: spl-fidelities} is the relation
inverted in Sec.~\ref{sec: model-fitting}: taking the logarithm gives
$-\log(f_a)/2 = \sum_k \langle a,k\rangle_{\textsf{sp}}\lambda_k$, which
is the linear system in Eq.~\eqref{eq: nnls-problem}. Two properties make this model the natural target for a
hardware deployment: the fidelities $f_a$ are accessible through cycle
benchmarking (CB) in a way that is robust to state-preparation and
measurement (SPAM) errors \cite{erhard_characterizing_2019,
chen_learnability_2023}, and the number of measurement bases needed to
reach all of them does not grow with the system size
\cite{berg_probabilistic_2023, berg_techniques_2024}. The model applies
to Pauli noise, a form that is imposed on the hardware by randomised
compiling \cite{wallman_noise_2016}.

\subsection{Learning protocol}\label{sec: learning-protocol}
\par
\begin{figure*}[!t]
    \centering
    \includegraphics[width=1\linewidth]{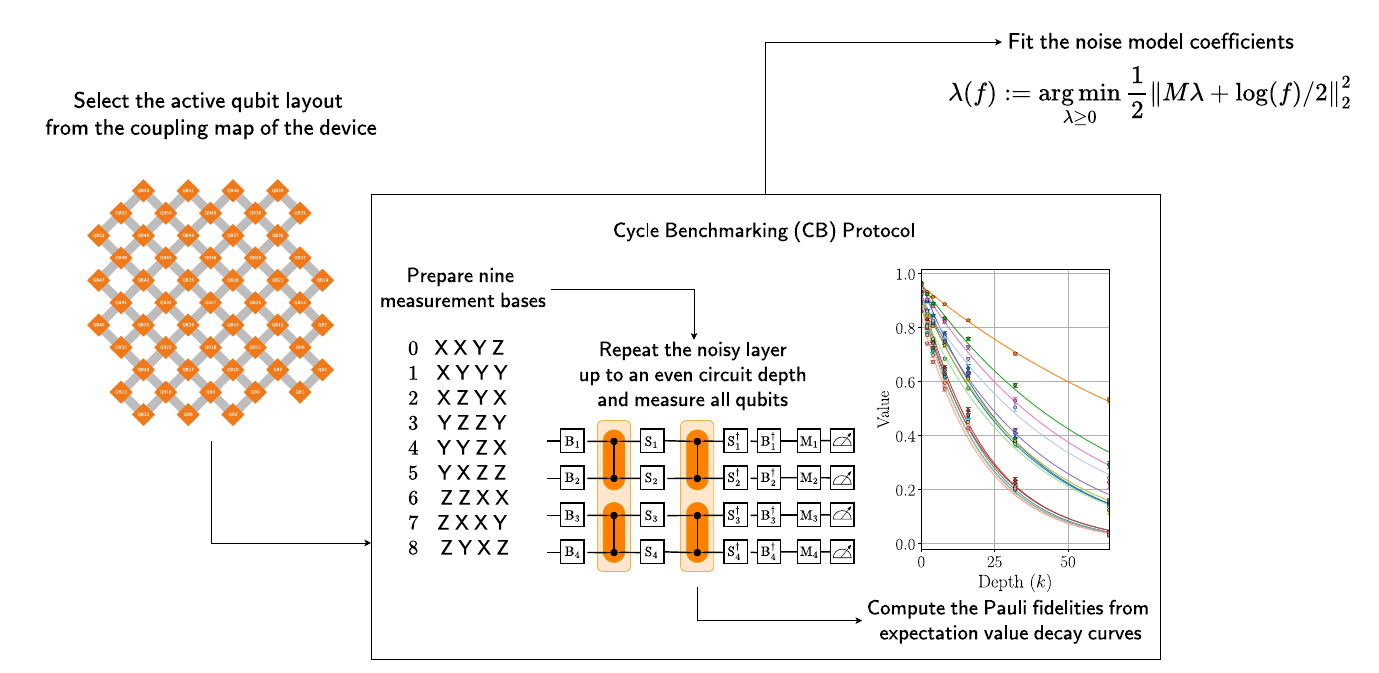}
    \caption[Noise learning protocol]{\textbf{Noise learning protocol.} After selecting the active qubit layout for the characterisation experiment, the cycle benchmarking (CB) protocol is used to determine the Pauli fidelity estimates required for fitting the noise model parameters.}
    \label{fig: cb-protocol-diagram}
\end{figure*}

 In practice, the SPL characterisation method is divided into four
stages: experimental preparation, cycle benchmarking experiment, fidelity
estimation, and model fitting. As seen from the diagram in
Fig.~\ref{fig: cb-protocol-diagram}, the first stage sets up the
device, which includes selecting the physical qubits to be used,
preparing the circuits to be run, and designing an efficient way to save
the collected data. In the second stage we choose the measurement bases,
run the benchmarking circuits at different circuit depths, and compute the
corresponding Pauli expectation values at every depth. In the third
stage the channel fidelities are estimated from the resulting decay
curves in a manner that mitigates the effects of SPAM errors. Finally,
the model coefficients are determined from \eqref{eq: nnls-problem},
typically via a nonnegative least-squares fit
\cite{berg_probabilistic_2023, calzona_multi-layer_2026}.

\par
\begin{figure}[t!]
    \centering
    \includegraphics[width=\linewidth]{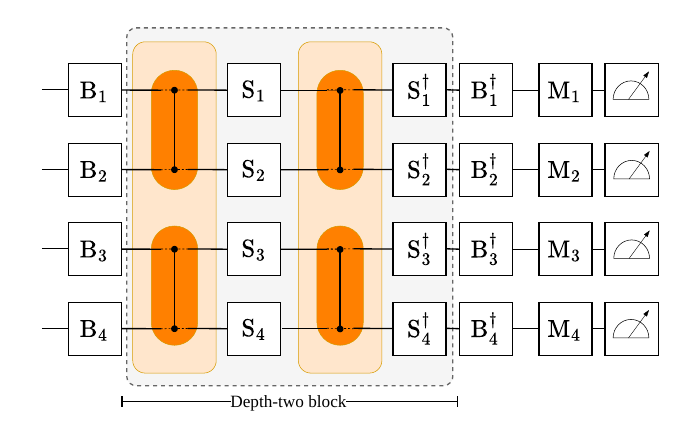}
    \caption[Structure of a depth-two cycle benchmarking circuit]{Structure of the cycle benchmarking circuit for a depth-two block consisting of two layers of concurrent noisy \(\mathrm{CZ}\) gates interleaved by single-qubit gates and preceded by state-preparation gates $B_i$. In general, after a depth-$d$ block, the inverse of the state preparation basis gates is applied, followed by measurement twirling gates $M_i$, randomly sampled from identity $I$ and $X$ gates. Finally, all qubits are measured.}
    \label{fig: cycle-benchmarking-protocol}
\end{figure}

 The central idea of cycle benchmarking is to repeat the noisy
process to be characterised up to a fixed, even circuit depth $d$ in such
a way that the overall logical circuit remains unchanged
\cite{berg_probabilistic_2023, erhard_characterizing_2019}. We focus on
circuit layers composed of $\mathrm{CZ}$ gates, the native entangling
gate of the device; non-Clifford gates can also be characterised with
this method, at the price of a more involved protocol
\cite{erhard_characterizing_2019, wallman_noise_2016}. The CB protocol
used for the experiments is as follows.
\begin{enumerate}
    \item \textit{Basis selection}: Select the measurement bases given by the benchmark-Pauli set $\mathcal{B}$. For each basis, prepare the associated eigenstate with eigenvalue $+1$.
    \item \textit{Layer Cycle}: Repeat the noisy layer for different even circuit depths $d$. Each layer is twirled using randomly sampled $n$-qubit Pauli operators.
    \item \textit{Readout mitigation}: Apply single-qubit readout-mitigation operations $M_i$, consisting of gates randomly sampled from identity $I$, and $X$ gates.
    \item \textit{Measurement}: Apply the inverse-basis preparation gates and measure all qubits in the computational basis.
\end{enumerate}
Step 3 implements the model-free readout mitigation of
\cite{vandenberg2022modelfree}: a bit-flip is applied prior to measurement
and the corresponding correcting bit-flip after the noisy measurement, as
illustrated in Fig.~\ref{fig: cycle-benchmarking-protocol}.
\par
\begin{figure}[!t]
    \centering
    \includegraphics[width=\linewidth]{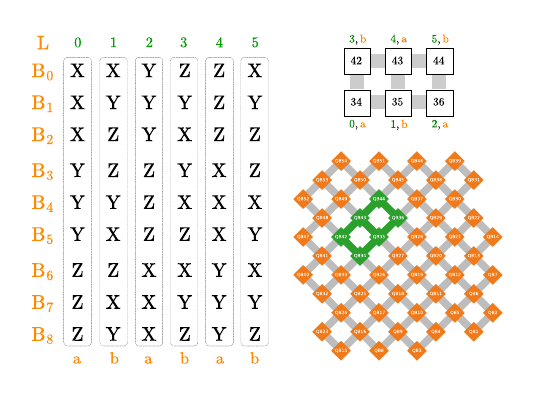}
    \caption[Basis selection protocol]{\textit{Basis selection protocol:} Given the processor coupling map and the selected physical-qubit layout, if the qubit ordering satisfies the condition of \cite[Theorem SIV.1]{berg_probabilistic_2023}, then there exist nine bases such that for each pair of physically connected qubits, the entire set $\{X, Y, Z\}^{\otimes2}$ can be covered. The procedure consists of relabelling the qubit layout in such a way that each edge corresponds to two distinct labels, given by $a$ and $b$ in the example above, and is detailed in Appendix~\ref{app: basis-construction}. Once all labels have been assigned, a $9 \times n$ matrix is constructed following the chosen layout order and the corresponding label vector, where $n$ is the number of qubits. The rows of this matrix are the nine measurement bases required for the experiment.}
    \label{fig: basis-selection}
\end{figure}

 For the local SPL model \eqref{eq: spl-model} it is enough to
measure in nine bases, irrespective of the number of qubits, provided the
active layout satisfies an ordering condition on the qubit graph
\cite[Theorem SIV.1]{berg_probabilistic_2023}. The fidelities of all
weight-one and weight-two Paulis on neighbouring qubits are then
accessible from a basis set $\mathcal{B}$ whose substrings exactly cover
$\{X, Y, Z\}^{\otimes2}$ on every pair of physically connected qubits.
The square-lattice coupling map of VTT Q50, Fig.~\ref{fig:
vtt-coupling-map}, satisfies the ordering condition.
Fig.~\ref{fig: basis-selection} shows the construction for one of the
layouts used in this work; the general recipe, its equivalence to a
covering array $\text{CA}(N;2,k,3)$, and the fact that nine measurement bases suffice
are given in Appendix~\ref{app: basis-construction}.
\par Casting the physical noise into the Pauli form assumed by
\eqref{eq: spl-model} is achieved by Pauli twirling of the Clifford layer
\cite{wallman_noise_2016, geller_efficient_2013}. The bare ideal layer
$\mathcal{C}$ is replaced by the dressed layer
$P_a^c\,\mathcal{C}\,P_a$, where $P_a$ is sampled from the Pauli group
and $P_a^c=\mathcal{C}P_a\mathcal{C}^{\dagger}$ is again a Pauli
operator. Averaging over $N$ such instances suppresses the off-diagonal
elements of the Pauli transfer matrix of the physical channel
$\tilde{\Lambda}$ proportionally to $1/\sqrt{N}$, and yields in the limit
$N\to\infty$ the average channel
\begin{equation}
\Lambda(\rho)
=
\mathbb{E}_{a}
\left[
P_a
\tilde{\Lambda}
\left(
P_a^{\dagger}\rho\, P_a
\right)
P_a^{\dagger}
\right]
=
\sum_i c_i P_i \rho\, P_i^\dagger,
\label{eq: twirled-channel}
\end{equation}
a Pauli channel with $c_i \geq 0$ and $\sum_i c_i = 1$, whose Pauli
transfer matrix is diagonal and therefore easily inverted. This feature
of twirling is commonly referred to as \emph{noise tailoring}
\cite{hashim_practical_2025}. The diagonal elements are the Pauli
fidelities $f_a=2^{-n}\operatorname{tr}\bigl[P_a^\dagger\Lambda(P_a)\bigr]$,
related to the channel coefficients by the symplectic Walsh--Hadamard
transformation \cite{berg_probabilistic_2023, flammia_efficient_2020}
\begin{equation}
c_b
=
\frac{1}{4^n}
\sum_{a}
(-1)^{\langle a,b\rangle_{\textsf{sp}}} f_a .
\label{eq: WH-transform}
\end{equation}
The derivation of the dressed layer, the transpilation of the
state-preparation, twirling and interleaving gates into a single layer of single-qubit gates, and the comparison with Clifford twirling are
collected in Appendix~\ref{app: pauli-twirling}.

\subsection{Fidelity estimation and learnability}\label{sec: fidelity-estimation}
\par Since the twirled channel \eqref{eq: twirled-channel} is diagonal in
the Pauli basis, its eigenvectors are the $n$-qubit Pauli operators,
\begin{equation}
\Lambda(P_a) = f_a\, P_a, \qquad \forall P_a \in \mathsf{P}^n,
\label{eq: pauli-eigenvalue}
\end{equation}
with the corresponding eigenvalues given by the Pauli fidelities $f_a=\sum_{b}c_b\,(-1)^{\langle a,b\rangle_{\textsf{sp}}}$. Consequently, learning the noise model amounts to estimating these fidelities, from which the Pauli coefficients can be recovered via 
$c_a=4^{-n}\sum_{b}f_b\,(-1)^{\langle a,b\rangle_{\textsf{sp}}}$.
Each fidelity $f_a$ is estimated by preparing the $+1$ eigenstate of
$P_a$, applying the noisy layer $d$ times, measuring the expectation
value of $P_a$, and averaging over random Pauli twirling sequences
\cite{erhard_characterizing_2019, chen_learnability_2023}. For a Pauli
operator that is invariant under conjugation by the Clifford gates, such as
$\mathrm{IZ}$ under $\mathrm{CZ}$, the decay is a pure exponential,
\begin{equation}
    \mathbb{E}\big[\mathrm{IZ}\big] = \alpha_{\mathrm{IZ}}\cdot f_{\mathrm{IZ}}^d,
    \label{eq: learnable-eval}
\end{equation}
where $\alpha_{\mathrm{IZ}}$ is a depth-independent scalar that accounts
for SPAM errors \cite{chen_learnability_2023, erhard_characterizing_2019}
and therefore does not influence the decay fit. In cycle benchmarking, individual exponential
decays are referred to as \textit{Pauli decays} labelled by the Pauli
operator $P_a$ specifying the state-preparation and measurement basis
\cite{hashim_practical_2025}. In the following we assume that the noise
channel acts before each layer of Clifford gate.
\begin{figure}[!t]
    \centering
    \includegraphics[width=\linewidth]{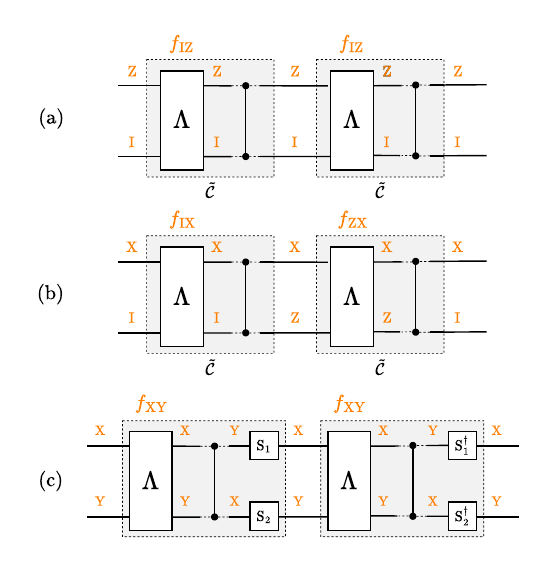}
    \caption[Fidelity learnability description]{\textbf{Fidelity learnability.} The learnability of Pauli fidelities is determined by the action of the Clifford gate on the corresponding Pauli operator. Fidelities whose associated Pauli operator remains invariant under the action of the Clifford gate (a) are individually learnable. In contrast, when a Pauli operator is not invariant and its support changes after the application of the gate, for example $\mathrm{IX} \rightarrow \mathrm{ZX}$, the corresponding fidelity becomes unlearnable (b). Finally, for Pauli operators that are not invariant but whose support is preserved under the Clifford gate, one can always map the resulting operator back to the original one by applying appropriate interleaved single-qubit gates (c).}
    \label{fig: learnability}
\end{figure}
\par A second set of fidelities is associated with Pauli operators that
are not invariant under the Clifford gate but whose support
$\mathfrak{s}$, that is, the set of qubits on which the operator acts
nontrivially, is preserved. Consider $\mathrm{XY}$, which is mapped to
$\mathrm{YX}$ under conjugation by $\mathrm{CZ}$. Interleaving the noisy
$\mathrm{CZ}$ operation with additional layers of single-qubit gates
$\mathrm{S}$ maps $\mathrm{YX}$ back to $\mathrm{XY}$, as shown in
Fig.~\ref{fig: learnability}. The required interleaving phase gates are compiled
together with the twirling layers. Both $f_{\mathrm{XY}}$ and
$f_{\mathrm{YX}}$ are therefore learnable. The interleaved layers add
circuit overhead and are themselves a source of error, but since single-qubit
gate infidelities on this device are about one order of magnitude smaller
than two-qubit ones (Fig.~\ref{fig: calibration-data}), in this protocol we treat the single-qubit gates as noiseless.
\par The remaining fidelities correspond to the unlearnable degrees of
freedom of the model. Consider $\mathrm{XZ}$, which is mapped to
$\mathrm{XI}$ by $\mathrm{CZ}$. The interleaving strategy is inapplicable
here, since the identity is invariant under all single-qubit gates. Therefore, at even circuit depth $d$ the measured expectation value reads
\begin{equation}
\mathbb{E}\big[\mathrm{XZ}\big]
=
\alpha_{\mathrm{XZ}}\cdot
f_{\mathrm{XZ}}\,f_{\mathrm{XI}}\,f_{\mathrm{XZ}}\,f_{\mathrm{XI}}\cdots
=
\alpha_{\mathrm{XZ}}
\left(f_{\mathrm{XZ}}f_{\mathrm{XI}}\right)^{\frac{d}{2}},
\label{eq:unlearnable-eval}
\end{equation}
so that the fit yields $\sqrt{f_{\mathrm{XZ}}f_{\mathrm{XI}}}$ instead of
the individual fidelity estimates. The unlearnability thus arises from changes in
the weight~\footnote{The weight of a Pauli operator $P \in \{I, X, Y, Z\}^{\otimes n}$ is the number of non-identity components.}
of the Pauli operators under conjugation by the Clifford gate: an
identity term cannot be mapped to a non-identity term, or vice versa.
These results are summarised in the following theorem from
\cite[Theorem 1]{chen_learnability_2023}.
\begin{theorem}[Learnability of Pauli noise]
Let $\mathcal{C}$ be an $n$-qubit Clifford gate. The Pauli fidelity $f_i$ relative to the Pauli operator $P_i$ is learnable if and only if the support $\mathfrak{s}_i$ of $P_i$ is preserved under conjugation by $\mathcal{C}$.
\label{theo: pauli-learnability}
\end{theorem}
\begin{table}[!t]
\centering
\begin{tabular}{|c|c|}
\hline
\textbf{Gate} & \textbf{CZ} \\
\hline
\textbf{Learnable DOF} &
\makecell[c]{
  $f_{\mathrm{II}}, f_{\mathrm{IZ}}, f_{\mathrm{ZI}}, f_{\mathrm{ZZ}},$ \\
  $f_{\mathrm{XY}}, f_{\mathrm{XX}}, f_{\mathrm{YY}}, f_{\mathrm{YX}},$ \\
  $f_{\mathrm{IX}}f_{\mathrm{ZX}}, f_{\mathrm{IY}}f_{\mathrm{ZY}}, f_{\mathrm{XI}}f_{\mathrm{XZ}}, f_{\mathrm{YZ}}f_{\mathrm{YI}},$ \\
  $f_{\mathrm{IX}}f_{\mathrm{ZY}}, f_{\mathrm{XI}}f_{\mathrm{YZ}}$
} \\
\hline
\textbf{Unlearnable DOF} & $f_{\mathrm{XI}},\, f_{\mathrm{IX}}$ \\
\hline
\end{tabular}
\caption{Learnable and unlearnable degrees of freedom for a single \(\mathrm{CZ}\) gate under the cycle benchmarking procedure.}
\label{tab:cz-learnability}
\end{table}
\par This limited learnability of Pauli noise results from the gauge freedom in the data collected during CB, rather than from a shortcoming of the protocol. Rescaling $f_{\mathrm{IX}},f_{\mathrm{XI}}$
by $\kappa$ and $f_{\mathrm{ZX}},f_{\mathrm{XZ}}$ by $\kappa^{-1}$ leaves
every CB decay invariant, so two noise models related by this
transformation produce identical output statistics for all experiments
within this framework and no such experiment can witness it
\cite{chen_disambiguating_2026, chen_learnability_2023}.
Table~\ref{tab:cz-learnability} lists the resulting degrees of freedom
for a single $\mathrm{CZ}$ gate. Although neither $f_{\mathrm{IX}}$ nor
$f_{\mathrm{ZX}}$ is individually learnable, their product is, leaving
one unlearnable degree of freedom within the pair
$\{f_{\mathrm{IX}},f_{\mathrm{ZX}}\}$. The graph-theoretical formulation
of these statements, in terms of the cycle and cut spaces of the pattern
transfer graph, is given in Appendix~\ref{app: gauge-ambiguity}.
\par Since our experiments involve systems with more than two qubits, the
amount of information to be learnt is considerably greater than in the
two-qubit setting, and assumptions must be introduced for the fidelities
that cannot be learnt individually. These assumptions, together with the
fitting procedure, are discussed next.

\subsection{Model fitting}\label{sec: model-fitting}
\par Once the individual fidelities and the pairwise products of
fidelities have been determined, the model coefficients
$\{\lambda_i\}$ are obtained by inverting \eqref{eq: spl-fidelities},
that is by solving the nonnegative least-squares problem
\begin{equation}
\boldsymbol{\lambda}(\boldsymbol{f})
:=
\operatorname*{argmin}_{\boldsymbol{\lambda}\geq 0}
\frac{1}{2}
\left\|
\mathbf{M}\boldsymbol{\lambda}
+
\log(\boldsymbol{f})/2
\right\|_2^2,
    \label{eq: nnls-problem}
\end{equation}
where $\boldsymbol{f}$ denotes a vector of fidelities $f_i$. However, not all
individual fidelities are accessible; instead, we have products of
fidelities, represented by the elementwise product of two fidelity
vectors $\boldsymbol{f}_1$ and $\boldsymbol{f}_2$. Depending on the Pauli
fidelity entries of each vector we define two binary matrices
$\mathbf{M}_1$ and $\mathbf{M}_2$ such that
$\mathbf{M}_i\boldsymbol{\lambda} = -\log(\boldsymbol{f}_i)/2,\ i=1,2$,
and adding the two expressions gives
\begin{equation}
    (\mathbf{M}_1 + \mathbf{M}_2)\boldsymbol{\lambda} = -\frac{1}{2}\log(\boldsymbol{f}_1\cdot\boldsymbol{f}_2),
    \label{eq: pair-wise-nnls}
\end{equation}
where $\cdot$ and the logarithm are elementwise. The model parameters are
then found by solving \eqref{eq: nnls-problem} with
$\mathbf{M} = \mathbf{M}_1 + \mathbf{M}_2$ and
$\boldsymbol{f} = \boldsymbol{f_1}\cdot\boldsymbol{f_2}$. The uniqueness
of the resulting coefficients is ensured provided that $\mathbf{M}$ has
full rank \cite{berg_probabilistic_2023}. As discussed in
Sec.~\ref{sec: regularised-fitting}, provided access to all individual single-qubit fidelities, in addition to the measured fidelity products, the resulting symplectic matrix $\mathbf{M}$ is full rank, whereas the
system built from the measured products alone is rank deficient. These two cases therefore lead to different fitting procedures.
\par With $\mathcal{K}$ the set of weight-one and weight-two Pauli
operators $K_i$ introduced above and $\mathcal{B}$ the set of benchmark
Pauli operators $B_i$, it can be shown that if
$\mathcal{B}\supseteq\mathcal{K}$ then the matrix
\[
\mathbf{M} = \mathbf{M}(\mathcal{B},\mathcal{K})
\quad \text{such that} \quad
M_{i,j} =
\begin{cases}
0 & [B_i, K_j] = 0, \\
1 & [B_i, K_j] \neq 0,
\end{cases}
\]
has full rank \cite[Theorem SIV.2]{berg_probabilistic_2023}.
Choosing $\mathcal{B} = \mathcal{K}$ therefore ensures uniqueness of the
fitted coefficients. Including additional Pauli operators in
$\mathcal{B}$ does not affect the column rank of $\mathbf{M}$ and leads
instead to an overdetermined system, which reduces the sensitivity of
the fit to noisy fidelity estimates \cite{berg_techniques_2024}.
\par Fitting the model therefore requires access to all weight-one Pauli fidelities
$f_i$ associated with the set $V_i = \{X_i, Y_i, Z_i\}$ for each qubit
$i \in [0, n-1]$, together with estimates of the products $f_if_j$. The
latter are associated with Pauli operators
$P^{(1)} \in \mathcal{P}_{i,j} = \{ X_i X_j,\, X_i Y_j,\, \dots,\, Z_i Z_j \}$
and $P^{(2)} \in \mathcal{P}'_{i,j}$ for each pair $(i,j)$ of physically
connected qubits, where $\mathcal{P}'_{i,j}$ denotes the set of Pauli
operators obtained after conjugation by the Clifford layer.
By Theorem~\ref{theo: pauli-learnability}, a Pauli whose support is
preserved under conjugation is associated with a learnable fidelity, as the insertion of interleaved single-qubit gates maps the operator back to its original form. For a given pair $(i,j)$ the terms of $\mathcal{P}'_{i,j}$ therefore either coincide with those of $\mathcal{P}_{i,j}$ or have one component replaced by the identity, lowering their weight. For pairs $(i,j)$ without a gate but with gates acting on qubits $i$ and $j$ individually, $\mathcal{P}'_{i,j}$ can reach weight four, whereas the weight of the operators in $\mathcal{P}_{i,j}$ is always two.

\subsubsection{Fidelity augmentation}
\par We define two sets of benchmark Pauli operators
$\mathcal{B}_k = V \cup \mathcal{P}^{(k)},\ k=1,2$, where
$V=\bigcup_i V_i$ and $\mathcal{P}^{(k)}$ collects the operators
$P^{(k)}$ over all connected pairs. Choosing $\mathcal{B}_1$ to coincide
with the set of model coefficients, $\mathcal{K}=\mathcal{B}_1$,
guarantees uniqueness of the fitted parameters. Not all fidelities
associated with the operators in these sets are individually learnable:
some weight-one fidelities in $V$, such as $\mathrm{IX}$ in the
two-qubit case, cannot be estimated independently through cycle
benchmarking.
\par Since individual weight-one fidelities are needed to fit the
model coefficients, some augmentation strategy must be employed. According to
\textcite{berg_probabilistic_2023}, there are three physically motivated
approaches for dealing with unlearnable fidelities that appear only in
pairs.

(i) \textit{Direct estimation}: The fidelity is estimated by applying the noisy gate only once. This method is not robust against SPAM errors, since the initial and final Pauli components may differ, so SPAM errors cannot be removed by normalising with the zero-depth fidelity.

(ii) \textit{Symmetry assumption}: The two fidelities within a pair are assumed to be equal, so that $f_\mathrm{IX}$ is inferred as the square root of the learnable product $f_\mathrm{IX}f_\mathrm{ZX}$.

(iii) \textit{Gate-independent noise}: The $\mathrm{CZ}$ and $\mathrm{CX}$ gates are assumed to be affected by the same noise process. Since the $\mathrm{CZ}$ gate is equivalent to a $\mathrm{CX}$ conjugated by $I\otimes H$
    \begin{align*}
    C_X\Lambda
    &= \tilde{C}_X \\
    &= (I \otimes H)\,\tilde{C}_Z\,(I \otimes H) \\
    &= (I \otimes H)\,C_Z\Lambda\,(I \otimes H) \\
    &= \underbrace{(I \otimes H)\,C_Z\,(I \otimes H)}_{C_X}
       (I \otimes H)\,\Lambda\,(I \otimes H),
    \end{align*}
    it follows that $\Lambda = (I\otimes H)\Lambda(I\otimes H)$, implying that the Pauli fidelities $f_\mathrm{P_1X_2}$ and $f_\mathrm{P_1Z_2}$ are the same for any Pauli operator $P$ acting on the first qubit. Since the fidelities $f_\mathrm{IZ}$, $f_\mathrm{XX}$, $f_\mathrm{YX}$, and $f_\mathrm{ZZ}$ are directly learnable, this assumption can be used to infer the corresponding unlearnable fidelities $f_\mathrm{IX}$, $f_\mathrm{XZ}$, $f_\mathrm{YZ}$, and  $f_\mathrm{ZX}$.

\par Of the three, the symmetry assumption is the one used in practice,
and it is the baseline against which we compare below. This assumption has several implications for the model fitting, as it is imposed rather than experimentally validated, thereby introducing bias in the resulting coefficients.

\subsubsection{Reconstruction bias of the symmetry assumption}\label{sec: bias}
\par To quantify the bias we simulate the protocol on a system whose
noise model is known. A local Pauli--Lindblad channel is applied before
the ideal Clifford layer, with generators covering all nodes and all
edges of the coupling map of the active layout, and with rates drawn from
a normal distribution of mean $\mu = 2\times10^{-3}$ and standard
deviation $5\times10^{-4}$, consistent with two-qubit error rates on
current hardware. A readout channel with conditional probabilities
$p(0|1) = 0.025$ and $p(1|0) = 0.010$ is included on every qubit. The
simulated experiment requires nine bases, as discussed in
Sec.~\ref{sec: learning-protocol}, $256$ shots per randomisation, $100$
randomisations per depth, and depths $d = 0, 2, 4, 8, 16, 32, 64$. The
simulated characterisation of the individual layers is reported in
Appendix~\ref{app: simulation-results}; here we use it only to compare
fitting procedures.
\begin{figure}[!t]
    \centering
    \includegraphics[width=1\linewidth]{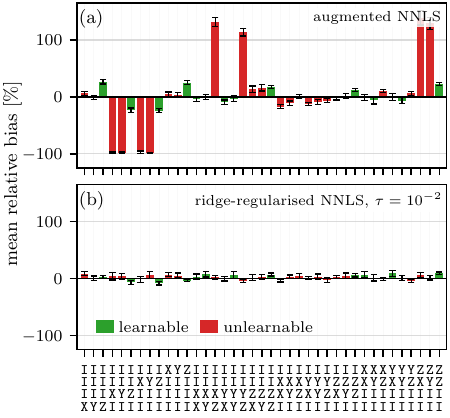}
    \caption[Sparse Pauli-Lindblad reconstruction bias]{\textbf{SPL reconstruction bias.} Mean relative bias of each reconstructed noise parameter over $20$ simulations with known ground-truth model coefficients. The symmetry-based augmentation strategy (a) overestimates the rates associated with unlearnable Pauli fidelities, such as $\mathrm{IIXZ}$, while suppressing the corresponding single-qubit conjugated Pauli term $\mathrm{IIXI}$. After applying $\mathrm{L}_2$ regularisation (b), these misestimated coefficients are corrected, since the penalty distributes the noise contribution across correlated generators instead of concentrating it in a few large coefficients.}
    \label{fig: bias-comparison}
\end{figure}
\par Fig.~\ref{fig: bias-comparison} panel (a) shows the mean relative bias of
each reconstructed coefficient over $20$ such simulations. The
coefficients associated with unlearnable fidelity pairs, for instance
$\mathrm{IIXI}$--$\mathrm{IIXZ}$ and $\mathrm{IXII}$--$\mathrm{ZXII}$,
carry by far the largest bias: within each pair the weight-two
coefficient is consistently overestimated, in some cases by more than a
factor of two, while the corresponding weight-one conjugated rate is suppressed and, in some cases, driven to zero. The origin of this issue lies in the symmetry-based augmentation strategy itself. Taking the square root of a measured product assigns half of the product to each member of the pair, and the resulting values enter the nonnegative least-squares fit as if they had been directly measured.

\subsubsection{Regularised fitting}\label{sec: regularised-fitting}
\par We therefore drop the augmentation and fit using only quantities
that follow from the data, namely the measured products
$f_{P_i}f_{P_i'}$. These define the linear problem
\begin{equation}
\mathbf{M}\boldsymbol{\lambda} \approx \mathbf{b},
\qquad
b_i = -\frac{1}{2}\log\left(f_{P_i}f_{P_i'}\right),
\label{eq: pair-only-system}
\end{equation}
with $\boldsymbol{\lambda}\in\mathbb{R}^m$ the vector of Pauli rates and
$\mathbf{M}$ the anticommutation matrix defined above. Without
augmentation, $\mathbf{M}$ is rank deficient,
$\operatorname{rank}(\mathbf{M}) < m$, so $\ker(\mathbf{M})$ is
nontrivial and any $\boldsymbol{\lambda}' = \boldsymbol{\lambda} +
\mathbf{z}$ with $\mathbf{z}\in\ker(\mathbf{M})$ and
$\boldsymbol{\lambda}'\geq\mathbf{0}$ reproduces the same predicted
products, since $\mathbf{M}\boldsymbol{\lambda} =
\mathbf{M}\boldsymbol{\lambda}'$. This is the same gauge freedom as in
Sec.~\ref{sec: fidelity-estimation}, seen at the level of the fit.
\par To select a solution from this family we add a quadratic penalty,
that is, we use ridge regularisation \cite{willoughby_solutions_1979},
\begin{equation}
\hat{\boldsymbol{\lambda}}_{\tau}
=
\operatorname*{argmin}_{\boldsymbol{\lambda}\ge 0}
\left(
\left\|\mathbf{M}\boldsymbol{\lambda}-\mathbf{b}\right\|_2^2
+
\tau\left\|\boldsymbol{\lambda}\right\|_2^2
\right),
\label{eq: regularized-nnls}
\end{equation}
where $\tau > 0$ is the regularisation strength. The first term enforces
agreement with the measured products and the second penalises solutions
of large Euclidean norm, $\|\boldsymbol{\lambda}\|_2^2 =
\sum_{j=1}^{m}\lambda_j^2$. Under a fixed total error-rate constraint ($\sum_k \lambda_k = {\rm const}$), this Euclidean penalty coincides with maximising the entropy $-\sum_k\lambda_k\log\lambda_k$, so the regularisation can be interpreted as a maximum-entropy selection among the gauge-equivalent noise models compatible with the data.

\begin{figure}[!t]
    \centering
    \includegraphics[width=\linewidth]{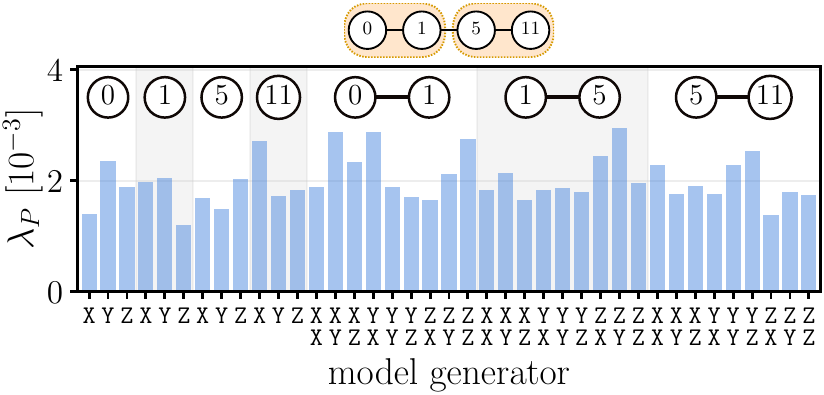}
    \caption[Ridge-regularised simulation results for the first Clifford layer]{\textbf{Ridge-regularised fit.} Noise model coefficients obtained by combining $\mathrm{L}_2$ regularisation with the nonnegative least-squares fit. The inferred rates are distributed around the ground-truth value $\mu = 2\times10^{-3}$, with no overestimated or vanishing coefficients associated with unlearnable degrees of freedom.}
    \label{fig: ridge-coefficients-simulation}
\end{figure}
\par Since the penalty grows quadratically with the coefficient
magnitude, the fit favours solutions in which the noise contribution is
spread over correlated generators rather than concentrated in a few large
coefficients. This shrinkage yields more stable estimates while
maintaining agreement with the data. We fixed $\tau$ by minimising the
mean residual coefficient error across simulations with known ground truth, which
gives $\tau = 10^{-2}$.
\par The effect is shown in Fig.~\ref{fig: bias-comparison} panel (b): using
only learnable data, the bias on the coefficients associated with
unlearnable degrees of freedom is substantially reduced relative to the
augmented baseline of Fig.~\ref{fig: bias-comparison} panel (a). The
corresponding rates, Fig.~\ref{fig: ridge-coefficients-simulation}, are
distributed around the ground-truth mean with neither overestimated nor
vanishing entries.
\par{} Regularisation does not make unlearnable degrees of freedom
learnable. It selects one particular solution from the family compatible
with the measured fidelity products, and the selection is made on a
stated criterion rather than by an assumption about the noise. All
coefficients reported for hardware in Sec.~\ref{section-experimental-results}
are obtained from \eqref{eq: regularized-nnls}.

\section{\label{section-tn} Noise learning through channel tomography}
Tensor-networks provide a scalable parametrisation for quantum process tomography (QPT).
At first, one may think of performing QPT by directly optimising the full Choi matrix $\Lambda$ of a channel \cite{d2001quantum}, a $4^n\times4^n$ matrix. But this has an exponential scaling with the number of qubits $n$, and the required measurement shots also scale exponentially in $n$. We can reduce these complexities by instead variationally training a structured tensor-network that has a smaller set of parameters. 
This was proposed by Torlai \emph{et al.}~\cite{torlai-2023}, where they learn the \emph{full} channel of a circuit, unitary and noise together, as a Choi matrix represented by a locally purified density operator (LPDO) with a one-dimensional virtual-bond topology.
Mangini \emph{et al.}~\cite{mangini2024tensor} further developed the protocol to work directly with error mitigation: given a known ideal layer $U$, they learn only the accompanying noise channel $\mathcal{N}$ in $\mathcal{E} = \mathcal{N}\circ U$, again on a chain of qubits.
The learnt $\mathcal{N}$ is exactly the input required by tensor-network error mitigation (TEM)~\cite{filippov2023scalable}.

Here we characterise the noise of the two-qubit gate layers of the VTT Q50 processor~\cite{noauthor_vtt_2026,marxer_long-distance_2023}, which dominate the error budget of the circuits of Sec.~\ref{section-circuits}.
Our contribution is an LPDO whose virtual-bond graph \emph{is} the physical square lattice of the device, so that a genuinely two-dimensional brickwork layer of CZ gates is characterised without flattening the qubit array onto a chain. In addition, we deploy the protocol on hardware, combining it with Quantum Detector Tomography (QDT)~\cite{cattaneo_self-consistent_2023} to mitigate measurement errors, rather than just focusing on classical simulations.

\subsection{Tensor-network representation of the noise channel}

A noise channel is a completely positive and trace-preserving (CPTP) map, written in Kraus form as
\begin{equation}
    \mathcal{N}[\rho] = \sum_{\kappa} K_\kappa\, \rho\, K_\kappa^\dagger,
    \qquad
    \sum_{\kappa} K_\kappa^\dagger K_\kappa = \mathds{1}.
    \label{eq: kraus}
\end{equation}

\begin{figure*}
    \centering
\definecolor{tnpink}{HTML}{FFAAAA}
\definecolor{tngreen}{HTML}{AFE9AF}
\definecolor{tngrey}{HTML}{808080}

\tikzset{
  tnode/.style={circle,draw=black,line width=0.8pt,inner sep=0pt,
                minimum size=8mm,font=\footnotesize},
  Anode/.style={tnode,fill=tnpink},
  Rnode/.style={tnode,fill=tngreen},
  gbox/.style={draw=black,line width=0.7pt,minimum width=7mm,
               minimum height=7mm,inner sep=1pt,font=\footnotesize},
  bond/.style={line width=0.8pt,line cap=round},
  multi/.style={line width=1.8pt,line cap=butt},
  kraus/.style={line width=1.4pt,draw=tngrey,line cap=round},
  krausb/.style={line width=1.4pt,draw=tngrey,line cap=round},
  ix/.style={font=\footnotesize,inner sep=1.2pt},
  hd/.style={font=\bfseries\small},
  lhs/.style={font=\large},
}

\newcommand{\DS}{0.8}
\newcommand{\LEG}{0.85}

\begin{tikzpicture}[x=1cm,y=1cm]

\node[hd,anchor=base west] at (-3.9,0) {(a)};
\node[anchor=base west] (eqa) at (-3.3,0)
  {$\mathcal{N}(\rho)=\sum_{\kappa}K_{\kappa}\,\rho\,K_{\kappa}^{\dagger}\;=$};

\begin{scope}[shift={($(eqa.base east)+(0.75,0.09)$)}]
  \draw[kraus] (0,0.35) -- (0,0.7) -- (1.8,0.7) -- (1.8,0.35);
  \node[ix,above] at (0.9,0.72) {$\kappa$};
  \draw[multi] (-0.6,0) -- (2.4,0);
  \node[gbox,fill=tnpink]  at (0,0)   {$K_{\kappa}$};
  \node[gbox,fill=tngreen] at (0.9,0) {$\rho$};
  \node[gbox,fill=tnpink]  at (1.8,0) {$K_{\kappa}^{\dagger}$};
\end{scope}

\node[hd,anchor=base west] at (-3.9,-1.0) {(b)};
\begin{scope}[shift={(0,-2.0)},x=\DS cm,y=\DS cm]

\node[lhs,anchor=east] at (-3.35,-1.725) {$\Lambda_{\mathcal{N}}\;\cong$};

\coordinate (A11) at ( 0.0, 0.0);
\coordinate (A21) at ( 0.0,-2.3);
\coordinate (A12) at (-1.8,-1.15);
\coordinate (A22) at (-1.8,-3.45);
\coordinate (B11) at ( 4.5, 0.0);
\coordinate (B21) at ( 4.5,-2.3);
\coordinate (B12) at ( 2.7,-1.15);
\coordinate (B22) at ( 2.7,-3.45);

\draw[krausb] ($(A11)+(0, 0.5)$) .. controls +(0, 1.0) and +(0, 1.0) .. ($(B11)+(0, 0.5)$);
\draw[krausb] ($(A12)+(0, 0.5)$) .. controls +(0, 2.75) and +(0, 2.75) .. ($(B12)+(0, 0.5)$);
\draw[krausb] ($(A21)+(0,-0.5)$) .. controls +(0,-1.90) and +(0,-1.90) .. ($(B21)+(0,-0.5)$);
\draw[krausb] ($(A22)+(0,-0.5)$) .. controls +(0,-1.65) and +(0,-1.65) .. ($(B22)+(0,-0.5)$);
\node[ix,anchor=east] at ($(A11)+(-0.1, 0.72)$) {$\kappa_{11}$};
\node[ix,anchor=east] at ($(A12)+(-0.1, 0.72)$) {$\kappa_{12}$};
\node[ix,anchor=east] at ($(A21)+(-0.1,-0.72)$) {$\kappa_{21}$};
\node[ix,anchor=east] at ($(A22)+(-0.1,-0.72)$) {$\kappa_{22}$};

\draw[bond] (A11)--(A21);  \draw[bond] (A12)--(A22);
\draw[bond] (A11)--(A12);  \draw[bond] (A21)--(A22);
\draw[bond] (B11)--(B21);  \draw[bond] (B12)--(B22);
\draw[bond] (B11)--(B12);  \draw[bond] (B21)--(B22);

\foreach \n in {A11,A21,A12,A22,B11,B21,B12,B22}
  {\draw[bond] ($(\n)+(-\LEG,0)$) -- ($(\n)+(\LEG,0)$);}

\node[Anode] at (A11) {$A^{[11]}$};
\node[Anode] at (A21) {$A^{[21]}$};
\node[Anode] at (A12) {$A^{[12]}$};
\node[Anode] at (A22) {$A^{[22]}$};
\node[Anode] at (B11) {$A^{[11]\dagger}$};
\node[Anode] at (B21) {$A^{[21]\dagger}$};
\node[Anode] at (B12) {$A^{[12]\dagger}$};
\node[Anode] at (B22) {$A^{[22]\dagger}$};

\foreach \n/\l in {A11/11,A21/21,A12/12,A22/22}{
  \node[ix,anchor=east] at ($(\n)+(-\LEG,0)$) {$b_{\l}$};
  \node[ix,anchor=west] at ($(\n)+( \LEG,0)$) {$a_{\l}$};}
\foreach \n/\l in {B11/11,B21/21,B12/12,B22/22}{
  \node[ix,anchor=east] at ($(\n)+(-\LEG,0)$) {$a'_{\l}$};
  \node[ix,anchor=west] at ($(\n)+( \LEG,0)$) {$b'_{\l}$};}

\node[ix,anchor=west] at ($(A11)!0.5!(A21)+( 0.1,0)$) {$\mu^{x}_{1}$};
\node[ix,anchor=east] at ($(A12)!0.5!(A22)+(-0.1,0)$) {$\mu^{x}_{2}$};
\node[ix,anchor=west] at ($(B11)!0.5!(B21)+( 0.1,0)$) {$\nu^{x}_{1}$};
\node[ix,anchor=east] at ($(B12)!0.5!(B22)+(-0.1,0)$) {$\nu^{x}_{2}$};
\node[ix,anchor=east] at ($(A11)!0.5!(A12)+(-0.12,0.1)$) {$\mu^{y}_{1}$};
\node[ix,anchor=east] at ($(A21)!0.5!(A22)+(-0.12,0.1)$) {$\mu^{y}_{2}$};
\node[ix,anchor=east] at ($(B11)!0.5!(B12)+(-0.12,0.1)$) {$\nu^{y}_{1}$};
\node[ix,anchor=east] at ($(B21)!0.5!(B22)+(-0.12,0.1)$) {$\nu^{y}_{2}$};

\end{scope}

\node[hd,anchor=base west] at (5.6,-1.0) {(c)};
\begin{scope}[shift={(8.7,-2.0)},x=\DS cm,y=\DS cm]

\node[lhs,anchor=east] at (-3.35,-1.725) {$\rho\;\cong$};

\coordinate (R11) at ( 0.0, 0.0);
\coordinate (R21) at ( 0.0,-2.3);
\coordinate (R12) at (-1.8,-1.15);
\coordinate (R22) at (-1.8,-3.45);

\draw[bond] (R11)--(R21);  \draw[bond] (R12)--(R22);
\draw[bond] (R11)--(R12);  \draw[bond] (R21)--(R22);
\foreach \n in {R11,R21,R12,R22}
  {\draw[bond] ($(\n)+(-\LEG,0)$) -- ($(\n)+(\LEG,0)$);}

\node[Rnode] at (R11) {}; \node[Rnode] at (R21) {};
\node[Rnode] at (R12) {}; \node[Rnode] at (R22) {};

\foreach \n/\l in {R11/11,R21/21,R12/12,R22/22}{
  \node[ix,anchor=east] at ($(\n)+(-\LEG,0)$) {$a_{\l}$};
  \node[ix,anchor=west] at ($(\n)+( \LEG,0)$) {$a'_{\l}$};}

\node[ix,anchor=west] at ($(R11)!0.5!(R21)+( 0.1,0)$) {$\lambda^{x}_{1}$};
\node[ix,anchor=east] at ($(R12)!0.5!(R22)+(-0.1,0)$) {$\lambda^{x}_{2}$};
\node[ix,anchor=east] at ($(R11)!0.5!(R12)+(-0.12,0.1)$) {$\lambda^{y}_{1}$};
\node[ix,anchor=east] at ($(R21)!0.5!(R22)+(-0.12,0.1)$) {$\lambda^{y}_{2}$};

\end{scope}

\end{tikzpicture}
    \caption{Tensor-network representation of the noise channel $\mathcal{N}$ accompanying one circuit layer on a two-dimensional lattice of $n = L_x\times L_y$ qubits, shown here for a $2\times2$ patch with sites labelled $(i,j)$. (a) Kraus decomposition $\mathcal{N}(\rho) = \sum_\kappa K_\kappa\,\rho\,K_\kappa^\dagger$: the Kraus index $\kappa$ (thick grey) is contracted between $K_\kappa$ and $K_\kappa^\dagger$, while the thick black line bundles the $n$ physical indices into a single multi-index. (b) The channel written as a locally purified density operator $\Lambda_{\mathcal{N}}$: each site tensor $A^{[ij]}$ carries the physical indices $a_{ij},b_{ij}$, one Kraus index $\kappa_{ij}$ (grey), and one virtual bond per lattice link, $\mu^x_\ell$ along the vertical links and $\mu^y_\ell$ along the oblique ones; primed indices and $\nu^{x,y}_\ell$ label the conjugate layer $A^{[ij]\dagger}$. Unlike the one-dimensional case, every tensor carries up to four virtual bonds, of dimension $\chi_b$, and the Kraus indices have dimension $\chi_\kappa$. (c) The tomographic state $\rho$ as a two-dimensional matrix-product (PEPO) network with virtual bonds $\lambda^{x,y}_\ell$; it is contracted with $\Lambda_{\mathcal{N}}$ through the indices $a_{ij}$ and $a'_{ij}$. Line crossings do not denote contractions.}
    \label{fig: lpdo-2d}
\end{figure*}
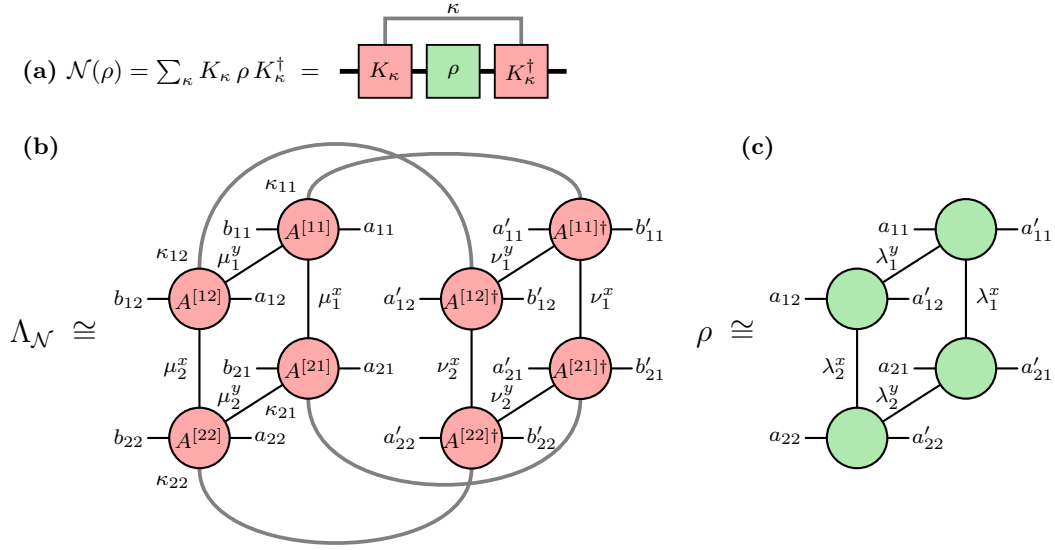

The LPDO ansatz $\mathcal{N}_{\bm\theta}$ writes each global Kraus operator $K_{\bm\kappa}$, for a multi-index $\bm\kappa = (\kappa_1,\dots,\kappa_n)$, as a matrix product operator of local tensors $A^{[j]}_{\mu_{j-1},\mu_j,\kappa_j}\in\mathbb{C}^{2\times2}$ acting on qubit $j$, with virtual bonds $\mu_j$ of dimension at most $\chi_b$ and Kraus indices $\kappa_j$ of dimension at most $\chi_\kappa$.
Sandwiching a state $\rho$ between $K_{\bm\kappa}$ and $K_{\bm\kappa}^\dagger$ and summing over $\bm\kappa$, as in Eq.~\eqref{eq: kraus}, gives the action of the ansatz directly,
\begin{equation}
\begin{split}
    \mathcal{N}_{\bm\theta}[\rho] = \sum_{\bm\kappa}\sum_{\bm\mu}\sum_{\bm\nu}\
    &\Big(A^{[1]}_{\mu_1,\kappa_1}\otimes\cdots\otimes A^{[n]}_{\mu_{n-1},\kappa_n}\Big)\\
    \rho\ &\Big(A^{[1]\dagger}_{\nu_1,\kappa_1}\otimes\cdots\otimes A^{[n]\dagger}_{\nu_{n-1},\kappa_n}\Big),
\end{split}
\label{eq: lpdo-1d}
\end{equation}
where $A^{[j]\dagger} := (A^{[j]})^\dagger$ are the conjugate transposes of the same tensors that build the upper branch, $\bm\mu,\bm\nu$ are two \emph{independent} families of virtual-bond indices (trivial boundary bonds $\mu_0=\mu_n=\nu_0=\nu_n=1$), and $\bm\kappa$ is shared between the two branches, as it must be since Eq.~\eqref{eq: kraus} contains a single sum over $\kappa$.
When $\rho$ is itself short-range correlated, as is the case for the tomographic states $\rho_\alpha$ used below, it admits an efficient MPO representation and Eq.~\eqref{eq: lpdo-1d} contracts to a single tensor-network.
Since the lower branch is by construction the conjugate transpose of the upper one, $\mathcal{N}_{\bm\theta}$ has the operator-sum form of Eq.~\eqref{eq: kraus} for \emph{every} $\bm\theta$, with $K_{\bm\kappa} = \sum_{\bm\mu} A^{[1]}_{\mu_1,\kappa_1}\otimes\cdots\otimes A^{[n]}_{\mu_{n-1},\kappa_n}$, and is therefore completely positive for any $\bm\theta$.
Trace preservation is not structural and must be imposed during the optimisation, see Eq.~\eqref{eq: tp-penalty}.
The model has $O(n\,\chi_\kappa\chi_b^2)$ parameters.
Fig.~\ref{fig: lpdo-2d} shows this construction for the two-dimensional lattice used below.

To extend Eq.~\eqref{eq: lpdo-1d} to a device of $n = L_x L_y$ qubits on a square lattice, with sites flattened row-major, $j = r L_y + c$, only the virtual-bond $\bm\mu$ and $\bm\nu$ graph changes, from a chain to the lattice edge index $(\mu^x_l, \mu^y_{l'})$ and $(\nu^x_l, \nu^y_{l'})$.
Each branch is thus a projected-entangled-pair-operator (PEPO) network~\cite{verstraete-2008}: an interior tensor carries four virtual bonds (up, down, left, right) and is a rank-$7$ object of shape $(\chi_b,\chi_b,\chi_b,\chi_b,\chi_\kappa,d,d)$, against rank-$5$ in one dimension, and the parameter count becomes $O(n\,\chi_\kappa\chi_b^4)$. Complete positivity remains structural.

Because only the bond graph changes, the likelihood, the trace-preservation
penalty, the trace, the batching of the data and the optimiser are reused
verbatim from the one-dimensional case; this is what makes the extension
economical. We contract the doubled network \emph{exactly}, with a
contraction path found by hypergraph
partitioning~\cite{cotengra,quimb}. 
The cost is then set by the minimal cut of the doubled network, whose bond
dimension is
\begin{equation}
    \chi_{\rm cut} = \big(\chi_b^{2}\big)^{\min(L_x,L_y)},
    \label{eq: mincut}
\end{equation}
e.g. $\chi_{\rm cut} = 64$ for a $3\times 3$ lattice at $\chi_b = 2$. This is
the scaling that ultimately limits the approach: it is mild in the lattice
size at $\chi_b = 2$ but grows sharply with the bond dimension.
This means that the method can still be used in cases where the noise can be described by a local model, and the non local correlations are weak.

\subsection{Data sampling}

A \emph{tomographic setting} consists of $n$ separable single-qubit input states and $n$ measurement bases $\beta_j \in \{X,Y,Z\}$, the latter realised by single-qubit rotations preceding a computational-basis readout. The measurement protocol used here relies on the methods introduced in Ref.~\cite{korhonen_practical_2025} and further developed in Ref.~\cite{korhonen_improving_2025}.
The input states are drawn from an informationally complete (IC) pool $\mathcal{P} = \{\rho^{\rm in}_{\alpha_j}\}$ with $|\mathcal{P}| \ge 4$; we use the four-element symmetric IC set
\begin{equation}
    \rho^{\rm in}_{\alpha} = \tfrac{1}{2}\big(\mathds{1} + \bm a_{\alpha}\cdot\bm\sigma\big),
    \qquad \bm\sigma = (X,Y,Z),
    \label{eq: sic}
\end{equation}
with $\bm a_{\alpha} = \tfrac{1}{\sqrt3}(\pm1,\pm1,\pm1)$ restricted to the four sign patterns of even parity.
Since the ideal layer $U$ is known, the experiment is regarded as the unknown $\mathcal{N}$ acting on the known tomographic state $\rho_\alpha = U[\rho^{\rm in}_\alpha]$, which is short-range correlated and therefore efficiently representable as a tensor-network.
The outcome probabilities follow from the Born rule,
\begin{equation}
    p(\bm\zeta \,|\, \alpha,\beta)
    = \mathrm{Tr}\Big[\mathcal{N}[\rho_\alpha]\bigotimes_{j=1}^{n}\zeta_j(\beta_j)\Big],
    \label{eq: born}
\end{equation}
where $\zeta_j(\beta_j)$ is the effect observed on qubit $j$.

Enumerating all settings is infeasible: an IC pool of $R$ states and three bases per qubit gives $(3R)^n$ of them.
Following Ref.~\cite{mangini2024tensor} we instead draw $N_{\rm set}$ settings at random and allocate $N_{\rm shots}$ shots to each, for a total of $N = N_{\rm set}N_{\rm shots}$ shots.
The split matters in practice: on cloud hardware the wall-time cost is dominated by compiling and loading distinct settings, so $N_{\rm set}$ is the bottleneck while shots are cheap.
In terms of sampling this is equivalent to shadow process tomography~\cite{huang-2020,innocenti_shadow_2023}; the difference lies entirely in the post-processing, a variational LPDO fit in place of linear inversion.

The protocol as stated assumes known preparations and known effects.
We correct the measurement side with QDT~\cite{cattaneo_self-consistent_2023}: the POVM is probed with a tomographically complete set of states and the reconstructed noisy effects $\tilde\zeta_j(\beta_j)$ replace the ideal ones in Eqs.~\eqref{eq: born} and \eqref{eq: kl}.
State-preparation errors are \emph{not} corrected and are assumed negligible; note that QDT itself assumes ideal preparation, so the reconstructed effects are biased when preparation and measurement errors are comparable.

\subsection{Optimisation}

The dataset $S = \{(\rho_{\alpha_m}, \zeta_m(\beta_m))\}_{m=1}^{N}$ contains one entry per shot.
The fit minimises the negative log-likelihood of the observed outcomes, a Monte Carlo estimate of the Kullback--Leibler divergence between the measured and the model outcome distributions,
\begin{equation}
    D_{\rm KL}(\bm\theta;S)
    = -\frac{1}{N}\sum_{m=1}^{N}
      \log \mathrm{Tr}\big[\mathcal{N}_{\bm\theta}[\rho_{\alpha_m}]\,\zeta_m(\beta_m)\big].
    \label{eq: kl}
\end{equation}
Trace preservation, which the ansatz does not enforce structurally, is added as a penalty,
\begin{equation}
    \delta_{\rm TP}(\bm\theta)
    = \big\|\,\mathrm{Tr}_{\rm out}[\mathcal{N}_{\bm\theta}] - \mathds{1}\,\big\|_F \big/ 2^{n/2},
    \label{eq: tp-penalty}
\end{equation}
so that the total loss reads
\begin{equation}
    \mathcal{L}(\bm\theta;S) = D_{\rm KL}(\bm\theta;S) + \eta\,\delta_{\rm TP}(\bm\theta),
    \qquad \eta = 1.2 .
    \label{eq: loss}
\end{equation}
The learnt map is therefore completely positive by construction but only approximately trace preserving; the residual $\delta_{\rm TP}$ at the end of training is reported as a diagnostic.

The tensor entries are initialised with independent Gaussian real and imaginary parts of variance $\sigma^2$ chosen so that $\mathds{E}\big[\mathrm{Tr}\,\mathcal{N}_{\bm\theta}\big] = 2^n$ \cite{mangini2024tensor}, followed by a pre-optimisation of $\delta_{\rm TP}$ alone that brings the ansatz onto the trace-preserving manifold before the likelihood term is switched on, using the optimizer L-BFGS-B \cite{doi:10.1137/0916069} provided in quimb.
The main loop uses Adam~\cite{kingma2017adammethodstochasticoptimization} with mini-batches of shots and an exponentially decaying learning rate $\eta_t = \eta_0\gamma^{t/T}$, with $T$ the number of batches per epoch; complex parameters are updated along conjugate gradients.
To evaluate the training we report in each epoch the loss computed on a held-out test set and the distance to a trace preserving channel, since the reconstruction error of Ref.~\cite{mangini2024tensor} requires the true noisy channel and is available only in simulations.

For physical interpretation we report the Pauli transfer matrix coefficients of the learnt channel,
\begin{equation}
    c_{ij} = \frac{1}{2^n}\,\mathrm{Tr}\big[P_i\,\mathcal{N}_{\bm\theta}[P_j]\big],
    \label{eq: ptm}
\end{equation}
whose zero-weight coefficient equals one is a necessary condition for the map to be trace preserving, making it a second, basis-resolved check of Eq.~\eqref{eq: tp-penalty}.

\subsection{Learning a layer from its repetitions}
\label{section-tn-repeated}

Here we present some theoretical results on a possible extension of this method. A characterisation experiment normally runs the layer to be learnt once, and the fit of Sec.~\ref{section-tn} then sees the channel $\mathcal{N}$ a single time.
On a dataset in which the same layer is applied $D$ times in a row, one channel can instead be fitted to the whole block and applied once per repetition with the same parameters,
\begin{equation}
    p_D(\bm\zeta\,|\,\alpha,\beta)
    = \mathrm{Tr}\Big[\big(\mathcal{N}_{\bm\theta}\circ U\big)^{D}\big[\rho^{\rm in}_\alpha\big]
      \bigotimes_{j=1}^{n}\zeta_j(\beta_j)\Big].
    \label{eq: repeated-born}
\end{equation}
The parameter count is that of a single layer, unchanged from the $D = 1$ fit, while the departure of $\mathcal{N}$ from the identity enters the outcome probabilities $D$ times: repetition amplifies the signal the fit has to work with, at no cost in parameters.

Eq.~\eqref{eq: repeated-born} alone does not identify $\mathcal{N}$, because a $D$-th root of a channel is not unique.
For a unitary $W$ obeying $W^{D} = \mathds{1}$ whose conjugation $\mathcal{W}[\,\cdot\,] = W \cdot W^\dagger$ commutes with $\mathcal{N}\circ U$, the dressed channel $\mathcal{W}\circ\mathcal{N}$ satisfies
\begin{equation}
    \big(\mathcal{W}\circ\mathcal{N}\circ U\big)^{D}
    = \mathcal{W}^{D}\circ\big(\mathcal{N}\circ U\big)^{D}
    = \big(\mathcal{N}\circ U\big)^{D},
    \label{eq: dressing}
\end{equation}
and therefore predicts the same distribution.
Since $W$ is unitary, $\mathcal{W}\circ\mathcal{N}$ is exactly trace preserving whenever $\mathcal{N}$ is, so the penalty \eqref{eq: tp-penalty} cannot discriminate against it either.

We remove the ambiguity by fitting a single channel to several datasets acquired at different numbers of repetitions $\mathcal{D} = \{D_1,\dots,D_{|\mathcal{D}|}\}$.
The dataset $S_D$ is recorded with the layer repeated $D$ times and holds $N_D$ shots, $S_D = \{(\rho^{\rm in}_{\alpha_m}, \zeta_m(\beta_m))\}_{m=1}^{N_D}$, sampled as in Sec.~\ref{section-tn}: each dataset is a separate acquisition, with its own settings drawn independently.
Writing $\ell_D$ for its negative log-likelihood under Eq.~\eqref{eq: repeated-born},
\begin{equation}
\begin{split}
    \ell_D(\bm\theta)
    &= -\frac{1}{N_D}\sum_{m=1}^{N_D}
       \log p_{D}\big(\bm\zeta_m\,|\,\alpha_m,\beta_m\big),\\
    \mathcal{L}_{\rm joint}(\bm\theta)
    &= \frac{1}{|\mathcal{D}|}\sum_{D\in\mathcal{D}} \ell_D(\bm\theta)
       + \eta\,\bar{D}\,\delta_{\rm TP}(\bm\theta),
\end{split}
\label{eq: joint-loss}
\end{equation}
with $\bar{D} = |\mathcal{D}|^{-1}\sum_{D\in\mathcal{D}} D$ the mean number of repetitions.
Each $\ell_D$ is Eq.~\eqref{eq: kl} evaluated through Eq.~\eqref{eq: repeated-born} in place of Eq.~\eqref{eq: born}, and the same $\bm\theta$ enters every term.
Including $D = 1$ in $\mathcal{D}$ forces $W = \mathds{1}$ in Eq.~\eqref{eq: dressing}, while the repeated datasets still contribute their amplified signal.
The trace-preservation weight is scaled by $\bar{D}$ because the ansatz does not normalise its own output: a channel whose partial trace is $\lambda\mathds{1}$ multiplies every probability predicted at $D$ repetitions by $\lambda^{D}$ and so lowers $\ell_D$ by $D\log\lambda$, whereas the penalty charges for the violation once.
With $\eta = 1.2$ as in Eq.~\eqref{eq: loss}, the effective weight $\eta\bar{D}$ is $1.2$, $1.8$ and $2.8$ for $\mathcal{D} = \{1\}$, $\{1,2\}$ and $\{1,2,4\}$.
The residual $\delta_{\rm TP}$ is always reported for the single channel, so that it remains comparable across $\mathcal{D}$.

\begin{figure}[!tb]
    \centering
    \includegraphics[width=\linewidth]{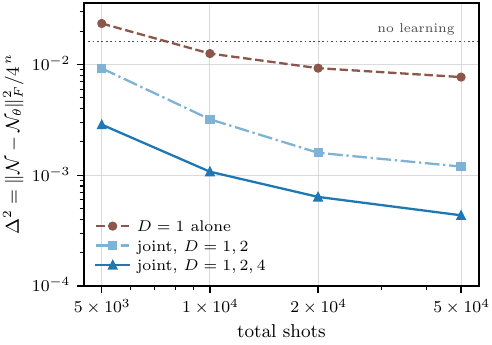}
    \caption[Repeated-layer learning on synthetic data]{\textbf{Learning one channel from repeated applications of the layer.} Squared Frobenius distance \eqref{eq: frob-sq} between the true channel $\mathcal{N}$ and the learnt one $\mathcal{N}_{\bm{\theta}}$, against the total number of shots shared by all the datasets a fit uses, for a correlated pair-dephasing channel on a ten-qubit chain. The three curves are a fit to $D=1$ alone and joint fits, Eq.~\eqref{eq: joint-loss}, over $\mathcal{D} = \{1,2\}$ and $\{1,2,4\}$; all three carry the same number of parameters. For reference, the dotted line is the distance with respect to the identity channel, that is  $\mathcal{N}_{\bm{\theta}}=\mathcal{I}$.}
    \label{fig: tn-repeated-synthetic}
\end{figure}

We validate the procedure in simulation, where the injected channel is known.
A correlated pair-dephasing channel is applied to a ten-qubit chain after the CZ layer, shots are sampled at each number of repetitions, and then we learn the channel with the procedure described above, at $\chi_b = \chi_\kappa = 2$, at which the ground truth is exactly representable.
The figure of merit is
\begin{equation}
    \Delta^{2} = \big\|\mathcal{N}-\mathcal{N}_{\bm\theta}\big\|_F^{2}\big/4^{\,n},
    \label{eq: frob-sq}
\end{equation}
the the normalised squared Frobenius distance between the true and learnt channel.
Fig.~\ref{fig: tn-repeated-synthetic} compares a fit to $\mathcal{D} = \{1\}$ alone with joint fits over $\mathcal{D} = \{1,2\}$ and $\{1,2,4\}$, at a total budget $N$ that the datasets share, $N_D = N/|\mathcal{D}|$.
At $5\times10^{4}$ total shots the $D = 1$ fit reaches $\Delta^2 = 7.7\times10^{-3}$ and the three-dataset joint fit $4.4\times10^{-4}$.
Therefore, from this simulated setting, the method shows improvement in the training using repeated layers. The actual implementation is limited by the current hardware, which accumulates noise in deeper circuits in ways the method may have difficulty reproducing, see Appendix~\ref{app: tn-repeated-hw}.

\section{\label{section-circuits} Benchmarks circuits and layers}
We benchmark the noise-characterisation methods described above on two families of circuits: an adaptation of a Loschmidt echo experiment \cite{filippov2025ole}, and a quantum-chemistry ansatz.
The two are complementary as validation targets.
The echo circuit has a known ideal signal, so any deviation from it is attributable to noise; the chemistry ansatz has no such reference, but is representative of the circuits that error mitigation is actually deployed on.

\subsection{Loschmidt echo experiment}
\label{section-echo-experiment}

Let $U$ be a unitary describing a forward evolution.
The Loschmidt echo is the probability that the initial state $\ket{0}^{\otimes n}$ returns to itself after a forward evolution, a perturbation $V_\theta$ of tunable strength $\theta$, and a backward evolution,
\begin{equation}
    S_\theta
    = \big|\bra{0}^{\otimes n} U^\dagger V_\theta\, U \ket{0}^{\otimes n}\big|^2 .
    \label{eq: loschmidt-ideal}
\end{equation}
On hardware the circuit implements a channel $\mathcal{E}$ rather than the ideal unitary.
The echo circuit can be read out in any observable $O$, and we write the measured signal as
\begin{equation}
    \tilde S_\theta[O]
    = \mathrm{Tr}\big[O\,\mathcal{E}\big(\big(\ket{0}\!\bra{0}\big)^{\otimes n}\big)\big],
    \label{eq: loschmidt-noisy}
\end{equation}
so that $O = \Pi_0 = \big(\ket{0}\!\bra{0}\big)^{\otimes n}$ recovers the Loschmidt echo itself, $\tilde S_\theta[\Pi_0]$, and the same circuit also gives the echo signal of the mean magnetisation $M_z = \tfrac{1}{n}\sum_i Z_i$ and of the parity $Z^{\otimes n}$, which decay at different rates under the same noise channel.
For $\theta = 0$ the perturbation is the identity, the two evolutions cancel exactly and $S_0 = 1$; the measured $\tilde S_0[\Pi_0] < 1$ is therefore a direct readout of the accumulated noise, with no reference simulation required.
As $\theta$ grows the ideal signal decays, quadratically at small $\theta$.
This makes the echo a stringent test of a learnt noise model: the model must reproduce both the offset at $\theta = 0$ and the decay.

\begin{figure}
    \centering
    \includegraphics[width=\linewidth]{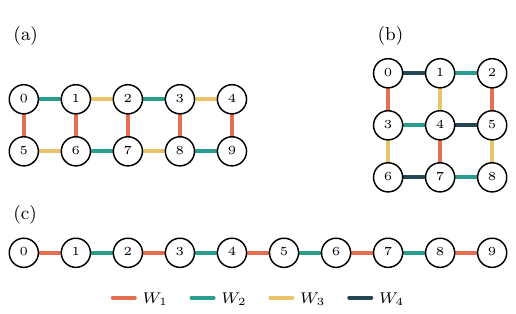}
    \caption{The qubit geometries considered, and the partition of their edges into the subsets $E_\ell$ of Eq.~\eqref{eq: floquet}, one colour to a subset, each executed as one layer $W_\ell$ of simultaneous $\mathrm{CZ}$ gates. (a) The $2\times5$ lattice and (b) the $3\times3$ lattice, used for the Loschmidt echo experiment. (c) The ten-qubit chain, which carries both the one-dimensional Loschmidt echo experiment and the chemistry ansatz of Sec.~\ref{section-chemistry-ansatz}, and whose bonds split into an even and an odd layer. Numbers label the qubits of the lattice, not the physical qubits of the device.}
    \label{fig: LE topology}
\end{figure}

The forward evolution $U$ is a Floquet circuit whose period alternates single-qubit rotations with layers of $\mathrm{CZ}$ gates that follow the qubit connectivity (Ref.~\cite{barron2026observable} gives the general construction).
Let $E$ be the set of edges $\langle j,j'\rangle$ of the qubit graph, partitioned into $L$ disjoint subsets $E_1,\dots,E_L$ such that no two edges of the same subset share a qubit.
These are the colours of Fig.~\ref{fig: LE topology}.
Then
\begin{align}
    U &= U_{\rm F}^{\,t}, \qquad
    U_{\rm F} = W_L R_L \cdots W_2 R_2\, W_1 R_1,
    \label{eq: floquet}\\
    W_\ell &= \prod_{\langle j,j'\rangle \in E_\ell} \mathrm{CZ}_{jj'},
    \nonumber\\
    R_\ell &= \bigotimes_{j \in Q_\ell} R_z(\tfrac{\pi}{4})\, R_x(b),
    \nonumber
\end{align}
where $t$ is the number of Floquet periods and $Q_\ell$ is the set of qubits touched by $E_\ell$.
The gates of each $W_\ell$ act on disjoint pairs and are executed simultaneously.
The perturbation is a global transverse rotation, $V_\theta = \bigotimes_{j=1}^{n} R_x(2\theta) = e^{-i\theta\sum_j X_j}$.

This is a digitised kicked Ising evolution~\cite{kim2023evidence,barron2026observable,leviatan2026prethermal,Haghshenas2026}: the $\mathrm{CZ}$ layers generate an Ising coupling along $z$, $R_z(\pi/4)$ acts as a longitudinal field, and the $R_x(b)$ rotations kick the system along $x$.
In the experiments below we have considered different combinations of $b$ and $t$.
The parameter $t$ simply dictates the depth of the circuit. With the $b$ parameter, on the other hand, we can control the physical properties of the system: in particular how fast it generates magic~\cite{filippov2025ole}, a resource in quantum information theory~\cite{veitch2014resource}.

We consider a one-dimensional chain and two lattice geometries, shown in Fig.~\ref{fig: LE topology}.
The number of CZ layers $L$ per Floquet period is set by the edge-colouring of the lattice, i.e. by the minimum number of layers of pairwise-disjoint gates needed to cover every connection: $L = 2$ for the chain, $L = 3$ for the $2\times5$ lattice and $L = 4$ for the $3\times3$ lattice.
Each of these layers is characterised independently by the procedure of Sec.~\ref{section-tn}, and the learnt channels are then composed to predict the noisy signal \eqref{eq: loschmidt-noisy}.

\subsection{Chemistry ansatz}
\label{section-chemistry-ansatz}

\par The second validation target is a quantum-chemistry ansatz. We
consider a linear chain of five hydrogen atoms, equally spaced along the
$z$ axis with an interatomic distance of $2.5$~\AA. At this separation
the chain is in the strongly correlated regime: each atom is close to
singly occupied, the spin correlations are antiferromagnetic, and the
natural-orbital occupancies of the active space approach unity for all
five orbitals, which is the signature of maximal static correlation. The
electronic structure was computed with PySCF
\cite{sun_recent_2020} in the cc-pVDZ basis. We use a
$\mathrm{CAS}(5,5)$ active space built from canonical Hartree--Fock
orbitals, one spatial orbital per hydrogen atom, with $(3,2)$ electrons
of $\alpha$ and $\beta$ spin, corresponding to the $M_S = 1/2$ doublet
ground state of the five-electron system. Under the Jordan--Wigner
mapping the five spatial orbitals give ten spin-orbitals, hence a
ten-qubit problem. The CASCI energy in this active space, also obtained
with PySCF, is $E_{\rm CASCI} = -2.46471$~Ha and serves as the classical reference.
\par The ansatz was generated with ADAPT-VMPE
\cite{chakraborty_scalable_2026}, which builds a fermionic circuit
iteratively using Majorana propagation to evaluate candidate gates
classically. The construction was stopped at a deliberately small number
of gates, giving a state of energy $E_{\rm ADAPT} = -2.45466$~Ha, that is
$\Delta E = |E_{\rm CASCI} -E_{\rm ADAPT}| =  1.005\times10^{-2}$~Ha above CASCI. Transpiled to the native gate set of
the device, $\{\mathrm{CZ}, \mathrm{PRX}\}$, on a linear qubit chain, the circuit
contains $36$ $\mathrm{CZ}$ and $67$ $u_3$ gates.

\par The ten qubits are mapped onto the $2\times5$ lattice of
Fig.~\ref{fig: LE topology}, the same layout used for the
Loschmidt echo experiment, with the linear chain of the transpiled
circuit embedded as a snake path through the lattice. Each layer of concurrent $\mathrm{CZ}$ gates appearing
in the circuit was characterised independently with the procedure of
Sec.~\ref{section-tn}, and the learnt channels were composed to predict the
noisy signal.
\par Unlike the echo circuit, this ansatz has no analytically known ideal
signal, which is why it is representative of the circuits that error
mitigation is deployed on in practice. The validation observable is the
active-space Hamiltonian mapped under Jordan--Wigner, so the quantity
compared with the noise-model prediction is the energy itself. The energy quoted above is the exact expectation value
of the ansatz, obtained with a Majorana-propagation cutoff beyond the
regime where truncation contributes; the $1.005\times10^{-2}$~Ha
separating it from CASCI reflects the small number of gates at which the
iterative construction was stopped. It is therefore the noiseless ansatz
value, not CASCI, against which the measured signal and the noise-model
prediction are compared.

\section{Experimental analysis of noise structure and stability}
\label{sec:noise analysis}
\par The hardware experiments reported in this section were run with the
same settings as the simulations of Sec.~\ref{section-cb}: nine measurement
bases, $256$ shots per randomisation, $100$ randomisations per depth, and even
depths up to $d = 64$. All noise model coefficients are obtained from the
regularised fit in Eq.~\eqref{eq: regularized-nnls}, without any fidelity
augmentation. We first characterise a four-qubit chain and a ten-qubit
$2\times5$ grid and test the learnt models against the measured Pauli
fidelities. We then ask whether the characterisation of a layer already
determines the noise of its sublayers, for both methods of
Secs.~\ref{section-cb} and \ref{section-tn}. We close with the stability of the
learnt model over time. Results for a ten-qubit one-dimensional chain and a
nine-qubit $3\times3$ grid are collected in
Appendix~\ref{app: additional-layouts}.

\subsection{Four-qubit one-dimensional chain}\label{section-4q-results}
\begin{figure}[!ht]
    \centering
    \includegraphics[width=1\linewidth]{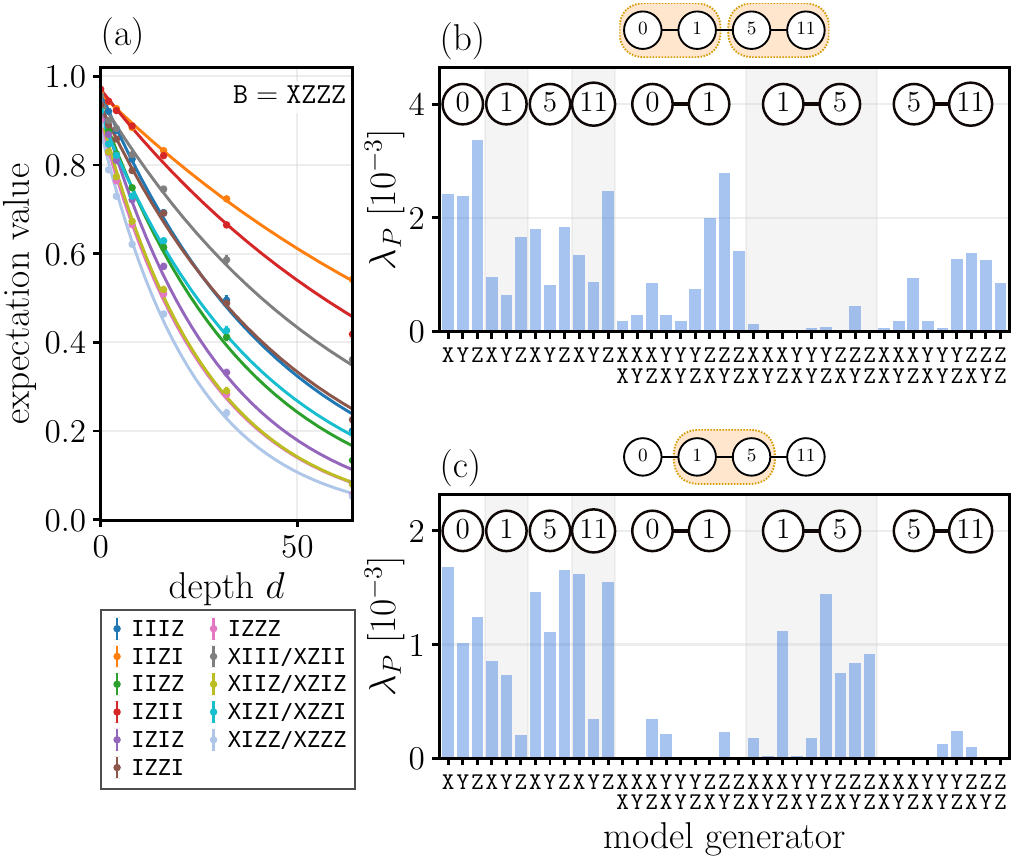}
    \caption[Four-qubit experiment results from quantum hardware]{\textbf{Four-qubit experiment.} Characterisation results for the two $\mathrm{CZ}$ layers. Panels (a) and (c) show the measured decay of all accessible Pauli expectation values $P_i$, listed in the legends, obtained from a single measurement basis, e.g. $\mathrm{B = XZZZ}$ for the first layer. The fidelity pairs $f_if_j$ are obtained by fitting exponentially decaying curves of the form $\alpha(f_if_j)^{d/2}$. The extracted products are then used to solve the regularised nonnegative least-squares problem \eqref{eq: regularized-nnls}. The resulting noise-model parameters for the first and second $\mathrm{CZ}$ layers are shown in panels (b) and (d). These are the rates associated with all weight-one and weight-two Pauli generators of the sparse Pauli-Lindblad model.}
    \label{fig: 4q-experiment-results-cz}
\end{figure}
\par We benchmark the same two layers considered in the simulations, on
the physical layout \texttt{[0, 1, 5, 11]}\footnote{On VTT's device the qubit indexing starts at $1$, so the physical qubits used in the experiments are obtained by subtracting one from the corresponding index. An additional shift is required for qubits ordered after a non-working qubit.}
describing a one-dimensional chain. Figs.~\ref{fig: 4q-experiment-results-cz}(a--b) show the results for the first layer, consisting of two concurrent $\mathrm{CZ}$ gates.
The decay curves follow the simulated ones: as the depth $d$ increases,
higher-weight Paulis decay faster. The learnt coefficients are largest
for the weight-two generators whose support coincides with the qubit
pairs on which the gates act.
\begin{figure}[!t]
    \centering
    \includegraphics[width=\linewidth]{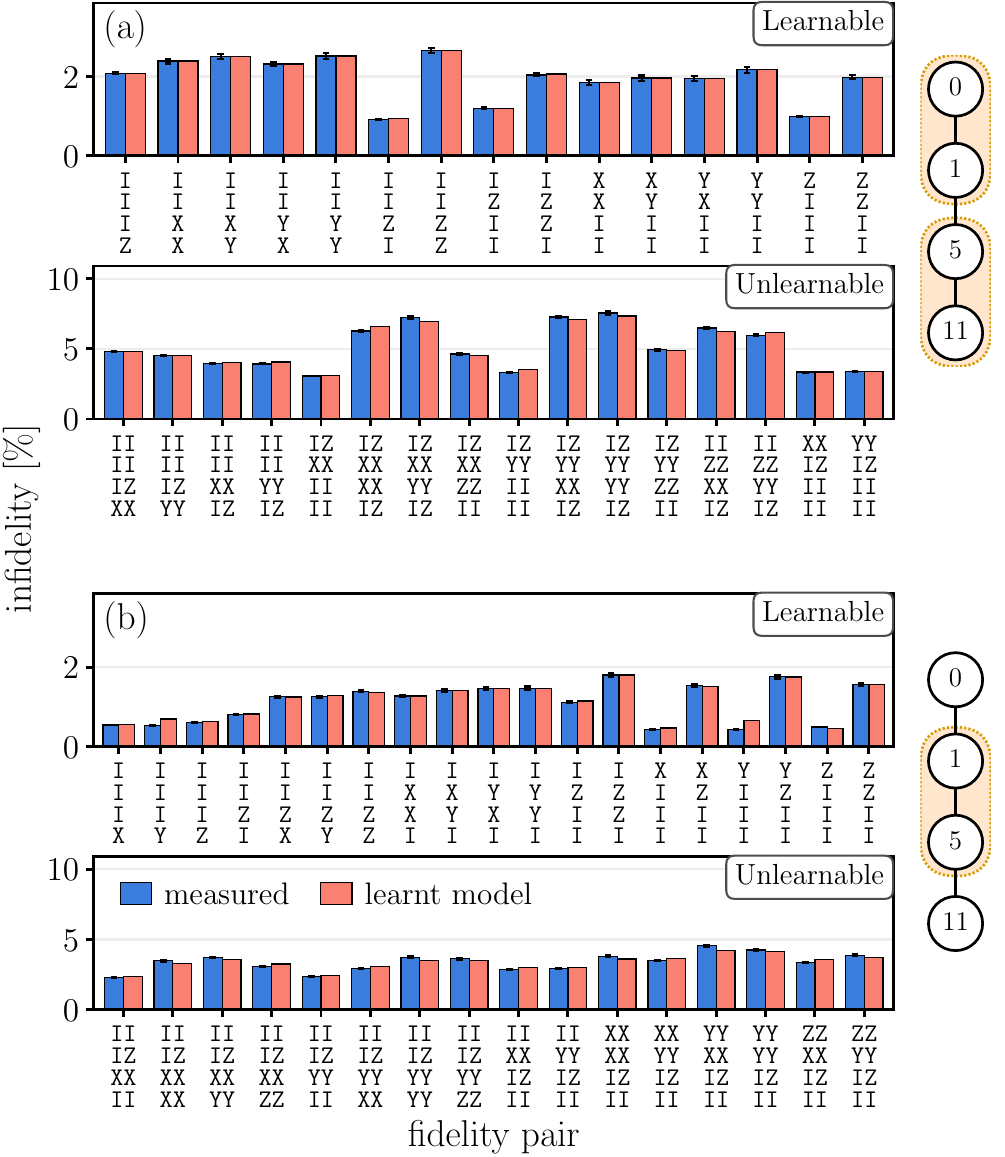}
    \caption[Four-qubit fidelity reliability test from experimental data]{\textbf{Four-qubit experiment.} Fidelity pairs fitted from the decay of Pauli expectation values during cycle benchmarking for both characterised layers. A reliability test for the noise model $\Lambda$ is to compare the measured fidelity pairs with the model-predicted products $f_if_j$, where $f_i = 2^{-n}\operatorname{tr}(P_i^{\dagger}\Lambda(P_i))$. Cycle benchmarking gives access not to all individual fidelities but only to the learnable ones and to learnable products. For each experiment, (a) and (b), the upper panel shows squared learnable fidelities, e.g. $f_{\mathrm{IIXX}}$ for the first layer, while the lower panel shows learnable products of individually unlearnable fidelities, e.g. $f_{\mathrm{IIXI}}\cdot f_{\mathrm{IZXI}}$ for the second layer.}
    \label{fig: experiments-results-cz-4q-fidelities}
\end{figure}
\par The second layer, a single $\mathrm{CZ}$ on the pair $(q_1,q_5)$, is
shown in Figs.~\ref{fig: 4q-experiment-results-cz}(c--d), with the
largest rates again on the edge where the gate acts. In both
characterisations the weight-one generators also carry sizeable rates
relative to the weight-two ones. These may capture local mechanisms such
as relaxation and dephasing, but their contribution is not restricted to
local errors, since such processes can also induce effective weight-two
parameters \cite{malekakhlagh_efficient_2025}. The learnt model is an
effective channel that accounts for the two-qubit errors together with
the single-qubit errors introduced by the twirling and interleaving gates
during benchmarking, and its behaviour can be read as a spreading of
incoherent noise between weight-one and weight-two generators at fixed
total error rate $\sum_i\lambda_i$ \cite{malekakhlagh_efficient_2025}.
\par Fig.~\ref{fig: experiments-results-cz-4q-fidelities} compares the
measured fidelity pairs with those predicted by the learnt model. The two
agree across both layers, which supports the predictive reliability of
the reconstruction.

\subsection{Ten-qubit two-dimensional grid}\label{section-10q-2d-results}
\begin{figure}[!t]
    \centering
    \includegraphics[width=\linewidth]{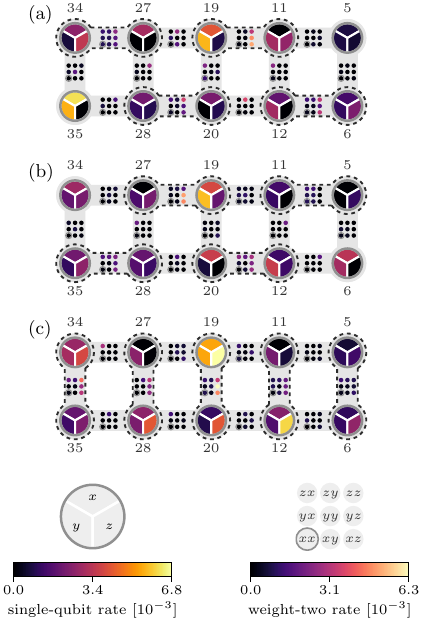}
    \caption[Ten-qubit two-dimensional experimental results]{\textbf{Ten-qubit two-dimensional experiment.} Learnt model coefficients for the three characterised Clifford layers. The qubit pairs on which the gates act are indicated by dotted lines. The layers are chosen so that every nearest-neighbour connection in the grid is covered. Single-qubit error rates are shown as pie-style circles at each qubit, and weight-two coefficients as smaller circles on the corresponding edges; the keys and the two colour scales at the bottom are shared by all three panels, so the layers can be compared directly.}
    \label{fig: experiment-results-10q-2d-coeff}
\end{figure}
\par We then characterise a two-dimensional layout, the $2\times5$ grid of Fig.~\ref{fig: experiment-results-10q-2d-coeff}. Three layers of $\mathrm{CZ}$ gates were characterised, chosen so that all connections between
neighbouring qubits are covered, which gives access to the crosstalk on every pair in the layout. The three layers are those into which the Floquet evolution of Sec.~\ref{section-circuits} decomposes.
\par The procedure is unchanged from
Sec.~\ref{section-cb}: the nine measurement bases are
generated for the two-dimensional layout, cycle benchmarking yields the
Pauli fidelities in a SPAM-free manner, and the coefficients are obtained from the regularised fit. Fig.~\ref{fig: experiment-results-10q-2d-coeff}
illustrates the results. As in the one-dimensional case the largest rates sit on the edges where the gates act. The single-qubit rates are largest on
qubit $19$, consistently across layers and in agreement with other
layouts, which identifies that vertex as the noisiest in the register.
\begin{figure}[!t]
    \centering
    \includegraphics[width=\linewidth]{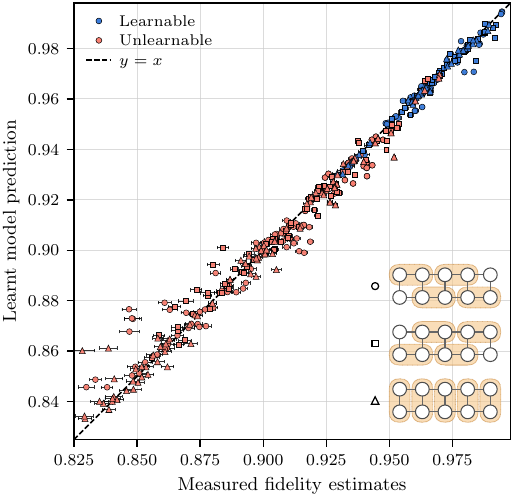}
    \caption[Ten-qubit two-dimensional fidelity reliability test from experimental data]{\textbf{Ten-qubit two-dimensional experiment.} Measured fidelities against the corresponding predictions of the learnt noise model, for the three characterised layers of $\mathrm{CZ}$ gates on the $2\times5$ grid, collected in a single panel. Each point is a measured fidelity and its model-reconstructed estimate: blue for the individually learnable fidelities, red for the learnable products of individually unlearnable ones. The dotted line is the ideal relation $y = x$.}
    \label{fig: experiment-results-10q-2d-fidelitities}
\end{figure}
\par As shown in Fig.~\ref{fig: experiment-results-10q-2d-fidelitities}, the measured fidelities of all three layers lie close to the model predictions, indicating that the model captures the noise on a two-dimensional qubit topology as well as on a chain. Together with the additional layouts presented in Appendix~\ref{app: additional-layouts}, these results show that the reconstructions behave as expected throughout: the inferred rates are largest on the qubits and pairs involved in the $\mathrm{CZ}$ gates, while the fidelities predicted by the learnt model agree with those measured by cycle benchmarking.

\par The two characterisations above each describe one fixed layer of
concurrent $\mathrm{CZ}$ gates. A deep circuit is built from many distinct
layers, and characterising each of them separately is the dominant
experimental cost. The next two subsections ask whether that cost can be
avoided, by testing on both methods whether the model of a layer can be cut
down to the model of a sublayer.

\subsection{Trimming a layer model learnt by cycle benchmarking}\label{sec: noise-trimming-results}
\par A layer of concurrent two-qubit gates contains sublayers, obtained
by retaining only a subset of its gates \footnote{Given a layer of non-overlapping two-qubit Clifford operations, a sublayer is a layer containing a subset of those operations.}. The question is whether the characterisation of the parent layer already determines the noise of its sublayers.
\begin{figure}[!t]
    \centering
    \includegraphics[width=\linewidth]{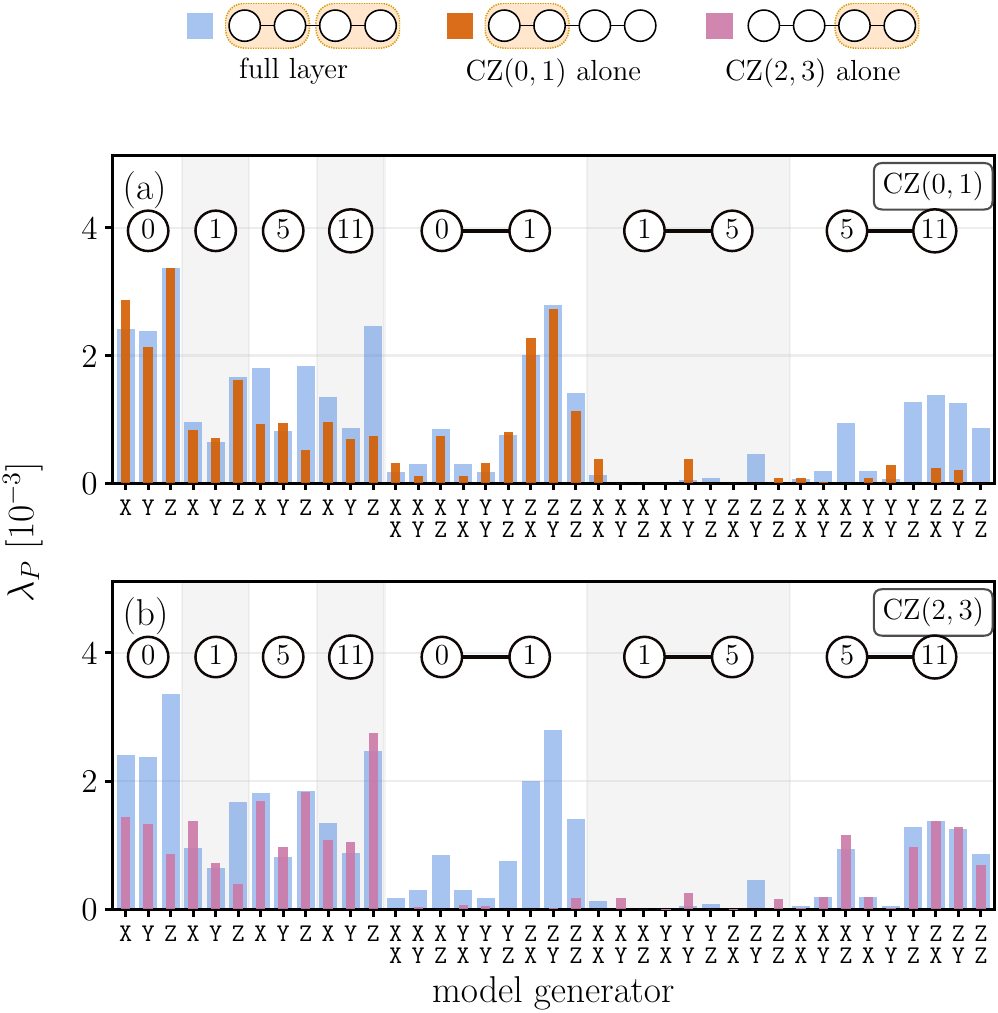}
    \caption[Noise trimming for concurrent Clifford layers]{%
    \textbf{Noise trimming.}
    Learnt noise model coefficients for the full layer of two concurrent
    $\mathrm{CZ}$ gates on the four-qubit register, compared with those
    learnt from layers containing a single $\mathrm{CZ}$ gate on either
    the first, (a), or the second, (b), qubit pair. Each colour of the
    legend is paired with the four-qubit chain it was characterised on,
    physical qubits $0,1,5,11$, in the order the bar groups name, with a
    dashed capsule over the $\mathrm{CZ}$ gates that characterisation
    drove: both gates for the full layer, one for each sublayer. The
    circled numbers above the bars give the support of each group of
    generators.}
    \label{fig:noise-trimming}
\end{figure}
\par We tested this on the four-qubit register by comparing the
coefficients learnt for the layer of two concurrent $\mathrm{CZ}$ gates
with those learnt from each single-gate sublayer, characterised
independently. Both the weight-one and the weight-two parameters
obtained from the parent layer reproduce the corresponding coefficients
of the stand-alone layers, as shown in
Fig.~\ref{fig:noise-trimming} for both independent characterisation experiments. The approximate form of noise model for a
sublayer can therefore be obtained at no additional experimental cost by nullifying the rates for irrelevant two-qubit generators and substituting qubit-wise decoherence rates (e.g., from the conventional $T_1$-$T_2$ noise model upon twirling) for irrelevant one-qubit generators.

\subsection{Trimming a layer channel learnt by tomography}\label{section-tn-trim}
\par The same question can be asked for the channel tomography of
Sec.~\ref{section-tn}, where the answer is not a subset of coefficients but a
cut through a tensor network. A channel learnt for a complete $\mathrm{CZ}$
layer is a single ten-qubit object. We ask whether it can be cut at the chain bonds that fall \emph{between} $\mathrm{CZ}$ pairs, so that the layer channel factorises into two-qubit blocks, and whether the block so obtained agrees with the channel learnt from an experiment that drives only that $\mathrm{CZ}$. Both questions are answered using the $16$ layer characterisations of the chemistry
ansatz described in Sec.~\ref{section-tn-chem}.
\begin{figure}[!tb]
    \centering
    \includegraphics[width=\linewidth]{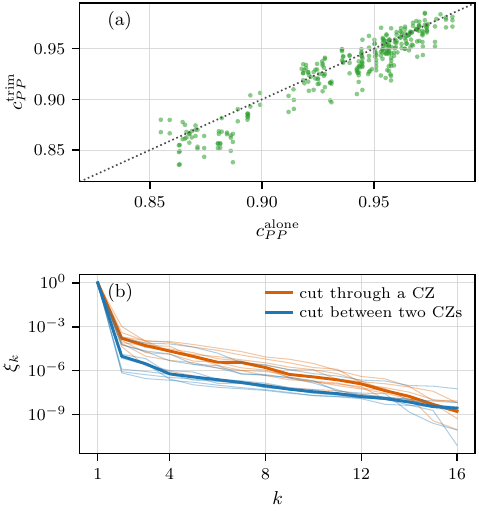}
    \caption[Trimming a learnt layer channel]{\textbf{Layer trimming.} (a) The $15$ non-trivial two-qubit diagonal Pauli transfer coefficients $c_{PP}$ of Eq.~\eqref{eq: ptm}, for $20$ (multi-$\mathrm{CZ}$ circuit, $\mathrm{CZ}$ pair) combinations. $c^{\rm trim}_{PP}$ is computed from the block obtained by cutting a multi-gate layer channel at the bonds between $\mathrm{CZ}$ pairs, and $c^{\rm alone}_{PP}$ from the channel learnt from an experiment driving that pair alone. The dotted line is $y = x$. (b) $\xi_k$, the normalised squared Schmidt spectrum across a bond, i.e. the fraction of the correlation carried by the $k$-th term of the Schmidt decomposition, with the terms ordered from strongest to weakest. It is shown at each of the nine chain bonds of the two complete layer channels, with the bonds separated into those a $\mathrm{CZ}$ straddles and those lying between two $\mathrm{CZ}$ pairs. Thin lines are individual bonds, thick lines their medians.}
    \label{fig: tn-trimming}
\end{figure}
\par Fig.~\ref{fig: tn-trimming}(a) compares the diagonal Pauli transfer
coefficients $c_{PP}$ of the trimmed block and of the independently learnt channel. The two-qubit channel learnt for each single $\mathrm{CZ}$ is compared with the block trimmed from every multi-gate layer channel that contains that $\mathrm{CZ}$, which gives $20$ combinations. The median absolute deviation over the $15$ non-trivial coefficients is $0.0090$, and the median relative Frobenius distance between the two channels is $0.0499$, ranging from $0.0208$ to $0.0774$. 
\par Fig.~\ref{fig: tn-trimming}(b) shows where the correlation sits. The
median value of the Schmidt coefficients is smaller on the bonds between
$\mathrm{CZ}$ pairs than on the bonds a $\mathrm{CZ}$ straddles. This suggests
that either the crosstalk is small, which would justify the trimming, or the
method is not sensitive to it.

\subsection{Noise model stability}\label{section-drift}
\par All the characterisations above are single snapshots of the device. The
validity of a learnt model over time bounds how often the
characterisation has to be repeated, and the drift of the noise
parameters is the systematic error common to all the mitigation
protocols discussed in Sec.~\ref{section-introduction}. We probe it by repeating
the characterisation of a layer of two concurrent $\mathrm{CZ}$ gates on
the four-qubit register over a window of about four hours, with
successive experiments separated by roughly one hour; each
characterisation takes close to $24$ minutes.
\begin{figure}[!t]
    \centering
    \includegraphics[width=1\linewidth]{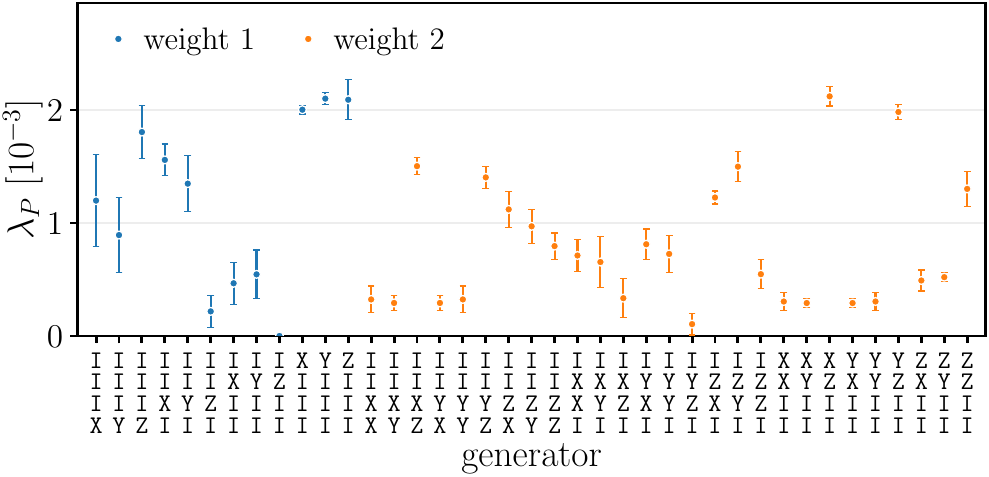}
    \caption[Drift study]{\textbf{Drift study.} Four SPL characterisations of the same layer over a window of $4.6$ hours. Each dot is the mean learnt rate $\lambda_P$ of one generator across the four experiments and the error bar its standard deviation. Blue: weight-one generators. Orange: weight-two generators.}
    \label{fig: drift-study-coeff}
\end{figure}
\par The mean coefficient per generator and its spread across the
experiments are shown in Fig.~\ref{fig: drift-study-coeff}. Two
features distinguish this dataset from the single characterisations of
Secs.~\ref{section-4q-results} and \ref{section-10q-2d-results}: several coefficients on the qubit pair carrying
no gate come out nonzero, and the single-qubit rates vary appreciably
between experiments while the weight-two rates stay close to their
averages.
\par Rather than tracking the coefficients themselves, it is convenient
to follow the accumulation factors
\begin{equation}
\begin{aligned}
\gamma_i &= \prod_{k \in \mathcal{K}_i} e^{2\lambda_k}, \\
\gamma_{i,j} &= \prod_{k \in \mathcal{K}_{i,j}} e^{2\lambda_k},
\end{aligned}
\label{eq: gammas}
\end{equation}
where $\mathcal{K}_i$ and $\mathcal{K}_{i,j}$ are the sets of weight-one
and weight-two generators. Their product over the whole generator set,
\begin{equation}
\gamma = \prod_{k \in \mathcal{K}} e^{2\lambda_k},
\label{eq: gamma-total}
\end{equation}
is the sampling overhead of probabilistic error cancellation
\cite{berg_probabilistic_2023}. An ideal implementation of the
$\mathrm{CZ}$ layer would give $\gamma_i$ and $\gamma_{i,j}$ close to
unity, and noisier gates give larger values, so these quantities track
single-qubit and pairwise fluctuations over time on a common scale.
\begin{figure}[!t]
    \centering
    \includegraphics[width=\linewidth]{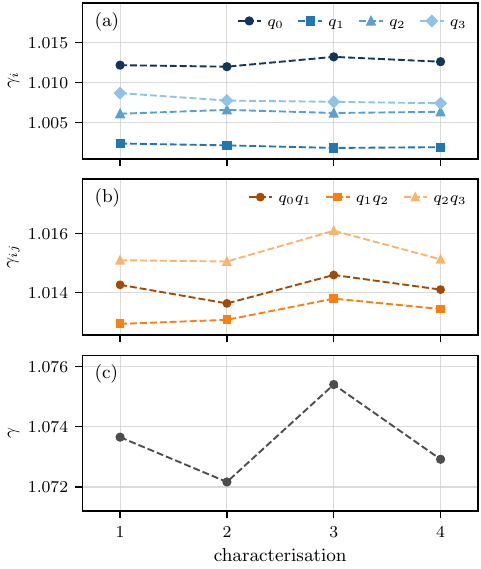}
    \caption[Noise accumulation factors]{Noise accumulation factors for each of the four repeated characterisations of the drift study, over a common characterisation axis: (a) the weight-one factors $\gamma_i$ and (b) the weight-two factors $\gamma_{i,j}$ of Eq.~\eqref{eq: gammas}, and (c) the overall factor $\gamma$ of Eq.~\eqref{eq: gamma-total}. Note the vertical scales, which differ from panel to panel: the four overall values span $3\times10^{-3}$ about a mean of $1.0735$.}
    \label{fig: noise-accumulation-factors}
\end{figure}
\par Panels (a) and (b) of
Fig.~\ref{fig: noise-accumulation-factors} show $\gamma_i$ and
$\gamma_{i,j}$ per characterisation. All values stay close to unity, which
is consistent with the calibration data used to select the register, where
both the single-qubit and the coupler metrics were favourable throughout.
The overall factor $\gamma$, panel (c), behaves likewise, with a slight
increase in the third characterisation relative to the average.
\par Finally we compute the error rate per qubit per layer, or error
density \cite{filippov2024scalability},
\begin{equation}
\varepsilon \equiv \sqrt[NL]{\gamma} - 1 ,
\label{eq: error-density}
\end{equation}
with $N$ the number of qubits and $L$ the number of layers, here $N = 4$
and $L = 1$. We obtain $\varepsilon = 0.0178 \pm 0.0003$, hence a noise
stability $\Theta = \Delta\varepsilon/\varepsilon = 1.59\%$ over four
hours. The drift is therefore small on this timescale, and a model learnt
now remains usable for several hours, which is the relevant statement for
error-mitigated experiments on registers of this size.

\section{Experimental results}\label{section-experimental-results}
\par We now use the noise models learnt in Sec.~\ref{sec:noise analysis} to
predict the outcome of circuits that were not part of any characterisation: the
Loschmidt echo circuit and the chemistry ansatz of
Sec.~\ref{section-circuits}. We start from the sparse Pauli--Lindblad models
of Sec.~\ref{section-cb} on the echo circuit, and then turn to the channels
learnt by the tomography of Sec.~\ref{section-tn}, which we apply to the echo
circuit on both a chain and a lattice and to the chemistry ansatz.

\subsection{Echo dynamics from the sparse Pauli--Lindblad model}\label{section-echo-validation}
\par The first target is the
Loschmidt echo circuit of Sec.~\ref{section-echo-experiment}, run with $t = 2$ Floquet period, magic parameter $b$, and
perturbation angles $\theta = \{0, 0.2, 0.3\}$.
A Pauli--Lindblad channel with the experimentally learnt rates is
attached to each layer of $\mathrm{CZ}$ gates, and we compute the echo
signal, namely the expectation value of the observable $|0\rangle\!\langle 0|^{\otimes n}$. The
prediction is then compared with the signal measured by running the same
circuit on hardware. The characterisation was run contemporaneously with
the circuit executions, so the comparison uses a reliable snapshot of the device noise rather than an earlier learnt noise model.
\par Single-qubit gates are treated as noiseless, and state-preparation
and measurement errors are corrected on the observable using TREX
coefficients built from the depth-zero cycle benchmarking data. For qubit
$q$, basis $\mathrm{B}$ and randomisation $r$, let $C_{b,r}(x)$ be the
depth-zero counts of bitstring $x$ and $N$ the number of shots. The
readout bit-flip rate is
\[
\varepsilon_{b,r}(q) = \frac{1}{N}\sum_{x:\, x_q = 1} C_{b,r}(x),
\]
and averaging over the $R_b$ randomisations and then over the $B$ bases
gives
\[
\bar{\varepsilon}_b(q) = \frac{1}{R_b} \sum_{r=1}^{R_b} \varepsilon_{b,r}(q),
\qquad
\varepsilon(q) = \frac{1}{B} \sum_{b=1}^{B} \bar{\varepsilon}_b(q).
\]
Averaging over the two characterisation experiments gives the per-qubit
SPAM error rate $p_q = \tfrac{1}{2}(\varepsilon^{(1)}(q) + \varepsilon^{(2)}(q))$
and the TREX coefficient
\begin{equation}
    \mathsf{trex}(q) = 1 - 2 p_q .
    \label{eq: trex-coeff}
\end{equation}
For a Pauli string $P$ with support $\mathfrak{s}_P = \{q : P_q \neq I\}$
the attenuation factor is $\prod_{q\in\mathfrak{s}_P}\mathsf{trex}(q)$,
so the corrected observable reads
\begin{equation}
    \tilde{O} = \sum_{P} c_P \left(\prod_{q \in \mathfrak{s}_P} \mathsf{trex}(q)\right) P .
    \label{eq: trex-observable}
\end{equation}
\begin{figure}[t]
    \centering
    \includegraphics[width=1\linewidth]{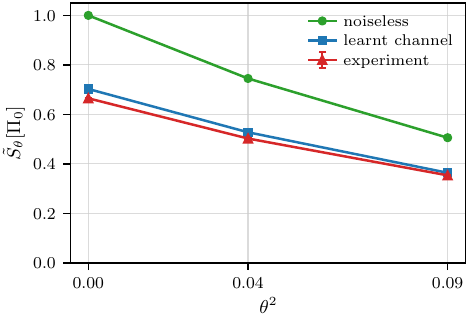}
    \caption[Echo signal experiment on four qubits]{\textbf{Echo signal on a four-qubit chain.} Signal $\tilde S_\theta[\Pi_0]$ of the Floquet circuit at $n = 4$, $t = 2$ and $b = 0.4\pi$, against $\theta^2$, at the three angles the device was run at. Green: noiseless value. Blue: simulation with the learnt sparse Pauli--Lindblad channel inserted before each Clifford layer. Red: hardware measurement, with its standard error.  Colours as in Fig.~\ref{fig: tn-1d-theta}.}
    \label{fig: 4q-echo-signal}
\end{figure}
\begin{figure}[t]
    \centering
    \includegraphics[width=1\linewidth]{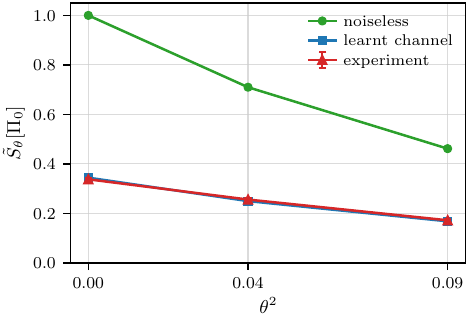}
    \caption[Echo signal experiment on a ten-qubit chain]{\textbf{Echo signal on a ten-qubit chain.} As Fig.~\ref{fig: 4q-echo-signal}, for the Floquet circuit at $n = 10$, $t = 2$ and $b = 0.1\pi$. The simulation with the learnt model and the hardware data both quantify the reduction of the signal from its noiseless value (green).}
    \label{fig: 10q-echo-signal}
\end{figure}
\par Figs.~\ref{fig: 4q-echo-signal} and \ref{fig: 10q-echo-signal}
show the echo signal against $\theta^2$ for the four- and ten-qubit
circuits, for which complete noise models were available. In both cases
the prediction from the learnt model tracks the measured signal, so the
sparse Pauli--Lindblad model captures the noise mechanisms that dominate
these dynamics. The agreement is closer for the ten-qubit circuit than
for the four-qubit one. A plausible reason is that on four qubits the
error contributions from interactions between the active register and
neighbouring spectator qubits are comparable in magnitude to the noise
acting within the register, whereas characterising a larger register
gives access to more Pauli fidelities and hence a more complete
description of the noise during gate execution. Characterisations at
intermediate sizes between four and ten qubits would test this directly.

\subsection{Ten-qubit one-dimensional chain}\label{section-tn-1d}
\par The remaining subsections repeat this experiment with the channel
tomography of Sec.~\ref{section-tn}, which can also address the two-dimensional layout. Each layer of concurrent
$\mathrm{CZ}$ gates is characterised independently, from $N_{\rm set} = 1000$ settings at $N_{\rm shots} = 200$ shots each, i.e. $N = N_{\rm set}N_{\rm shots}= 2\times10^{5}$ per layer, divided into training and test samples with an $80\%/20\%$ split. The effects entering the likelihood \eqref{eq: kl} are the noisy
POVMs reconstructed by QDT on the same acquisition, while the input states are considered noiseless, so that only the
$\mathrm{CZ}$ layers are characterised. Every acquisition reported here is
Pauli-twirled. Every figure and every number below is a single fit per layer,
with a random initialisation fixed by a single seed. The spread over two further seeds is reported in
Appendix~\ref{app: tn-seeds}.

\par The first target is the Floquet circuit of
Sec.~\ref{section-echo-experiment} on a ten-qubit chain, at
$b = 0.1\pi$, Floquet depth $t = 2$ and perturbation angles
$\theta = \{0, 0.2, 0.3\}$. The brickwork period decomposes into two layers,
Fig.~\ref{fig: LE topology}(c), each characterised
separately with an LPDO at $\chi_b = \chi_\kappa = 2$.
Details on the training can be found in Appendix~\ref{app:10q 1d training}.
\begin{figure}[!tb]
    \centering
    \includegraphics[width=\linewidth]{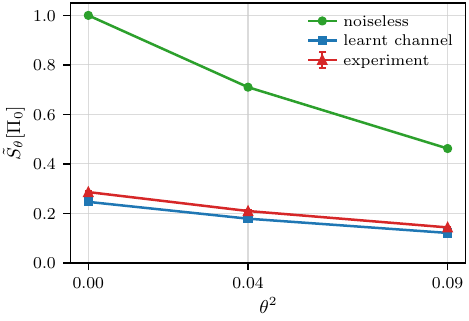}
    \caption[Echo signal on a ten-qubit chain, channel tomography]{\textbf{Ten-qubit chain, echo signal of $\Pi_0$.} The signal $\tilde S_\theta[\Pi_0]$ of Eq.~\eqref{eq: loschmidt-noisy}, i.e. the echo circuit read out in $\Pi_0 = |0\rangle\!\langle0|^{\otimes n}$, against $\theta^2$ for the Floquet circuit at $b = 0.1\pi$ and $t = 2$. All three series are given at the three angles the device was run at, joined by straight segments. Green: noiseless statevector value. Blue: prediction obtained by inserting the learnt channel after each completed $\mathrm{CZ}$ layer. Red: hardware measurement, with its standard error. The remaining two observables are in Appendix~\ref{app: tn-observables}.}
    \label{fig: tn-1d-theta}
\end{figure}
\par The learnt channels are then inserted after every completed $\mathrm{CZ}$
layer and $\tilde S_\theta[\Pi_0]$ is propagated in Pauli transfer form.
Fig.~\ref{fig: tn-1d-theta} compares the result with the measurement. At
$\theta = 0$ the noiseless signal is exactly $1$, the device returns
$0.286 \pm 0.003$ and the model $0.247$. The measured value at this point is
already a first estimate of the noise strength of the device, obtained from the
theoretical properties of the echo circuit alone and before any noise learning.

\begin{table}[!t]
    \centering
    \caption{Echo signal on the ten-qubit chain: hardware measurement against the
    channel-tomography prediction, at each perturbation angle $\theta$ of
    Fig.~\ref{fig: tn-1d-theta}.}
    \label{tab: tn-1d-theta}
    \begin{tabular}{cccc}
        \toprule
        $\theta$ & Measurement & Prediction & Deviation \\
        \midrule
        $0$   & $0.2861 \pm 0.0028$ & $0.2465$ & $13.8\%$ \\
        $0.2$ & $0.2095 \pm 0.0024$ & $0.1786$ & $14.8\%$ \\
        $0.3$ & $0.1432 \pm 0.0017$ & $0.1209$ & $15.6\%$ \\
        \bottomrule
    \end{tabular}
\end{table}

Table~\ref{tab: tn-1d-theta} summarises the comparison: the prediction is roughly $14.5\%$ below the measurement across the scan. 
The divergence increases with $\theta$: this is expected because the noise becomes more and more correlated, making it more difficult to learn it in this setting. The procedure described in Sec.~\ref{section-tn-repeated} could help improve this. 

\subsection{Ten-qubit two-dimensional lattice}\label{section-tn-2d}
\par The same circuit run on the $2\times5$ lattice of Fig.~\ref{fig: LE topology} requires three $\mathrm{CZ}$ layers per brickwork period, labelled A, B and C, and the LPDO bond graph is the lattice rather than a chain.
Each is characterised by an LPDO whose virtual-bond graph is the lattice itself, at $\chi_b = \chi_\kappa = 2$ and batch size $T=500$.
\begin{figure}[!tb]
    \centering
    \includegraphics[width=\linewidth]{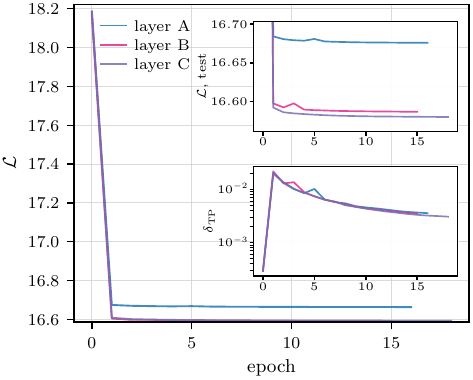}
    \caption[Training of the two-dimensional $2\times5$ channels]{\textbf{$2\times5$ lattice, training.} Total loss \eqref{eq: loss} per epoch for the three $\mathrm{CZ}$ layers A, B and C of the brickwork period, labelled as in Fig.~\ref{fig: LE topology}. Upper inset: $\mathcal{L}$ on the held-out test shots. Lower inset: the residual $\delta_{\rm TP}$ over the same runs, on a logarithmic scale. Both insets share the epoch axis of the main panel and are shown only to establish that they converge; the values they settle at are quoted in the text.}
    \label{fig: tn-2d-training}
\end{figure}
\par The three fits, Fig.~\ref{fig: tn-2d-training}, stop after $16$, $15$ and
$18$ epochs on A, B and C at $D_{\rm KL}$ between $16.587$ and $16.659$, and $\mathcal{L}$
between $16.580$ and $16.676$ on the test shots. The residual
$\delta_{\rm TP}$ settles at $3.1$--$3.6\times10^{-3}$, about twice its
one-dimensional value at the same shot count (Appendix~\ref{app: tn-additional}).
\begin{figure}[!tb]
    \centering
    \includegraphics[width=\linewidth]{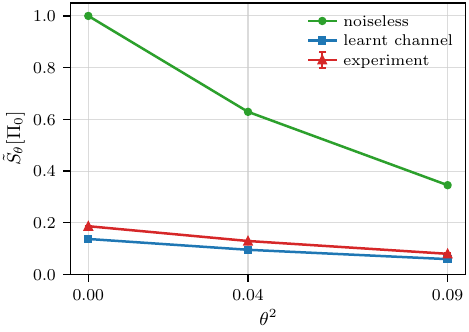}
    \caption[Echo signal on the $2\times5$ lattice, channel tomography]{\textbf{$2\times5$ lattice, echo signal of $\Pi_0$.} The signal $\tilde S_\theta[\Pi_0]$ against $\theta^2$, with the three learnt lattice channels inserted after their respective $\mathrm{CZ}$ layers. Colours as in Fig.~\ref{fig: tn-1d-theta}. The remaining two observables are in Appendix~\ref{app: tn-observables}, the nine-qubit $3\times3$ lattice in Appendix~\ref{app: tn-3x3}.}
    \label{fig: tn-2d-theta}
\end{figure}
\par Fig.~\ref{fig: tn-2d-theta} compares the predicted and measured echo
signal, with exact values in Table~\ref{tab: tn-2d-theta}.
\begin{table}[!t]
    \centering
    \caption{Echo signal on the $2\times5$ lattice: hardware measurement against
    the channel-tomography prediction, at each perturbation angle $\theta$ of
    Fig.~\ref{fig: tn-2d-theta}.}
    \label{tab: tn-2d-theta}
    \begin{tabular}{cccc}
        \toprule
        $\theta$ & Measurement & Prediction & Deviation \\
        \midrule
        $0$   & $0.1865 \pm 0.0021$ & $0.1372$ & $26.5\%$ \\
        $0.2$ & $0.1293 \pm 0.0016$ & $0.0953$ & $26.3\%$ \\
        $0.3$ & $0.0797 \pm 0.0011$ & $0.0591$ & $25.8\%$ \\
        \bottomrule
    \end{tabular}
\end{table}
We can see that the simulated signal with the learnt noise overshoots the actual experiment, for this setting and observables, as for the one dimensional case, showing the limitations of the method in this setting and the available quantum resources. 

\subsection{Chemistry ansatz}\label{section-tn-chem}
\par The last target is the chemistry ansatz, which is deeper and less regular
than the Floquet circuits above. The transpiled circuit of
Sec.~\ref{section-circuits} carries
$36$ $\mathrm{CZ}$ gates on ten qubits, whose layers of concurrent gates fall
into $16$ distinct patterns: nine carrying a single gate, five carrying two or
three, and the complete even and odd brickwork layers carrying five and four.
Each pattern was characterised independently, with $4\times10^{4}$ shots for the
first fourteen and $2\times10^{5}$ for the two complete layers, and fitted at
$\chi_b = \chi_\kappa = 4$. The $16$ channels are then composed in Pauli
transfer form to predict the energy.
\begin{figure}[!tb]
    \centering
    \includegraphics[width=\linewidth]{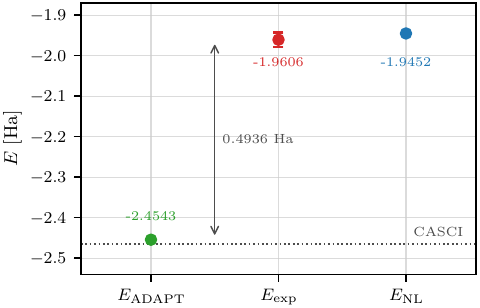}
    \caption[Energy of the chemistry ansatz, measured and predicted]{\textbf{Chemistry ansatz.} Energy of the ADAPT ansatz for the five-atom hydrogen chain at $2.5$~\AA{}: the noiseless ansatz energy $E_{\rm ADAPT}$, the energy $E_{\rm exp}$ measured on the device from $2\times10^{5}$ shots with its standard error, and the energy $E_{\rm NL}$ obtained by propagating the circuit through the $16$ learnt sublayer channels. The arrow marks $E_{\rm ADAPT} - E_{\rm exp}$ and the dotted line is the CASCI energy of the active space.}
    \label{fig: tn-chem-energy}
\end{figure}
\par The three energies are shown in Fig.~\ref{fig: tn-chem-energy}. The
noiseless ansatz value is $E_{\rm ADAPT} = -2.4543$~Ha, the device returns
$E_{\rm exp} = -1.9606 \pm 0.0178$~Ha and the computed energy with the learnt noise
is $E_{\rm NL} = -1.9452$~Ha. Device noise therefore accounts for
$E_{\rm ADAPT} - E_{\rm exp} = 0.4936$~Ha, while the learnt model accounts for
$E_{ADAPT} - E_{NL} = 0.5091$~Ha of it, $3.1\%$ more; the predicted and measured 
energies differ by $0.0155$~Ha, which is $0.87$ standard errors of the 
measurement. The experimental energy was estimated using classical shadows, namely with single-qubit Pauli measurements, which constitute an informationally-complete POVM, post-processed with canonical duals~\cite{huang-2020, malmi2024enhanced, korhonen_improving_2025}.
\par This characterisation cost $16$ separate experiments. The trimming result 
of Sec.~\ref{section-tn-trim} indicates that the two complete-layer 
characterisations alone can supply blocks for the sublayers, which would let the same total shot budget be concentrated on those two layers.

\section{Discussion and conclusions} \label{section-conclusions}

In this work, we have developed and implemented two noise-characterisation techniques for a superconducting processor with a square-lattice topology. We have also addressed the questions of noise structure, temporal noise stability, and validation of the learnt noise models against experimentally measured noisy observables in benchmark circuits for physical and chemistry problems.

The first noise-characterisation technique is a revised cycle-benchmarking protocol for the sparse Pauli–Lindblad model, in which we substantially reduce a systematic bias (induced by the unlearnable degrees of freedom) by exploiting the regularised cost function. We have used this technique in scrutinising the noise structure and the drift of noise parameters over time (Sec.~\ref{sec:noise analysis}). 

Our findings show that although the device calibration provides adequate estimates of the noise intensity (Sec.~\ref{section-device}), the internal structure of the noise is much more complicated, with the error rates associated with specific one- and two-qubit generators varying significantly. While the cross-talk noise components (on qubit pairs not directly coupled by the layer's gates) are nonzero, they are typically much smaller than the average error rate. This observation underpins the noise trimming method, which creates an approximate noise model for a sublayer from a parent noise model for an extended layer of gates by nullifying the rates for irrelevant two-qubit generators and substituting qubit-wise decoherence rates for irrelevant one-qubit generators. 

Regarding temporal noise stability, we have analysed the high-quality four-qubit subset and identified the average error rate $\varepsilon = 0.0178 \pm 0.0003$ with the noise stability factor (relative drift) $\Theta = \Delta\varepsilon/\varepsilon = 1.59\%$ over four hours. Projecting this favourable four-qubit snapshot to a larger scale provides an estimate of the best expected device performance. The parameters $\varepsilon$ and $\Theta$ are exactly the inputs to the systematic-error term $\delta^{(1)}_{\rm sys} = \tfrac{1}{2}\varepsilon N\sqrt{L}\,\Theta$ common to PEC, ZNE, and TEM in the resource framework of Ref.~\cite{filippov2024scalability}. Using them at $N = 50$, $L = 15$ to $30$ keeps this systematic floor at $3$ to $5\%$, well below the random-error term in noise mitigation, which for TEM, scaling as $\exp(\varepsilon NL/2)$, falls under $10\%$ already at $M = 10^8$ shots, while PEC, scaling as $\exp(\varepsilon NL)$, remains impractical over the same range. Under the assumption that the best observed device performance scales to the full processor, this suggests that circuits containing approximately $400$ CZ gates can be mitigated to below $13\%$ total error with $10^8$ shots using TEM.

The second noise-characterisation technique is a two-dimensional extension of tensor-network channel tomography, in which the locally purified density operator's virtual-bond graph is taken to coincide with the physical square-lattice layout of the device rather than being flattened onto a chain. We have deployed this technique on hardware together with the quantum detector tomography that corrects for measurement errors. The benefit of the tensor-network noise characterisation is that it captures the full CZ-layer channel, including those correlations beyond the sparse Pauli-restricted noise assumption. Unlike the SPL learning, the tensor-network noise characterisation is not restricted to the twirled Pauli noise. A further advantage is its implementation cost, which scales mildly with the lattice size for fixed ansatz bond dimension. We have shown that the information from repeated layer applications can improve the reconstruction at no extra parameter cost. The learnt noisy channel in the form of a tensor network can further be cut at inter-pair bonds to approximate noise models for sublayers, which is relevant for scenarios with large-scale circuits and numerous unique entangling layers. 

Leveraging the greater expressivity of tensor-network noise characterisation, we have applied it to a 10-qubit chemistry-ansatz circuit with $16$ unique noisy channels and estimated the energy of the prepared state via informationally complete measurements. The predicted noisy estimate based on $16$ learnt noise layers is $E_{\rm NL} = -1.9452$~Ha, which is in good agreement with the experimentally measured noisy estimate $E_{\rm exp} = -1.9606 \pm 0.0178$~Ha. This result confirms the predictive adequacy of the method for deeper and less regular circuits as well.

The comparison of predicted and experimentally observed noisy observables essentially constitutes the only readily accessible tool for validation of the multiple noise models learnt for individual layers of a specific circuit. In fact, the cross-validation between the two presented techniques is hardly possible due to the different parametrisations and implementations (e.g., twirling reshapes the noise) as well as the noise drift over time. By contrast, executing circuits whose noise structure matches that assumed by the learnt model is experimentally more affordable. The effect of the noise drift in this case can be partially mitigated via alternating executions of the target circuit and the auxiliary circuits for noise learning, effectively reshaping the noise to its time average. The structured Loschmidt-echo circuits across one- and two-dimensional layouts are well suited for validation, as they require characterisation of $2$ to $4$ unique entangling layers, which is affordable for both techniques. Both the revised cycle-benchmarking noise characterisation and the tensor-network noise characterisation adequately reproduce the Loschmidt echo signal measured on hardware and capture the major effects of noise. The residual discrepancy in both cases can be possibly attributed to the restricted number of measurement shots and model expressivity as well as to the single-qubit, idle, state-preparation and non-Markovian errors that are not fully captured by either of the methods but can accumulate during the circuit execution. 

Identifying the mechanisms responsible for the remaining mismatch between predicted and observed noisy estimates is an important direction for future work. Such an analysis could lead to further developments in both the device performance and the noise mitigation tools. Once the mismatch is essentially eliminated and the noise level is brought down to the minimum measured intensities across all the qubits, a $53$-qubit processor with square-lattice topology would be able to estimate observables with accuracy $\sim 10\%$ in circuits with effective volume of $400$ CZ gates and $10^8$ measurement shots, provided it is equipped with resource-optimal noise mitigation.

\section*{Code Availability}

Parts of the code base developed for the implementation of the noise learning methods are available at \url{https://github.com/Algorithmiq/TensorNetworksNoiseLearning}.

\begin{acknowledgements}
The authors thank Ricardo Rivera Cardoso, Joachim Favre, Julianne Eckert, Joonas Malmi, Anton Nyk\"{a}nen, Kalman Szenes, Leander Thiessen, Fabio Tarocco, and Zolt\'{a}n Zimbor\'{a}s for useful discussions, helpful comments and suggestions, and technical assistance. The authors acknowledge the access to the quantum computer and technical support provided by VTT QX Quantum Cloud Computing services. The work is supported by Business Finland.
\end{acknowledgements}

\bibliography{bibliography}

\appendix

\section{Construction of the nine measurement bases}\label{app: basis-construction}
\par The claim used in Sec.~\ref{sec: learning-protocol}, that nine
measurement bases suffice to obtain the Pauli fidelities required to fit
the local sparse Pauli--Lindblad model irrespective of the number of
qubits, holds only when the ordering of the subgraph induced by the
active qubits satisfies the following theorem
\cite[Theorem SIV.1]{berg_probabilistic_2023}
\begin{theorem}
    Let $(\mathcal{V}, \mathcal{E})$ be a qubit topology whose vertices are ordered in such a way that no vertex $v\in\mathcal{V}$ has more than two adjacent vertices of lower order. Then there exist nine Pauli strings such that, for each edge $(v_i, v_j) \in \mathcal{E}$, the substrings describing those edges exactly cover $\{X, Y, Z\}^{\otimes2}$.
    \label{theo: nine-basis}
\end{theorem}
\par
\begin{figure}[tbp]
    \centering
    \begin{subfigure}{0.32\linewidth}
        \centering
        \includegraphics[width=\linewidth]{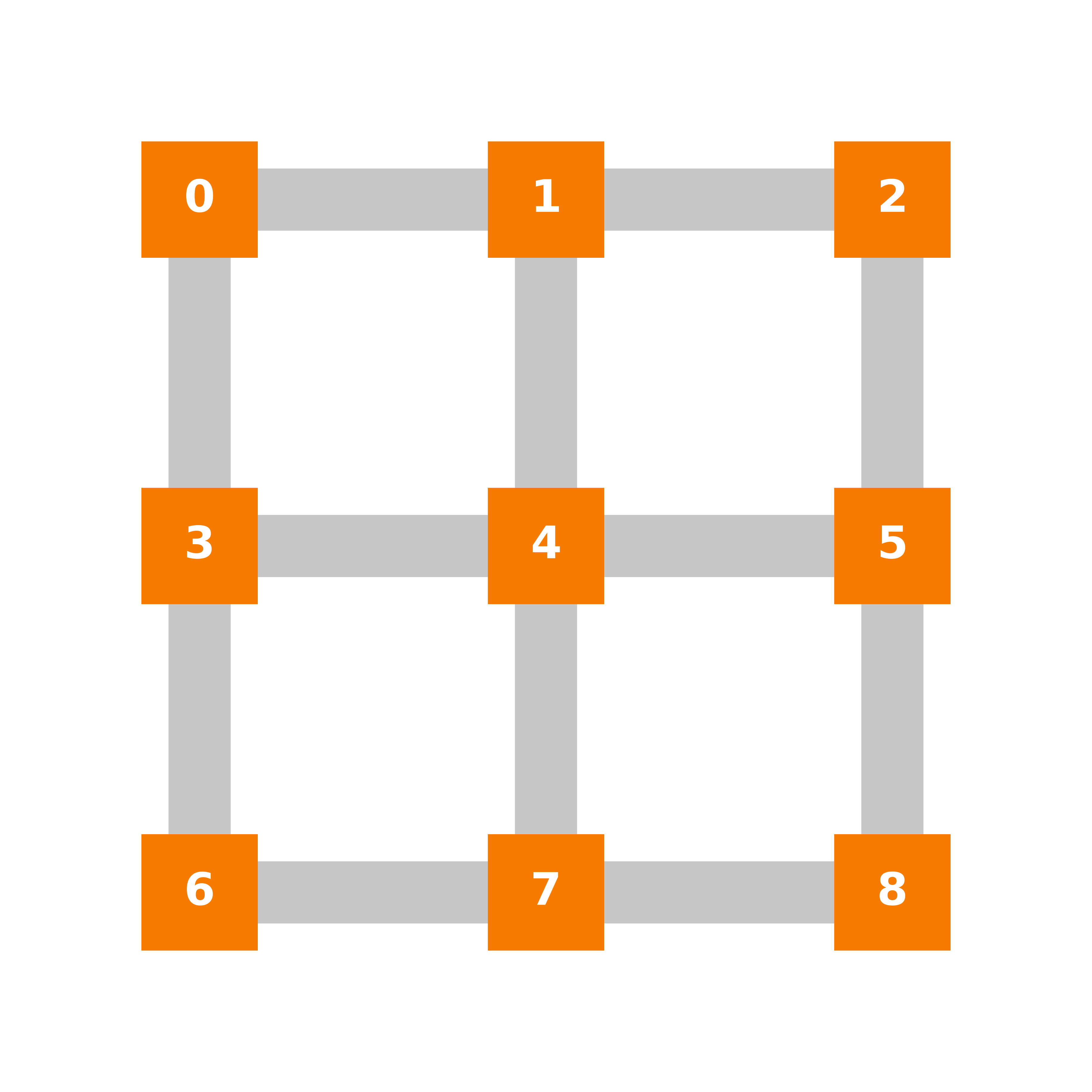}
        \caption{Left-to-right.}
        \label{fig: lattice-left-to-right}
    \end{subfigure}
    \hspace{2cm}
    \begin{subfigure}{0.32\linewidth}
        \centering
        \includegraphics[width=\linewidth]{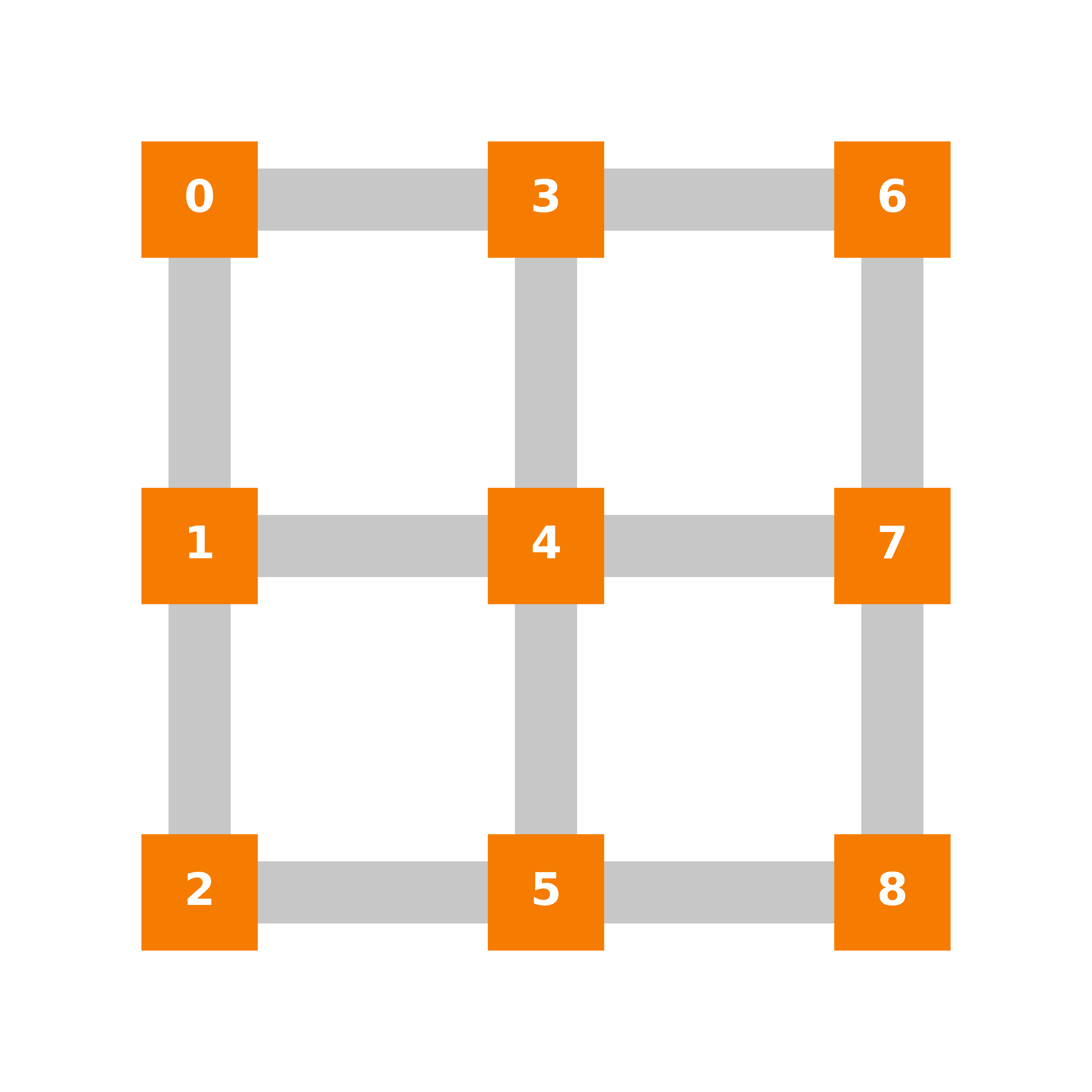}
        \caption{Top-to-bottom.}
        \label{fig: lattice-top-to-bottom}
    \end{subfigure}
    \caption[Examples of vertex ordering in a square lattice]{Examples of vertex orderings on a \(3\times 3\) square lattice. In both orderings no vertex has more than two adjacent vertices of lower order, so Theorem~\ref{theo: nine-basis} applies.}
    \label{fig: lattice-orderings}
\end{figure}

 For common two-dimensional grids, the conditions on qubit topology
required by Theorem~\ref{theo: nine-basis} are automatically satisfied
if the ordering is performed in a left-to-right and top-to-bottom fashion
as shown in Fig.~\ref{fig: lattice-orderings} for a $3\times3$ square
lattice. The coupling map of VTT's device,
Fig.~\ref{fig: vtt-coupling-map}, admits such an ordering.

\subsection{Relabelling recipe}\label{app: relabelling}
\par In practice, to construct the nine measurement bases, it is useful
to classify the nodes based on the number of nearest neighbours. For
instance, given an isolated node, i.e. a node with no nearest
neighbours, it is enough to randomly select nine Paulis from
$\{X,Y,Z\}$. For qubits with two or more connections, the procedure used
goes as follows.
\par First, note that the whole set of Pauli strings
$\{X, Y, Z\}^{\otimes2}$ can be generated by considering three
instances of the same Pauli, say $XXX$. For each Pauli tuple of this
form, one then samples a random permutation of the three Paulis, say
$YXZ$. Repeating this procedure for each Pauli generates the desired set,
as illustrated in Fig.~\ref{fig: basis-selection}.
\par If the ordering condition stated in Theorem \ref{theo: nine-basis}
is met, the basis construction reduces to an appropriate relabelling of
the qubits. In particular, let $a$ denote a Pauli-string column obtained
from a random permutation of the three uniform Pauli strings
$\{XXX, YYY, ZZZ\}$, and let $b$ denote the corresponding column obtained
by assigning, for each instance, a random permutation of $\{X, Y, Z\}$.
The relabelling is chosen such that every edge
$(v_i, v_j) \in \mathcal{E}$ is associated with a tuple $(a, b)$,
thereby covering exactly $\{X, Y, Z\}^{\otimes2}$. For a square topology
the assignment is straightforward, since it can be done qubit-wise by
giving distinct labels to nearest-neighbour pairs. Then, given the active
qubit layout, and the Pauli column vectors $a$ and $b$, one constructs a
$9 \times n$ matrix, where $n$ corresponds to the number of active
qubits, whose nine rows coincide with the nine measurement bases required
for the experiment.
\par Suppose that, in a given experiment, the active qubit layout is
specified by qubits $[\texttt{34, 35, 36, 42, 43, 44}]$, as illustrated
in Fig.~\ref{fig: basis-selection}, and satisfies the ordering
constraints imposed by Theorem~\ref{theo: nine-basis}. Then, the basis
construction protocol described above enables the generation of all nine
measurement bases required for the cycle benchmarking experiments. For
instance, the 4th-indexed measurement basis in
Fig.~\ref{fig: basis-selection} is given by
$\mathrm{B_4= YYZXXX}$.

\subsection{Equivalence to a covering array}\label{app: covering-array}
\par The columns $a$ and $b$ may also be represented by colours. Indeed,
the combinatorial aspect underlying the basis-selection protocol is
equivalent to constructing a covering array $\text{CA}(N; t,k,v)$. This
is an $N\times k$ array over an alphabet of size $v$, with the property
that, for every $N \times t$ subarray, each $t$-tuple occurs at least
once. The parameters $t$ and $k$ denote, respectively, the strength or
degree of interactions and the number of components
\cite{torres-jimenez_covering_2021, colbourn_combinatorial_2004}.
\par In the present context, since the aim is to characterise crosstalk
between pairs of qubits, one sets \(t=2\). The alphabet is then given by
the Pauli set \(\{X,Y,Z\}\), so that \(v=3\). Once the graph induced by
the active qubits is specified, we can apply graph colouring and find the
covering array $\text{CA}(N ; 2, k, 3)$, where $k$ denotes the number of
qubits (components), and the rows of the array correspond to the
measurement bases \cite{berg_techniques_2024}. In this setting, the
minimal number of rows, for which the covering array
$\text{CA}(N ; 2, k, 3)$ exists is $N=9$.
\section{Pauli twirling of the entangling layer}\label{app: pauli-twirling}
\par A generic noise channel $\tilde{\Lambda}$ acting on $n$ qubits,
with $\{P_i\}_{i=0}^{4^n - 1}$ the Pauli basis of the corresponding
Hilbert space, can be represented in the Pauli transfer matrix (PTM) form
\(\mathcal{T}_{\tilde{\Lambda}}\) as
\begin{equation}
\mathcal{T}_{\tilde{\Lambda}}[a,b]
=
\frac{1}{2^n}
\operatorname{Tr}
\left[
P_a^\dagger
\left(
\tilde{\Lambda}(P_b)
\right)
\right],
\label{eq: noise-channel-ptm}
\end{equation}
which in general describes a dense matrix. The number of parameters
therefore scales exponentially in the number of qubits, so learning the
full channel is impractical. This limitation is further exacerbated by
the fact that these parameters are typically estimated through
tomography, thereby increasing the experimental resources required
\cite{berg_probabilistic_2023}. Pauli twirling reduces the channel to the
diagonal form assumed by the sparse Pauli--Lindblad model
\cite{wallman_noise_2016, geller_efficient_2013}.

\begin{figure}[tbp]
    \centering
    \includegraphics[width=\linewidth]{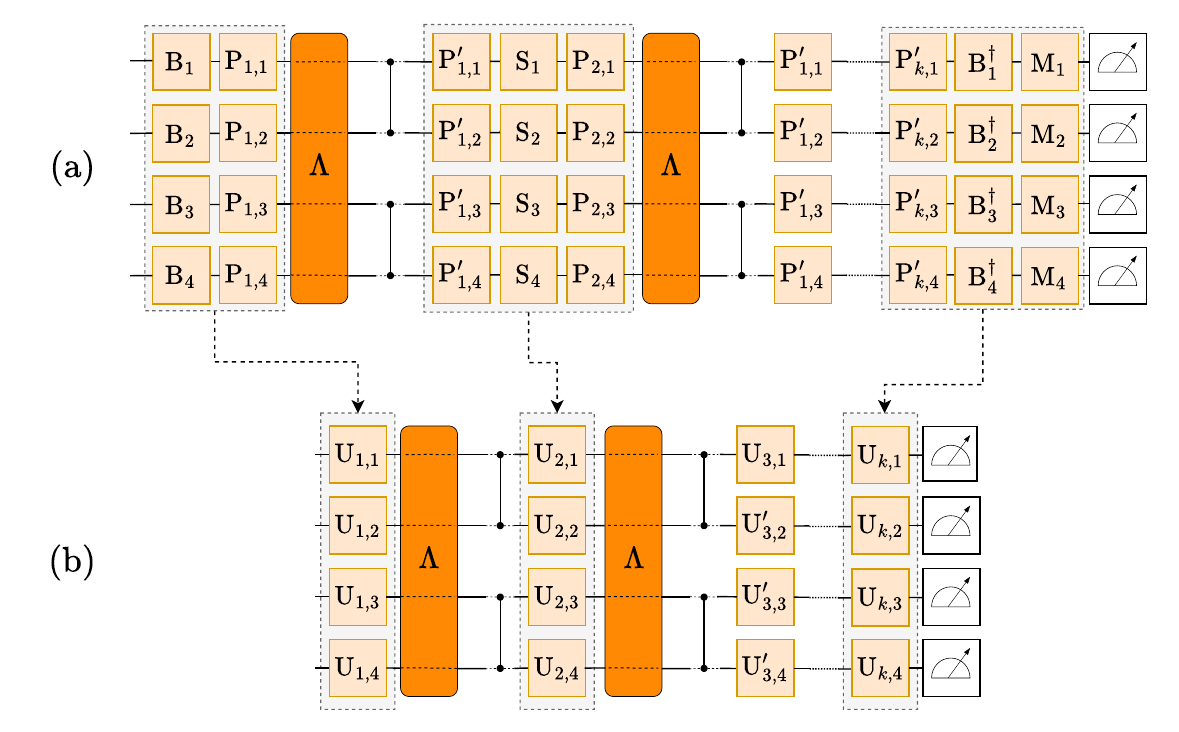}
    \caption[Single-qubit gate optimisation]{
    \textbf{Single-qubit gate optimisation:}
    The circuit overhead introduced by the state-preparation $B_i$, twirling $P_i$, and single-qubit interleave $S_i$ gates during cycle benchmarking (a), can be reduced by compiling consecutive layers of these gates into a single unitary layer during transpilation. The resulting circuit (b) is logically equivalent to the original one, while requiring fewer single-qubit gates between successive Clifford layers.}
    \label{fig: pauli-twirling}
\end{figure}
\subsection{Dressed layers}\label{app: dressed-layers}
\par In practice, one considers various instances of the circuit, each
with randomly sampled Pauli twirling layers before and after the noisy
layer of Clifford gates. However, since we only have access to a noisy
implementation $\tilde{\mathcal{C}}_k = \mathcal{C}_k \circ
\tilde{\Lambda}$ of the ideal gate $\mathcal{C}_k$, and we only aim to
twirl the noise channel $\tilde{\Lambda}$, one must push the Pauli
operator through the Clifford gate, that is
\begin{align*}
\mathcal{C} P_a^\dagger \tilde{\Lambda}
\!\left(
P_a \,\cdot\, P_a^\dagger
\right)
P_a
&=
\mathcal{C} P_a^\dagger (\mathcal{C}^\dagger \mathcal{C})
\tilde{\Lambda}
\!\left(
P_a \,\cdot\, P_a^\dagger
\right)
 P_a
\\
&=
P_a^{c \dagger}\,
\tilde{\mathcal{C}}
\!\left(
P_a \,\cdot\, P_a^\dagger
\right)
P_a,
\end{align*}
where $P_a^c = \mathcal{C}P_a\mathcal{C}^{\dagger}$ is also a Pauli
operator since, by definition, the Clifford gate $\mathcal{C}$ maps Pauli
operators to Pauli operators.
Therefore, Pauli twirling can be efficiently implemented by applying
$P_a$ and $P_a^c$ before and after the noisy gate as
$P_a^c\,\tilde{\mathcal{C}}P_a$, respectively.
\par In general, $n$-qubit Pauli operators can be represented either by
strings in $\{I, X, Y, Z\}^{\otimes n}$ or as a binary vector of length
$2n$ in the symplectic representation
\cite{sayginel_fault-tolerant_2025}. From a computational perspective,
the latter representation is more efficient for computing the twirling
pairs $(P_a,P_a^c)$ associated with a given Clifford operation
$\mathcal{C}$ \cite{berg_probabilistic_2023}.

\subsection{Single-qubit gate overhead}\label{app: twirling-overhead}
\par The dressed layer, the state-preparation gates $B_i$, the
interleaving gates $S_i$ used to restore non-invariant Paulis, and the
readout-mitigation gates $M_i$ all consist of single-qubit operations.
The circuit overhead they introduce can be reduced during transpilation
by compiling the dressed noisy Clifford gates into a single round of
elementary gates rather than three separate layers of single-qubit
operations \cite{wallman_noise_2016}. This optimisation step is
illustrated in Fig.~\ref{fig: pauli-twirling}.

\subsection{Comparison with Clifford twirling}\label{app: clifford-twirling}
\par Twirling a quantum channel is not restricted to the Pauli group
\cite{berg_probabilistic_2023}. In particular, the Clifford group can
also be used to transform a generic noise channel $\mathcal{E}$ into a
stochastic Pauli channel, producing the same suppressing effect on
off-diagonal PTM elements. However, unlike Pauli twirling, Clifford
twirling does not preserve the eigenvalues of the original channel.
Instead, it averages all diagonal entries of the PTM, except for the
leading identity element, to the same value. Consequently, the effective
noise channel is tailored into a depolarising channel
\cite{hashim_practical_2025}. Since the quantities we set out to learn
are precisely the individual Pauli fidelities, only Pauli twirling is
used in this work.

\section{Gauge ambiguity and the pattern transfer graph}\label{app: gauge-ambiguity}
\par This appendix collects the ideas underlying the proof of
Theorem~\ref{theo: pauli-learnability}. First, the sufficiency of the
condition follows directly from the cycle benchmarking protocol. The
necessary condition, corresponding to the case in which
$\mathfrak{s}_i \ne \mathfrak{s}_i'$, where $\mathfrak{s}_i'$ denotes the
support of the Pauli operator $\mathcal{C}P_i\mathcal{C}^{\dagger}$, is
closely related to the notion of gauge ambiguity. Table~(\ref{tab:
cz-conjugation}) lists the action of the $\mathrm{CZ}$ gate on the
two-qubit Pauli operators, from which the three individually learnable
fidelities and the effect of the interleaved single-qubit layer can be
read off.
\begin{table}[!t]
\centering
\resizebox{0.48\textwidth}{!}{
\begin{tabular}{c|ccccccccccccccc}
$P_{a}$
& IX & IY & IZ & XI & XX & XY & XZ & YI & YX & YY & YZ & ZI & ZX & ZY & ZZ \\
\hline
$U_C P_{a} U_C^\dagger$
& ZX & ZY & \textbf{IZ} & XZ & YY & YX & XI & YZ & XY & XX & YI & \textbf{ZI} & IX & IY & \textbf{ZZ} \\
$\bar{U}_C P_{a} \bar{U}_C^\dagger$
& ZY & ZX & \textbf{IZ} & YZ & \textbf{XX} & \textbf{XY} & YI & XZ & \textbf{YX} & \textbf{YY} & XI & \textbf{ZI} & IY & IX & \textbf{ZZ} \\
\end{tabular}
}
\caption[Learnability of two-qubit Pauli fidelities for the $\mathrm{CZ}$ gate]{The learnability of Pauli fidelities depends on whether the associated Pauli operator is invariant under conjugation by the Clifford gate. The second row illustrates the Pauli operators obtained after conjugation by the Clifford gate. Only three of these operators (shown in bold) remain invariant, and therefore correspond to learnable fidelities. Furthermore, if the support of the transformed Pauli operator is preserved under the action of the gate, the original Pauli operator can be recovered by introducing an additional interleaved layer of single-qubit gates after the Clifford operation. For example, the $\mathrm{CZ}$ gate maps the operator $\mathrm{XX}$ to $\mathrm{YY}$, which can subsequently be mapped back to $\mathrm{XX}$ by applying phase gates to both qubits. The resulting operators obtained from these dressed gates (i.e., the Clifford gate with the interleaved single-qubit gates) are shown in the third row.}
\label{tab: cz-conjugation}
\end{table}

\begin{figure}[tbp]
    \centering
    \includegraphics[width=1\linewidth]{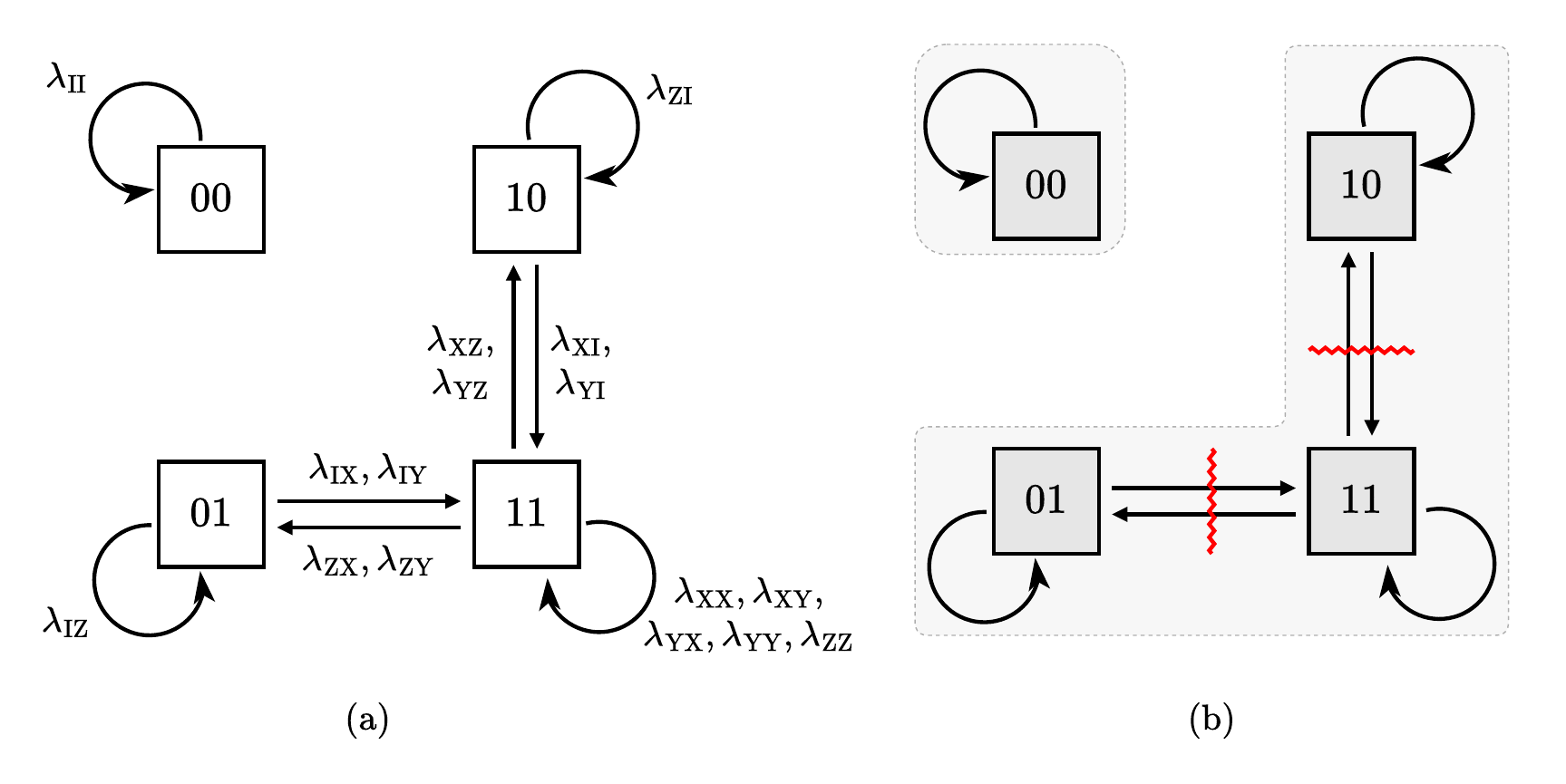}
    \caption[Pattern Transfer Graph of the $\mathrm{CZ}$ gate.]{\textit{Pattern Transfer Graph} of the $\mathrm{CZ}$ gate. The graph consists of $4$ vertices and $16$ edges (a). Self-loops correspond to fidelities that are individually learnable, whereas cycles correspond to learnable products of fidelities. The graph contains two connected components (b), implying that the associated cut space, shown in red, is two-dimensional.}
    \label{fig: pattern-transfer-graph}
\end{figure}
\subsection{Gauge transformations}\label{app: gauge-transformations}
\par To illustrate the unlearnability of individual fidelities, consider
the gauge transform
\begin{equation}
\begin{aligned}
f_{\mathrm{IX}},\, f_{\mathrm{XI}} &\longrightarrow \kappa f_{\mathrm{IX}},\, \kappa f_{\mathrm{XI}}, \\ f_{\mathrm{ZX}},\, f_{\mathrm{XZ}} &\longrightarrow \kappa^{-1} f_{\mathrm{ZX}},\, \kappa^{-1} f_{\mathrm{XZ}}.
\end{aligned}
\label{eq: gauge-transformation}
\end{equation}
where $\kappa$ is a real number close to unity. This gauge
transformation also induces changes in the SPAM coefficients $\alpha_i$,
which are omitted here for simplicity. Since
$\frac{1}{\kappa}f_\mathrm{IX}\cdot\kappa f_\mathrm{ZX} =
f_\mathrm{IX}f_\mathrm{ZX}$ holds for any nonzero value of $\kappa$, the
cycle benchmarking results remain invariant under such transformation.
Consequently, the unlearnability of certain fidelities manifests itself
in the appearance of pairwise products of fidelities, which prevents
individual fidelities from being extracted in a SPAM-free manner.
\par From a practical perspective, this implies that different noise
models can yield distinct values for an unlearnable fidelity, e.g.
$\mathrm{XZ}$, while remaining experimentally indistinguishable. In
other words, two distinct noise models may lead to identical output
statistics for all quantum experiments within this framework.
Consequently, it is possible to rearrange or redistribute the
unlearnable fidelities in such a way that they reproduce the output of
all cycles, and no experiment based on these protocols can witness such a
transformation \cite{chen_disambiguating_2026, chen_learnability_2023}.
This feature is not a limitation of the cycle benchmarking protocol, but
a manifestation of the limited learnability of Pauli noise.

\subsection{Cycle and cut spaces}\label{app: pattern-transfer-graph}
\par To gain a clearer understanding of what information is learnable,
consider the pattern transfer graph in Fig.~\ref{fig:
pattern-transfer-graph} which illustrates how the $\mathrm{CZ}$ gate
changes the weight (or pattern) of two-qubit Pauli operators. In this
graph, consisting of $4$ nodes and $16$ edges, individual fidelities
corresponding to self-loops are directly learnable, while products of
Pauli fidelities $f_1f_2\cdots f_k$ are learnable in a SPAM-free manner
if the corresponding edges form a cycle in the pattern transfer graph
\cite{dahlhauser_benchmarking_2022, chen_learnability_2023}.
\par More generally, define the log-Pauli fidelities $l_i = \log{f_i}$
for all Pauli operators $P_i$. The space of linear functions $f$ of
$\{l_i\}$
\[
\mathcal{F} = \left\{ f : \mathbb{R}^n \to \mathbb{R} \;\mid\; f(\boldsymbol{l}) = \sum_{i} v_i l_i, \quad v_i \in \mathbb{R} \right\},
\]
is isomorphic to the edge space of the pattern transfer graph. In
particular, for the $\mathrm{CZ}$ gate, the orbit of the Pauli operator
$\mathrm{IX}$ during benchmarking forms the cycle
\[ \mathrm{IX} \overset{\scriptscriptstyle \mathrm{CZ}}{\longrightarrow} \mathrm{ZX} \overset{\scriptscriptstyle \mathrm{CZ}}{\longrightarrow} \mathrm{IX} \overset{\scriptscriptstyle \mathrm{CZ}}{\longrightarrow} \cdots
\]
\par Since the product of Pauli fidelities
$f_\mathrm{IX}f_\mathrm{ZX}$ can be estimated by fitting exponentially
decaying curves, the corresponding linear function
$l_\mathrm{IX}+l_\mathrm{ZX}$ is learnable. As a result, the learnable
linear functions form a subspace of $\mathcal{F}$, that is naturally
identified with the cycle space of the pattern-transfer graph
\cite{dahlhauser_benchmarking_2022}. Therefore, for a given cycle
$\mathrm{C}$, we can construct a CB experiment to estimate the products
of fidelities along the cycle, or, equivalently, the cycle function of
the form $f_\mathrm{C} = \sum_{e_i\in \mathrm{C}}\log f_i$ is learnable.
It follows that any linear combination of cycle functions
\[
\left\{ \sum_C \alpha_C \log f_C \;\mid\; \alpha_C\in\mathbb{R} \right\},
\]
is also learnable. In fact, this space comprises all learnable
information about the Pauli noise model \cite{chen_learnability_2023}.
\par To establish that information outside of the cycle space is
fundamentally unlearnable, it suffices to show that every cut in the
pattern transfer graph induces a gauge transformation of the form given
in (\ref{eq: gauge-transformation}). Indeed, every vector in the cut
space corresponds to a gauge transformation. Since every learnable
function is orthogonal to such transformations, the cut space must
contain precisely the unlearnable degrees of freedom. By definition, the
edge space is the orthogonal direct sum of the cycle space and the cut
space \cite{bollobas_modern_1998}. Their complementarity therefore
defines the boundary between the learnable and unlearnable degrees of
freedom of the Pauli noise model \cite{chen_learnability_2023}.
\par This picture is consistent with the counting of
Table~(\ref{tab:cz-learnability}). For a directed graph
$G(\mathcal{V},\mathcal{E})$, the cycle space has dimension
$|\mathcal{E}| - |\mathcal{V}| + c$, while the dimension of the cut space
is $|\mathcal{V}| - c$, where $c$ denotes the number of connected
components. In the present case, $c = 2$, implying that the cut space is
two-dimensional, in agreement with the two independent unlearnable
parameters appearing in the two-qubit case.
\section{Simulated characterisation of the four-qubit layers}\label{app: simulation-results}
\begin{figure}[tbp]
    \centering
    \includegraphics[width=\linewidth]{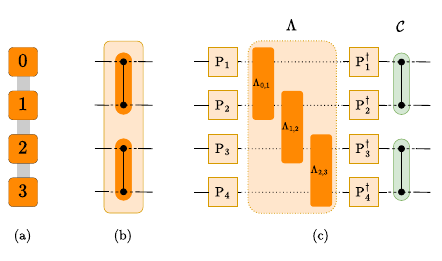}
    \caption[Noise characterisation simulation protocol]{To simulate the noise model during the cycle benchmarking experiments, a local Pauli-Lindblad noise channel was applied before the ideal Clifford gates $\mathcal{C}$. The noisy layer (b) is then described by the twirled noise channel preceding the ideal $\mathrm{CZ}$ gates (c). The model includes only weight-one and weight-two Pauli operators whose supports cover all qubits and all pairs of physically connected qubits (a). Owing to the sparsity of the model it admits a decomposition into local two-qubit noise channels, since the generators act nontrivially on at most pairs of connected qubits.}
    \label{fig: cb-simulation}
\end{figure}
\begin{figure}[tbp]
    \centering
    \includegraphics[width=1\linewidth]{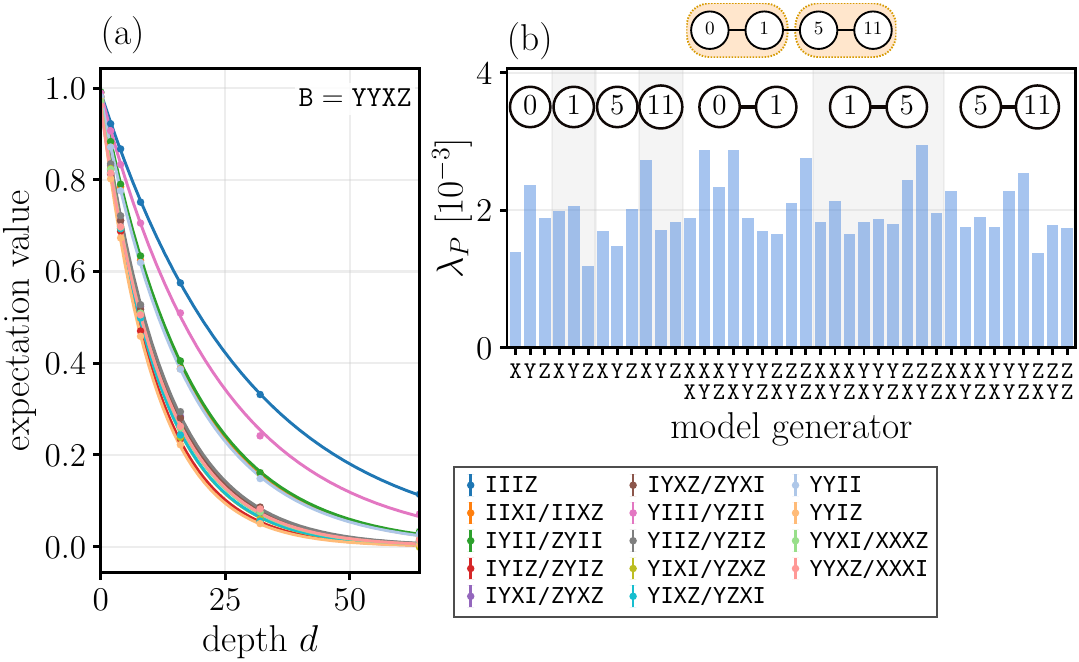}
    \caption[Four-qubit noise characterisation simulation results for the first characterised Clifford layer]{\textbf{Four-qubit noise characterisation simulation using ridge regularisation.} Results for the first characterised Clifford layer. The decay of Pauli expectation values (a), obtained by repeated application of the noisy layer of Fig.~\ref{fig: cb-simulation}, gives the fidelity pairs $f_if_j$ required to fit the noise parameters. The learnt coefficients (b) are the rates of all weight-one and weight-two Pauli generators. The channel was constructed by sampling the rates from a normal distribution of mean $\mu = 2\times10^{-3}$ and standard deviation $5\times10^{-4}$. The fidelity pairs of (a) are listed in the box below (b). Above (b) is the characterised layer, the four-qubit chain with a dashed capsule over each $\mathrm{CZ}$ it drives; inside, the shading and the circled qubit numbers mark the support each group of generators acts on.}
    \label{fig: simulation-decay-coeff-cz-q01q23}
\end{figure}
\par This appendix reports the simulated cycle benchmarking experiments
used in Sec.~\ref{sec: bias} to compare fitting procedures. The target
noise was implemented with Qiskit's \texttt{PauliLindbladError} object \cite{noauthor_paulilindbladerror_nodate}
, applying an $n$-qubit
Pauli--Lindblad channel before the layer of Clifford gates, with
generators covering all nodes and edges of the coupling map of the active
layout. The simulation settings are those of Sec.~\ref{sec: bias}.
\par The first characterised layer consists of the two concurrent
$\mathrm{CZ}$ gates of Fig.~\ref{fig: cb-simulation}. The simulated
experiment comprises nine runs, one per measurement basis, each
benchmarking the layer at increasing even depth and averaging over
randomly twirled instances.
\par Fig.~\ref{fig: simulation-decay-coeff-cz-q01q23} (a) shows the
decay curves for the basis $\mathrm{YYXZ}$. The noise channel of
Fig.~\ref{fig: cb-simulation} produces a progressive decrease of the
expectation values with depth, and the higher-weight Paulis decay faster
than the lower-weight ones. Fitting curves of the form
$\alpha(f_1f_2)^{d/2}$ gives the products needed to determine the
coefficients, shown in Fig.~\ref{fig: simulation-decay-coeff-cz-q01q23}(b);
the learnt parameters agree with the underlying model, whose rates are
distributed around $\mu = 2\times10^{-3}$.
\begin{figure}[tbp]
    \centering
    \includegraphics[width=1\linewidth]{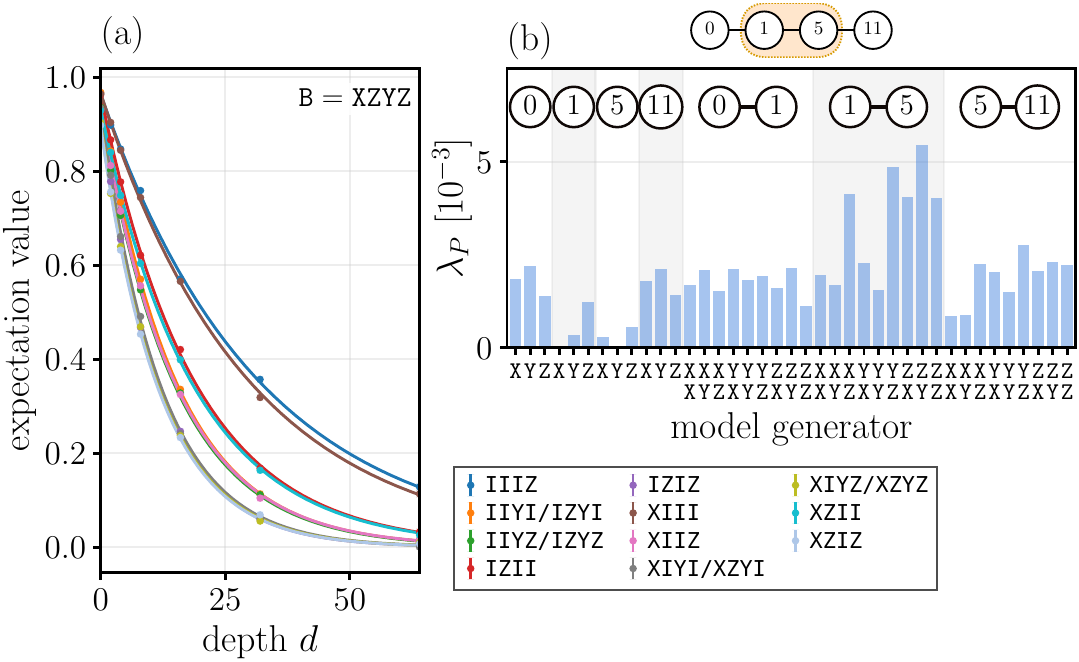}
    \caption[Four-qubit noise characterisation simulation results for the second characterised Clifford layer.]{\textbf{Four-qubit noise characterisation simulation using the symmetry-based augmentation strategy} Results for the second characterised Clifford layer. Pauli expectation values for a representative measurement basis as a function of depth $d$ (a), from which the fidelity pairs are extracted, and the fitted model coefficients (b).}
    \label{fig: simulation-decay-coeff-cz-q12}
\end{figure}
\par Fig.~\ref{fig: simulation-decay-coeff-cz-q12} shows the second
layer, a single $\mathrm{CZ}$ on qubits \texttt{[1,5]}, with the same
noise model. Most learnt parameters again lie around $\mu$. However, the coefficients are larger for generators whose support coincides with the qubits carrying the Clifford gates. The coefficients associated with unlearnable fidelities, such as
$f_{\mathrm{IZXI}}$ for the second layer, are considerably overestimated
by the symmetry-based augmentation, in some cases exceeding twice the
expected mean, while the corresponding single-qubit coefficients are
driven to zero. This is the bias analysed in Sec.~\ref{sec: bias} and
removed by the regularised fit of Sec.~\ref{sec: regularised-fitting}.
\begin{figure}[tbp]
    \centering
    \includegraphics[width=1\linewidth]{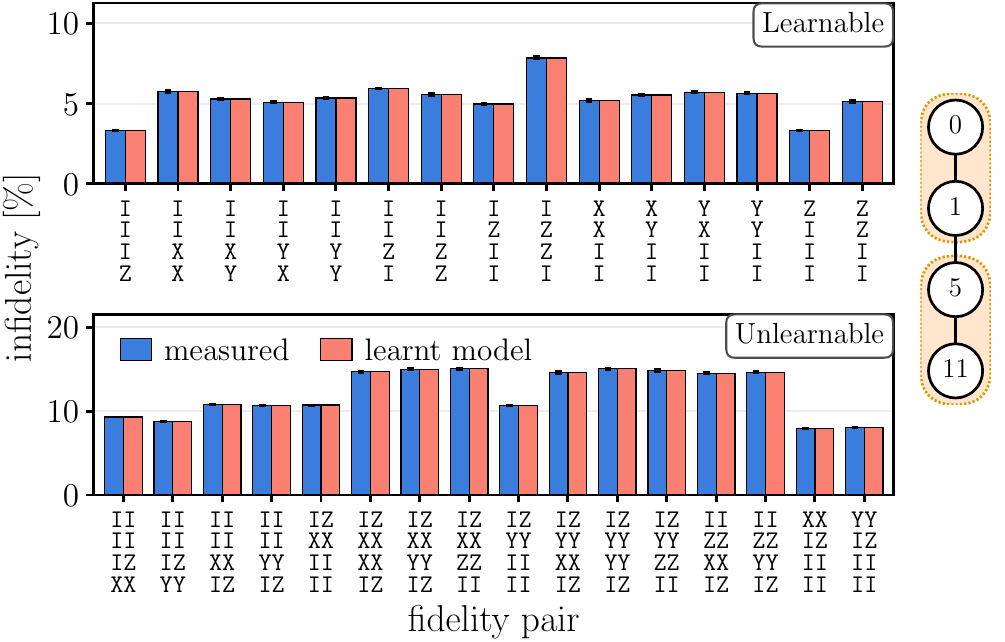}
    \caption[Fidelity reliability test from simulation results]{\textbf{Fidelity reliability test.} Measured Pauli infidelities $1 - f_P$ from the cycle benchmarking simulation against the infidelities predicted by the learnt noise model. The upper panel holds the individual learnable infidelities and the lower one the learnable products of individually unlearnable fidelities; blue is the simulated measurement and red the prediction of the learnt model. The characterised layer is drawn to the right, the chain with a dashed capsule over each $\mathrm{CZ}$ it drives.}
    \label{fig: simulation-fidelities}
\end{figure}
\begin{figure}[tbp]
    \centering
    \includegraphics[width=\linewidth]{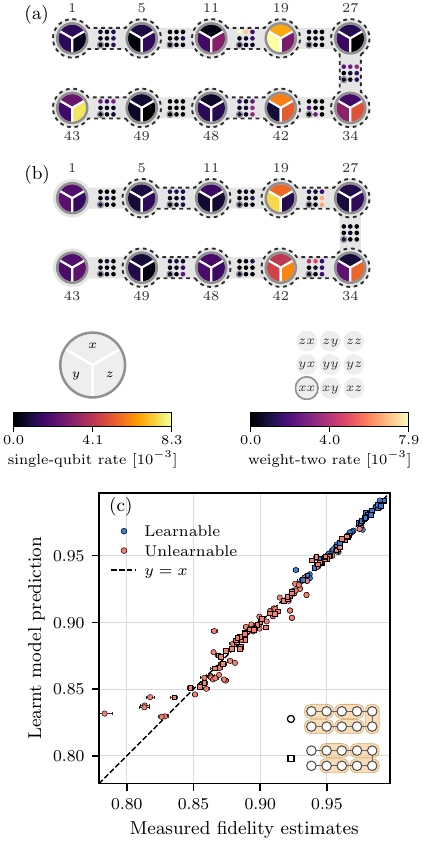}
    \caption[Ten-qubit one-dimensional experimental results]{\textbf{Ten-qubit one-dimensional experiment.} (a,~b) Learnt model coefficients for the two characterised layers of $\mathrm{CZ}$ gates, indicated by dotted lines around the corresponding qubit pairs. The fitting procedure is the one used for the four-qubit experiments. Pie-style circles represent single-qubit error rates, and the smaller circles between neighbouring qubits the rates of the weight-two generators; both panels share the keys and the two colour scales at the bottom. (c) Fidelity pairs measured from the cycle benchmarking data against the corresponding predictions of the learnt noise model, for both layers collected in a single panel, one marker shape per layer. The key in the lower right draws each layer on the folded chain of (a,~b), with a shaded capsule over each pair it drives. Blue is an individually learnable fidelity, red a learnable product of individually unlearnable ones, and the dashed line the ideal relation $y=x$.}
    \label{fig: experiments-results-10q-1d}
\end{figure}
\par A practical consistency check is to compare the fidelities
$f_P = 2^{-n}\operatorname{tr}(P^{\dagger}\Lambda(P))$ of the learnt
model with those extracted from the simulation. Fig.~\ref{fig: simulation-fidelities}
shows the comparison of the pairwise infidelities $1 - f_{P_1}f_{P_2}$
for the first layer; the two agree, which supports the ability of the
model to reproduce the simulated noise.

\section{Additional layouts and sublayers}\label{app: additional-layouts}

\subsection{Ten-qubit one-dimensional chain}\label{app: 10q-1d}
\begin{figure}[tbp]
    \centering
    \includegraphics[width=0.9\linewidth]{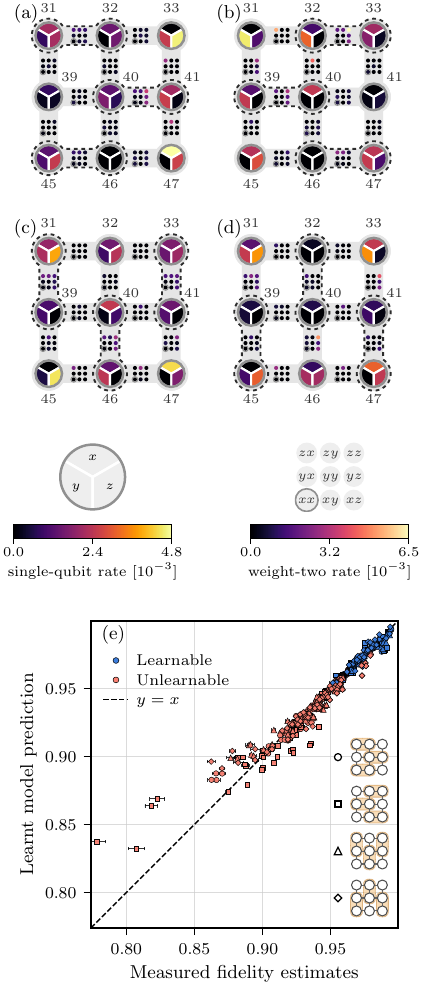}
    \caption[Nine-qubit two-dimensional experimental results]{\textbf{Nine-qubit two-dimensional experiment.} (a--d) Learnt noise model coefficients for the four characterised layers of $\mathrm{CZ}$ gates on a $3\times3$ layout. The four layers together cover all nearest neighbours. Single-qubit rates and weight-two coefficients use the same representation as in the other experiments, with the keys and the two colour scales at the bottom shared by all four panels. (e) Measured fidelity estimates against the corresponding predictions of the learnt noise model for the four layers, collected in a single panel, one marker shape per layer as in the key in the lower right. Colours as in Fig.~\ref{fig: experiments-results-10q-1d}(c); the dashed line is the ideal relation $y = x$.}
    \label{fig: experiment-results-9q-2d}
\end{figure}
\par Two layers of $\mathrm{CZ}$ gates were also characterised on a
ten-qubit one-dimensional layout. As in the four-qubit case one layer
consists of concurrent gates on all qubits, while in the other the gates
act only on internal pairs, leaving the first and last qubits idle. Both
settings are shown in Fig.~\ref{fig: experiments-results-10q-1d}(a,~b)
together with the learnt coefficients. The weight-two rates are again
larger on the pairs carrying the gates, and the single-qubit rates are
larger overall than in the four-qubit register, with qubit $19$ reaching
values near $8\times10^{-3}$.
\par Fig.~\ref{fig: experiments-results-10q-1d}(c)
compares the measured fidelity estimates, both individual learnable
fidelities and learnable products of unlearnable ones, with the model
predictions for both layers. The points lie close to the $y=x$ line.

\subsection{Nine-qubit two-dimensional grid}\label{app: 9q-2d}
\par A nine-qubit characterisation was performed on a $3\times3$ layout,
with the layers again chosen so that all qubit pairs are covered. The
learnt coefficients are shown in
Fig.~\ref{fig: experiment-results-9q-2d}(a--d) and follow the same
pattern, with larger coefficients on the pairs carrying a $\mathrm{CZ}$
gate. The fidelity estimates agree with the model predictions,
Fig.~\ref{fig: experiment-results-9q-2d}(e), although these
layers show a few products deviating more noticeably from their
model-estimated values than in the other layouts.

\section{Additional channel-tomography results}\label{app: tn-additional}

\subsection{Training details for the 10 qubits 1 dimensional case}
\label{app:10q 1d training}
\begin{figure}[tbp]
    \centering
    \includegraphics[width=\linewidth]{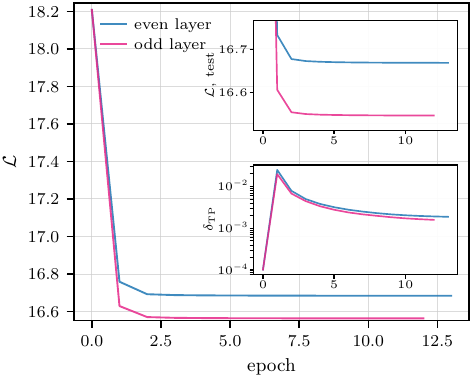}
    \caption[Training of the one-dimensional ten-qubit channels]{\textbf{Ten-qubit chain, training.} (a) Total loss \eqref{eq: loss} per epoch for the two $\mathrm{CZ}$ layers of the Floquet period; the first point is the initialisation after the trace-preservation warm start. The inset shows the loss on the test shots, on which the reported epoch is selected. (b) Residual trace-preservation violation $\delta_{\rm TP}$ of Eq.~\eqref{eq: tp-penalty} over the same runs.}
    \label{fig: tn-1d-training}
\end{figure}
\par Fig.~\ref{fig: tn-1d-training} shows the optimisation. The early-stopping
tolerance of $10^{-4}$ on $\mathcal{L}$ is met after $13$ epochs on the even
layer and $12$ on the odd one. The two fits end at $D_{\rm KL} = 16.682$ and
$16.562$, against the value $n\ln 6 = 17.92$ that Eq.~\eqref{eq: kl} takes for a
model assigning every outcome the same probability, and at $\mathcal{L} = 16.669$
and $16.547$ on the test shots. The trace-preservation warm start leaves $\delta_{\rm TP}$ below
$10^{-4}$; after the first epoch, the LPDO loses trace preservation, which is slowly restored during training. 

\subsection{Spread over the random-initialisation seeds}\label{app: tn-seeds}
\par
\begin{figure}[tbp]
    \centering
    \includegraphics[width=\linewidth]{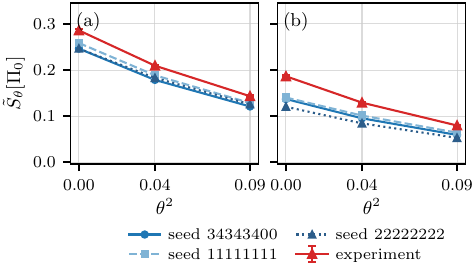}
    \caption[Spread of the prediction over seeds]{\textbf{Seed spread.} Echo signal $\tilde S_\theta[\Pi_0]$ predicted by each of the three fits, against the measurement, for (a) the ten-qubit chain and (b) the $2\times5$ lattice with the raw end-of-training channels, on a common $y$ axis. The solid blue curve is the seed used everywhere else in the paper.}
    \label{fig: tn-seed-spread}
\end{figure}

 The channels of Sec.~\ref{section-experimental-results} are the fits at seed
$34343400$. Every layer was also fitted at the seeds $11111111$ and $22222222$,
on the same shots and with the same hyperparameters, to check that the
non-convex optimisation lands in comparable places, Fig~\ref{fig: tn-seed-spread}.
\par On the chain the three fits end within $0.036$ of each other in
$\mathcal{L}$ on the test shots, and the predicted $\tilde S_\theta[\Pi_0]$ at $\theta = 0$ spans
$0.2465$ to $0.2586$, a standard deviation of $2.3\%$ of its mean, rising to
$2.5\%$ at the largest angle. On the $2\times5$ lattice the test values of
$\mathcal{L}$ again agree to $0.036$ per layer, and the predicted $\tilde S_\theta[\Pi_0]$ at $\theta = 0$
spans $0.1203$ to $0.1404$, a standard deviation of $6.7\%$, rising to $7.8\%$
across the scan. For the chemistry
ansatz, refitting all $16$ patterns at the three seeds moves the median
$D_{\rm KL}$ by $0.007$--$0.015$, with a largest per-layer standard deviation of
$0.021$.

\subsection{Remaining observables}\label{app: tn-observables}
\par Sec.~\ref{section-tn-1d} and Sec.~\ref{section-tn-2d} read the echo circuit
out in $\Pi_0$ only. The same two scans were evaluated for the mean
magnetisation $M_z = \frac{1}{n}\sum_i Z_i$ and the all-$Z$ parity
$Z^{\otimes n}$, which decay at different rates under the same channel.
Fig.~\ref{fig: tn-theta-appendix} collects them, together with the $3\times3$
scans of Appendix~\ref{app: tn-3x3}.
\begin{figure}[tbp]
    \centering
    \includegraphics[width=\linewidth]{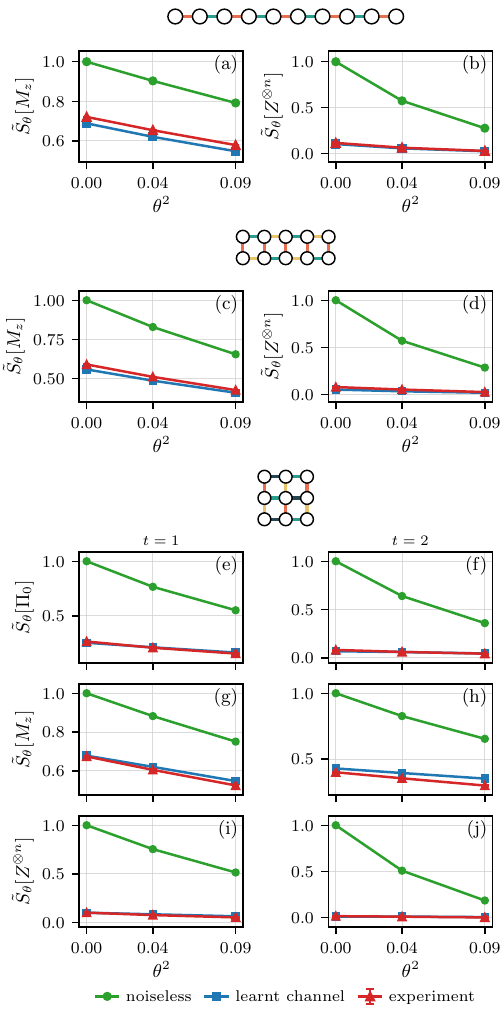}
    \caption[Appendix echo scans]{\textbf{Echo scans of the appendix.} Echo signal against $\theta^2$ for every arm the main text does not show. Three blocks, each headed by its geometry, with the edges coloured by the $\mathrm{CZ}$ layer they belong to as in Fig.~\ref{fig: LE topology}. In the first two an observable runs across the columns: the echo circuit of Fig.~\ref{fig: tn-1d-theta} on the ten-qubit chain (a,~b) and that of Fig.~\ref{fig: tn-2d-theta} on the $2\times5$ lattice (c,~d), read out in $M_z$ (left) and in $Z^{\otimes n}$ (right). In the third, the $3\times3$ lattice, an observable runs down the rows, $\Pi_0$ (e,~f), $M_z$ (g,~h) and $Z^{\otimes n}$ (i,~j), and a Floquet depth across the columns, $t = 1$ on the left and $t = 2$ on the right.}
    \label{fig: tn-theta-appendix}
\end{figure}
\par On the chain, Fig.~\ref{fig: tn-theta-appendix}(a,~b), the prediction is
$4.5$--$5.4\%$ below the measurement on $M_z$ and $11.0$--$14.4\%$ below it on
$Z^{\otimes n}$. On the $2\times5$ lattice,
Fig.~\ref{fig: tn-theta-appendix}(c,~d), it is $4.0$--$5.4\%$ below on $M_z$ and
$30.5$--$36.9\%$ below on $Z^{\otimes n}$, the observable that carries the
smallest absolute signal, $0.0769$ at $\theta = 0$ against a measurement standard
error of $0.0037$.

\subsection{Repeated-layer learning on hardware}\label{app: tn-repeated-hw}
\par Sec.~\ref{section-tn-repeated} validates the repeated-layer fit against a
known channel in simulation. The same procedure was run on the device, on the
two ten-qubit acquisitions that carry $D = 1$, $2$ and $4$
repetitions of the same $\mathrm{CZ}$ layer at $2\times10^{5}$ shots each,
one Pauli-twirled and one not. Each $\mathrm{CZ}$ layer of the Floquet period
was characterised twice: once from the $D = 1$ dataset alone, and once jointly
from $\mathcal{D} = \{1,2,4\}$ with the parameters tied, Eq.~\eqref{eq: joint-loss}. The two
channels carry the same number of parameters. They are then composed into the
echo circuit of Sec.~\ref{section-echo-experiment}, at $b = 0.1\pi$ and $t = 2$
as in Sec.~\ref{section-tn-1d}, which applies each of them four times, and the
predicted signal is compared with the measurement.
\begin{figure}[tbp]
    \centering
    \includegraphics[width=\linewidth]{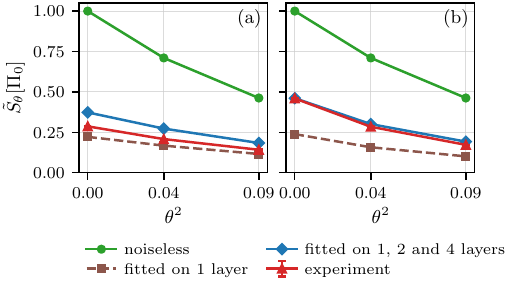}
    \caption[Repeated-layer learning on the device]{\textbf{Repeated-layer learning on hardware.} Echo signal $\tilde S_\theta[\Pi_0]$ against $\theta^2$ for the acquisition (a) with and (b) without Pauli twirling, on a common $y$ axis, predicted by a channel fitted to the $D = 1$ dataset alone (brown, dashed) and by one fitted jointly over $\mathcal{D} = \{1,2,4\}$ (blue, solid). Green: noiseless. Red: hardware, with its standard error. Both channels carry the same number of parameters and are applied four times each in this circuit.}
    \label{fig: tn-repeated-hw}
\end{figure}
\par Fig.~\ref{fig: tn-repeated-hw} shows the comparison on $\tilde S_\theta[\Pi_0]$.
The channel fitted to the single dataset predicts less surviving signal than the
device shows on both acquisitions, and the shortfall is large: at $\theta = 0$ it
gives $0.2214$ against a measured $0.2865 \pm 0.0029$ on the twirled acquisition
and $0.2385$ against $0.4593 \pm 0.0037$ on the untwirled one, i.e. $23\%$ and
$48\%$ below. The jointly fitted channel gives $0.3728$ and $0.4606$ at the same
two points, $30\%$ and $0.3\%$ above respectively.
\par Taking the root-mean-square deviation from the measurement over the three
measured angles, the joint fit is the better model on the untwirled acquisition.
\par While for the untwirled scenario the multi-layer optimisation seems to provide a reliable characterisation, neither model accurately reproduces the experimental data on the twirled acquisition. Nonetheless, training on multiple layers does reduce the overshooting that has often been empirically observed for the original tensor-network-based characterisation method. Overall, the evidence is encouraging for multi-layer training, and leaves open the question of the actual capabilities of this approach in broader scenarios, especially when quantum resources allow for more shots $N_{\text{shots}}$ and more settings $N_{\text{set}}$.

\subsection{Nine-qubit $3\times3$ lattice}\label{app: tn-3x3}
The same protocol was run on the $3\times3$ lattice, whose brickwork period needs four $\mathrm{CZ}$ layers, labelled A to D in
Fig.~\ref{fig: LE topology}, again at $2\times10^{5}$ shots per layer and one fit per layer. Two acquisitions were taken,
differing only in the Floquet depth, $t = 1$ and $t = 2$.

The results are collected in the bottom block of Fig.~\ref{fig: tn-theta-appendix}, panels (e), (g) and (i) at $t = 1$ and
(f), (h) and (j) at $t = 2$. Unlike the $2\times5$ lattice, where the prediction sits below the measurement at every angle, here the residual changes sign along the scan. On $\tilde S_\theta[\Pi_0]$ the
model is below the measurement at $\theta = 0$ and above it at $\theta = 0.3$ at both depths: $0.2554$ against a measured $0.2667 \pm 0.0025$ and $0.1658$ against $0.1566 \pm 0.0017$ at $t = 1$, and $0.0683$ against $0.0817 \pm 0.0011$ and $0.0451$ against $0.0430 \pm 0.0008$ at $t = 2$. On $M_z$ the model is above the measurement throughout, by $0.6$--$4.3\%$ at $t = 1$ and by $7.2$--$17.6\%$ at $t = 2$, where it gives $0.4298$ against a measured $0.4009 \pm 0.0016$ at $\theta = 0$.

\end{document}